\documentclass[%
 reprint,
 amsmath,amssymb,
 aps
prb,
]{revtex4-2}

\usepackage{graphicx}
\usepackage{dcolumn}
\usepackage{bm}
\usepackage{bbm}
\usepackage{comment}
\usepackage{color}
\usepackage{hyperref}
\usepackage{diagbox}
\usepackage{tabularx}
\usepackage{physics}

\begin{document}

\title{Vertex corrections to conductivity in the Holstein model:\\A numerical--analytical study}
\author{Veljko Jankovi\'c}
 \email{veljko.jankovic@ipb.ac.rs}
 \affiliation{Institute of Physics Belgrade, University of Belgrade, Pregrevica 118, 11080 Belgrade, Serbia}
\author{Petar Mitri\'c}
 \email{petar.mitric@ipb.ac.rs}
 \affiliation{Institute of Physics Belgrade, University of Belgrade, Pregrevica 118, 11080 Belgrade, Serbia}
\author{Darko Tanaskovi\'c}
 \email{darko.tanaskovic@ipb.ac.rs}
 \affiliation{Institute of Physics Belgrade, University of Belgrade, Pregrevica 118, 11080 Belgrade, Serbia}
 \author{Nenad Vukmirovi\'c}
 \email{nenad.vukmirovic@ipb.ac.rs}
 \affiliation{Institute of Physics Belgrade, University of Belgrade, Pregrevica 118, 11080 Belgrade, Serbia}

\begin{abstract}
The optical-conductivity profile carries information on electronic dynamics in interacting quantum many-body systems.
Its computation is a formidable task that is usually approached by invoking the single-particle (bubble) approximation and neglecting the vertex corrections, the importance of which remains elusive even in model Hamiltonian calculations.
Here, we combine analytical arguments with our recent breakthroughs in numerically exact and approximate calculations of finite-temperature real-time correlation functions to thoroughly assess the importance of vertex corrections in the one-dimensional Holstein polaron model.
We find, both analytically and numerically, vanishing vertex corrections to optical conductivity in the limits of zero electron--phonon interaction, zero electronic bandwidth, and infinite temperature.
Furthermore, our numerical results show that vertex corrections to the electron mobility also vanish in many parameter regimes between these limits.
In some of these cases, the vertex corrections still introduce important qualitative changes to the optical-conductivity profile in comparison to the bubble approximation even though the self-energy remains approximately local.
We trace these changes back to the bubble approximation not fully capturing a time-limited slow-down of the electron on intermediate time scales between ballistic and diffusive transport.
We find that the vertex corrections are overall most pronounced for intermediate electron--phonon interaction and may increase or decrease the bubble-approximation mobility depending on the values of model parameters.
\end{abstract}

\maketitle

\section{Introduction}
Charge carrier transport in semiconducting materials is the key physical process behind the operation of many semiconductor electronic and optoelectronic devices~\cite{Rossibook,Jacoboni}.
Under typical operating conditions, the carrier density is low, and their transport is limited by the interaction with phonons~\cite{NatRevMater.6.560,AdvMater.33.2007057}.
Quantifying the ability of a carrier to cover long distances, phonon-limited electron (dc) mobility is the primary factor determining device performance.
On the other hand, the frequency profile of the dynamical (ac) mobility, which is proportional to the optical conductivity, carries information on charge dynamics on various time and length scales, thus providing fundamental insights into the mechanisms of carrier transport~\cite{RevModPhys.83.543,JPhysChemC.116.19665,AdvOpticalMater.8.1900623}.
While the dc and ac mobility are nowadays experimentally accessible~\cite{RevModPhys.83.543}, reliable theoretical results for these quantities are rather scarce.

Within the linear-response theory~\cite{Kubo-noneq-stat-mech-book}, transport properties are encoded in the two-particle current--current correlation function, which is, however, seldom calculated exactly.
One usually calculates it in the so-called independent-particle (single-particle or bubble) approximation~\cite{RepProgPhys.83.036501,Mahanbook}, which expresses it entirely in terms of single-particle quantities (the so-called bubble term), and neglects two-particle correlations (commonly referred to as vertex corrections).
The bubble term is, in principle, much easier to evaluate than the vertex corrections, and
the bubble approximation is commonly employed both in first-principles studies of transport properties~\cite{RepProgPhys.83.036501,PhysRevResearch.1.033138,npjComputMater.8.63} and in many model Hamiltonian calculations~\cite{PhysRevB.63.153101,PhysRevLett.91.256403,PhysRevB.74.075101,PhysRevLett.103.266601,arxiv.2212.13846}.

Understanding how the vertex corrections influence transport properties is an arduous task whose solution has been attempted in just a few instances.
Historically, the first instance is impurity scattering in metals, where the Green's functions in the Born approximation are used for the calculation of the ladder diagrams, whose contribution to the dc conductivity is comparable to the bubble term~\cite{Mahanbook,Bruus-Flensberg-book}.
These diagrams are responsible for the dominant contribution to dc resistivity from the large-angle scattering within the semiclassical Boltzmann approach in the presence of diluted impurities.
Another well-studied case is the disorder scattering in two dimensions, where the maximally-crossed diagrams give a divergent contribution to resistivity at low temperatures even in the presence of weak disorder~\cite{Rammer-book,JETPLetters.30.228}.
Concerning the electron--phonon models, the contribution of vertex corrections to conductivity~\cite{AnnPhys.20.157,AnnPhys.29.410,SovPhysJETP.14.886,PhysRev.133.A1070,PhysRev.142.366} has been studied so far only for weak interaction.
In classic papers from the 1960s, vertex corrections to conductivity stemming from scattering on acoustic phonons were calculated by summing the ladder diagrams~\cite{PhysRev.142.366}.
Within the Fr\"{o}hlich model, vertex corrections originating from scattering on optical phonons were shown to be negligible at low temperatures and for weak electron--phonon coupling~\cite{PhysRev.133.A1070,PhysRev.142.366}.
It is a challenge to determine the importance of vertex corrections outside the weak-coupling limit and, to the best of our knowledge, this has not been done before either for the Fröhlich model or for the Holstein model, on which we focus in this study.

Transport properties of interacting electron--phonon models have been commonly studied either by approximately calculating the current--current correlation function~\cite{jcp.128.114713,PhysRevB.79.235206,PhysRevB.69.064302,PhysRevLett.107.076403,JPhysChemLett.5.1335,JChemPhys.142.174103,AnnPhys.391.183,PhysRevB.99.104304,PhysRevX.10.021062} or by starting from (possibly approximate) single-particle spectral functions~\cite{PhysRevB.56.4494,PhysRevLett.97.036402,PhysRevB.74.075101,PhysRevB.105.224304,PhysRevB.105.224305,PhysRevLett.129.096401,arxiv.2212.13846} and using the bubble approximation~\cite{PhysRevB.63.153101,PhysRevLett.91.256403,PhysRevB.74.075101,PhysRevLett.103.266601,arxiv.2212.13846}.
However, elucidating the role of vertex corrections requires genuine numerically exact approaches that provide results on both single-particle and two-particle correlation functions at finite temperatures.
While imaginary-axis quantum Monte Carlo (QMC) approaches are formulated directly in the thermodynamic limit~\cite{PhysRevLett.114.086601,PhysRevLett.114.146401,PhysRevLett.123.076601,JChemPhys.156.204116}, the uncertainties associated with the procedure of numerical analytical continuation~\cite{JMathPhys.2.232,JLowTempPhys.29.179,PhysRevLett.75.517,PhysRep.269.133,PhysRevB.82.165125} cast doubts on the reliability of real-axis results thus obtained.
It is therefore of paramount importance that numerically exact methods used to study vertex corrections be formulated directly on the real-time or real-frequency axis~\cite{PhysRevLett.123.036601}.
Calculations of the dc mobility are particularly challenging for such methods because of finite-size effects (e.g., Lanczos diagonalization-based methods~\cite{PhysRevB.100.094307,PhysRevB.103.054304,PhysRevB.106.174303}) or maximum simulation time that is not sufficiently long to fully capture the carrier's diffusive motion (e.g., real-time QMC~\cite{PhysRevB.107.184315} or density matrix renormalization group~\cite{PhysRevB.60.14092,PhysRevB.106.155129}).

There have been three very recent advances that facilitate our present study.
First, we developed the numerically exact momentum-space hierarchical equations of motion (HEOM) method for calculating both single- and two-particle correlation functions~\cite{PhysRevB.105.054311,JChemPhys.159.094113}.
This method provides dc mobilities whose uncertainties due to finite-size effects are controllable and can be suppressed in a wide range of model parameters.
Second, we developed the real-axis path integral QMC method, which can provide a real-time current--current correlation function for weak and intermediate electron--phonon coupling at elevated temperatures~\cite{PhysRevB.105.054311,PhysRevB.107.184315}.
For lower temperatures or stronger coupling, valuable information can still be obtained for short time correlations, before the sign problem sets in.
Finally, we also have at our disposal the dynamical mean-field theory (DMFT), a computationally inexpensive method, producing close-to-exact results for the spectral functions of the Holstein polaron in the thermodynamic limit, even in one dimension, in the whole parameter space~\cite{PhysRevLett.129.096401}.
This can be used for reliable calculations of conductivity without vertex corrections, as all the results to be presented demonstrate that the DMFT ac mobility practically coincides with the one calculated in the bubble approximation within the HEOM method.

In this study, we first develop analytical arguments demonstrating that the vertex corrections vanish in the limits of vanishing electron--phonon interaction, vanishing electronic bandwidth (the atomic limit), and infinite temperature.
We then proceed to numerically analyze their importance in parameter regimes between these limits for three values of phonon energy (intermediate, low, and high with respect to the bare electron's kinetic energy) and at temperatures that are not too low for HEOM (to minimize finite-size effects) and QMC (to avoid severe sign problem) calculations.
For intermediate phonon frequency and moderate electron--phonon coupling, the numerically exact dynamical mobility assumes a two peak structure: the Drude-like peak is accompanied by another peak at finite frequency.
At higher temperatures, this peak is centered away from $\omega_0$, and it can be ascribed to a temporally limited slow-down of the carrier during the crossover between the ballistic and diffusive transport regimes.
This slow-down is not fully captured by the bubble approximation, and the corresponding dynamical-mobility profile features only the Drude-like peak.
On the other hand, at low temperatures, the finite-frequency peak is positioned exactly at $\omega_0$, and it is recovered also within the bubble approximation because it corresponds to the optical transitions between the quasiparticle and the satellite peaks in a single-particle spectral function.
In all these cases, the numerically exact dc mobility is somewhat larger than that in the bubble approximation.
For low phonon frequency and moderate interaction, we find that the dynamical-mobility profile bears qualitative similarities to that for intermediate phonon frequency.
Interestingly, the peak at zero frequency persists, and the numerically exact dc mobility is somewhat smaller than, yet comparable to, the one in the bubble approximation.
Only as $\omega_0$ is further decreased, is the dc mobility expected to decrease to the values much below the bubble-approximation result.
For high phonon frequency, available HEOM results do not indicate a large discrepancy in comparison to DMFT.

The paper is structured as follows.
Section~\ref{Sec:Model_method} provides an overview of the Holstein model and methods we use to study its transport properties.
Section~\ref{Sec:Analytical_insights} exposes the readers to the analytical results and illustrative numerical examples demonstrating vanishing vertex corrections in the above-listed limiting cases.
The in-depth analytical arguments are deferred to Appendices~\ref{App:g_to_0}--\ref{App:beta_to_0}.
We analyze our numerical results and present our main findings on the importance of vertex corrections in Sec.~\ref{Sec:Numerical_results}.
Section~\ref{Sec:Conclusions} summarizes our results.

\section{Model and methods}\label{Sec:Model_method}
\subsection{Formalism: Holstein model and its transport properties}
We examine the Holstein model on a one-dimensional (1D) lattice composed of $N$ sites with periodic boundary conditions.
In momentum space, its Hamiltonian reads
\begin{equation}
\label{Eq:def_H}
\begin{split}
    H&=H_\mathrm{e}+H_\mathrm{ph}+H_\mathrm{e-ph}\\&=\sum_k\varepsilon_k c_k^\dagger c_k+\omega_0\sum_q b_q^\dagger b_q\\&+\frac{g}{\sqrt{N}}\sum_{kq} c_{k+q}^\dagger c_k \left(b_q+b_{-q}^\dagger\right).
\end{split}
\end{equation}
The electronic and phononic wave numbers $k$ and $q$ may assume any of the $N$ allowed values $2\pi n/N$ ($n$ is an integer) in the first Brillouin zone $(-\pi,\pi]$.
The free-electron Hamiltonian $H_\mathrm{e}$ describes electrons in a band whose dispersion $\varepsilon_k=-2t_0\cos k$ originates from the nearest-neighbor hopping of amplitude $t_0$.
The operator $c_k^\dagger$ ($c_k$) creates (annihilates) an electron in the state with wave number $k$.
The free-phonon Hamiltonian $H_\mathrm{ph}$ describes dispersionless optical phonons of frequency $\omega_0$, with $b_q^\dagger$ ($b_q$) creating (annihilating) a phonon of momentum $q$.
The interaction term $H_{\mathrm{e-ph}}$ is characterized by its strength $g$. 
In the following, we set the lattice constant $a_l$, the elementary charge $e_0$, and physical constants $\hbar$, and $k_B$ to unity.
As a convenient measure of the electron--phonon interaction strength, we use the dimensionless parameter
\begin{equation}
    \lambda=\frac{g^2}{2t_0\omega_0}.
\end{equation} 

We consider the dynamics of a single spinless electron in the band, which is determined by the current--current correlation function (normalized to the electron number)
\begin{equation}
\label{Eq:def_C_jj_t}
    C_{jj}(t)=\frac{\langle j(t)j(0)\rangle_K}{\langle N_\mathrm{e}\rangle_K}.
\end{equation}
In Eq.~\eqref{Eq:def_C_jj_t}, $N_\mathrm{e}=\sum_k c_k^\dagger c_k$ denotes the electron-number operator, and the expectation values are evaluated in the grand-canonical ensemble defined by temperature $T=\beta^{-1}$ and chemical potential $\mu_\mathrm{F}$ ($K=H-\mu_\mathrm{F}N_\mathrm{e}$)
\begin{equation}
    \langle A\rangle_K=\frac{\mathrm{Tr}\{Ae^{-\beta K}\}}{\mathrm{Tr}\:e^{-\beta K}}.
\end{equation}
The current operator reads as
\begin{equation}
\label{Eq:def_j}
    j=\sum_k j_k c_k^\dagger c_k,
\end{equation}
with
\begin{equation}
\label{Eq:j_k}
    j_k=-2t_0\sin k.
\end{equation}
We assume that $\mu_\mathrm{F}$ lies far below the bottom of the band (formally, $\mu_\mathrm{F}\to-\infty$), i.e., the electron density is low.
A reasoning analogous to that in Refs.~\cite{CanJPhys.53.321,PhysRevB.105.224304,PhysRevLett.129.096401} shows that the dominant contributions to the expectation values entering Eq.~\eqref{Eq:def_C_jj_t} as $\mu_\mathrm{F}\to-\infty$ read as
\begin{equation}
    \langle j(t)j(0)\rangle_K=e^{\beta\mu_\mathrm{F}}\frac{\mathrm{Tr}_{1\mathrm{e}}\{e^{iHt}je^{-iHt}je^{-\beta H}\}}{Z_\mathrm{ph}},
\end{equation}
\begin{equation}\label{eq:NeK}
    \langle N_\mathrm{e}\rangle_K=e^{\beta\mu_\mathrm{F}}\frac{Z}{Z_\mathrm{ph}},
\end{equation}
where $Z_\mathrm{ph}=\mathrm{Tr}_\mathrm{ph}\:e^{-\beta H_\mathrm{ph}}$ denotes the free-phonon partition sum, the trace $\mathrm{Tr}_{1\mathrm{e}}$ is taken over states containing a single electron (and an arbitrary number of phonons), while
\begin{equation}
 Z=\mathrm{Tr}_{1\mathrm{e}}e^{-\beta H}   
\end{equation}
is the corresponding partition sum.
Equation~\eqref{Eq:def_C_jj_t} is then recast as
\begin{equation}
\label{Eq:C_jj_t_low_density_limit}
    C_{jj}(t)=\langle j(t)j(0)\rangle_{H,1}=\frac{\mathrm{Tr}_{1\mathrm{e}}\{e^{iHt}je^{-iHt}je^{-\beta H}\}}{Z}.
\end{equation}
The real part of the frequency-dependent mobility (for $\omega\neq 0$) is~\cite{Mahanbook}
\begin{equation}
\label{Eq:def_re_mu_ac_omega}
    \mathrm{Re}\:\mu(\omega)=\frac{1-e^{-\beta\omega}}{2\omega}\int_{-\infty}^{+\infty} dt\:e^{i\omega t}C_{jj}(t),
\end{equation}
and the corresponding dc mobility is
\begin{equation}
\label{Eq:mu_dc}
\begin{split}
    \mu_\mathrm{dc}&=\frac{1}{T}\int_0^{+\infty}dt\:\mathrm{Re}\:C_{jj}(t)\\
    &=-2\int_0^{+\infty}dt\:t\:\mathrm{Im}\:C_{jj}(t).
\end{split}
\end{equation}
While the dynamical-mobility profile encodes information on carrier dynamics on all time and length scales, a more intuitive understanding of carrier transport can be gained from the evolution of the carrier's spread,
\begin{equation}
    \Delta x(t)=\sqrt{\left\langle[x(t)-x(0)]^2\right\rangle_{H,1}},
\end{equation}
where $x$ is the carrier position operator.
The growth rate of the spread is determined by the time-dependent diffusion constant
\begin{equation}
    \mathcal{D}(t)=\frac{1}{2}\frac{d}{dt}[\Delta x(t)]^2=\int_0^t ds\:\mathrm{Re}\:C_{jj}(s),
\end{equation}
which varies from $0$ at short times to its long-time limit $\mathcal{D}_\infty$, for which the Einstein relation $\mathcal{D}_\infty=\mu_\mathrm{dc}T$ holds.
The carrier's dynamics then changes from short-time ballistic dynamics, when $\Delta x(t)\propto t$, to long-time diffusive dynamics, when $\Delta x(t)\propto\sqrt{t}$.

\subsection{Single-particle (bubble) approximation}
$C_{jj}(t)$ is a four-point (two-particle) correlation function, which can be expressed as [combine Eqs.~\eqref{Eq:def_C_jj_t} and~\eqref{Eq:def_j}]:
\begin{equation}
\label{Eq:C_jj_two_particle_cf}
    C_{jj}(t)=\frac{\sum_{k'k}j_{k'}j_k\left\langle c_{k'}^\dagger(t)c_{k'}(t)c_k^\dagger(0)c_k(0)\right\rangle_K}{\left\langle N_\mathrm{e}\right\rangle_K}.
\end{equation}
Its evaluation in the most general many-body setup is a formidable task.
This remains true even when we limit ourselves to a single electron in the system [Eq.~\eqref{Eq:C_jj_t_low_density_limit}], which represents an important case that has been successfully solved only very recently using the HEOM formalism~\cite{JChemPhys.156.244102,JChemPhys.159.094113}.

Quite generally~\cite{Kirabook}, the two-particle correlation function in Eq.~\eqref{Eq:C_jj_two_particle_cf} can be formally decomposed into the sum of products of two single-particle correlation functions plus a remainder containing genuine two-particle correlations (denoted as $\Delta_2$):
\begin{equation}
\label{Eq:cluster_expansion_jj}
\begin{split}
    C_{jj}(t)=\frac{1}{\left\langle N_\mathrm{e}\right\rangle_K}\sum_{k'k}j_{k'}j_k\left\{
    \left\langle c_{k'}^\dagger(t)c_{k'}(t)\right\rangle_K\left\langle c_k^\dagger(0)c_k(0)\right\rangle_K+ \right. \\ \left.
    \delta_{k'k}\left\langle c_{k}^\dagger(t)c_{k}(0)\right\rangle_K\left\langle c_{k}(t)c_{k}^\dagger(0)\right\rangle_K+ \right. \\ \left.
    \Delta_2\left[\left\langle c_{k'}^\dagger(t)c_{k'}(t)c_k^\dagger(0)c_k(0)\right\rangle_K\right]\right\}.
\end{split}
\end{equation}
The Kronecker delta in the second term on the RHS of Eq.~\eqref{Eq:cluster_expansion_jj} comes from momentum conservation.
The first term on the RHS of Eq.~\eqref{Eq:cluster_expansion_jj} is $O(e^{\beta\mu_\mathrm{F}})$, and is thus negligible in the $\mu_\mathrm{F}\to-\infty$ limit with respect to the remaining two terms, which are both $O(1)$.
The single-particle (or bubble) approximation additionally neglects the $\Delta_2$ term, so that the current--current correlation function in this approximation reads
\begin{equation}
\label{Eq:C_jj_bubble}
    C_{jj}^\mathrm{bbl}(t)=-\frac{1}{\langle N_\mathrm{e}\rangle_K}\sum_k j_k^2\mathcal{G}^>(k,t)\mathcal{G}^<(k,t)^*.
\end{equation}
Here, the greater and lesser single-particle Green's functions read
\begin{equation}
\label{Eq:G_gtr}
\begin{split}
    \mathcal{G}^>(k,t)&=-i\langle c_k(t)c_k^\dagger(0)\rangle_K\\
    &=-i\int_{-\infty}^{+\infty}d\omega\:e^{-i\omega t}\frac{\mathcal{A}(k,\omega)}{1+e^{-\beta\omega}},
\end{split}
\end{equation}
\begin{equation}
\label{Eq:G_less}
\begin{split}
    \mathcal{G}^<(k,t)&=i\langle c_k^\dagger(0)c_k(t)\rangle_K\\
    &=i\int_{-\infty}^{+\infty}d\omega\:e^{-i\omega t}\frac{\mathcal{A}(k,\omega)}{e^{\beta\omega}+1}.
\end{split}
\end{equation}
The first equalities in Eqs.~\eqref{Eq:G_gtr} and~\eqref{Eq:G_less} are the textbook definitions, while the second equalities use the fluctuation--dissipation theorem~\cite{Bruus-Flensberg-book} to express $\mathcal{G}^{>/<}$ in terms of the spectral function $\mathcal{A}(k,\omega)$, which is normalized so that $\int_{-\infty}^{+\infty}d\omega\:\mathcal{A}(k,\omega)=1$.
In the limit $\mu_\mathrm{F}\to-\infty$, we can ensure that the spectral weight occurs at finite frequencies by defining $A(k,\omega)=\mathcal{A}(k,\omega-\mu_\mathrm{F})$; see the Supplemental Material of Ref.~\onlinecite{PhysRevLett.129.096401}.
Equations~\eqref{Eq:G_gtr} and~\eqref{Eq:G_less} then become (we exploit $\mu_\mathrm{F}\to-\infty$)
\begin{equation}
\label{Eq:G_gtr_mu_F_neg_lrg}
    \mathcal{G}^>(k,t)=-ie^{i\mu_\mathrm{F}t}\int_{-\infty}^{+\infty} d\omega\:e^{-i\omega t} A(k,\omega),
\end{equation}
\begin{equation}
\label{Eq:G_less_mu_F_neg_lrg}
    \mathcal{G}^<(k,t)=ie^{\beta\mu_\mathrm{F}}e^{i\mu_\mathrm{F}t}\int_{-\infty}^{+\infty} d\omega\:e^{-i\omega t}e^{-\beta\omega}A(k,\omega).
\end{equation}
Remembering that $\langle N_\mathrm{e}\rangle_K=-i\sum_k \mathcal{G}^<(k,t=0)$, and performing the Fourier transformation of Eq.~\eqref{Eq:C_jj_bubble}, we obtain the well-known results~\cite{Mahanbook,PhysRevB.63.153101,arxiv.2212.13846} for the dynamical mobility
\begin{equation}
\begin{split}
    &\mathrm{Re}\:\mu^\mathrm{bbl}(\omega)=4\pi t_0^2\frac{1-e^{-\beta\omega}}{\omega}\times\\&\frac{\sum_k \sin^2 k\int_{-\infty}^{+\infty} d\nu\:e^{-\beta\nu}A(k,\omega+\nu)A(k,\nu)}{\sum_k \int_{-\infty}^{+\infty} d\nu\:e^{-\beta\nu}A(k,\nu)},
\end{split}
\end{equation}
and the dc mobility
\begin{equation}
    \mu_\mathrm{dc}^\mathrm{bbl}=\frac{4\pi t_0^2}{T}\frac{\sum_k \sin^2 k\int_{-\infty}^{+\infty} d\nu\:e^{-\beta\nu}A(k,\nu)^2}{\sum_k \int_{-\infty}^{+\infty} d\nu\:e^{-\beta\nu}A(k,\nu)}
\end{equation}
in the bubble approximation.
The computation of transport properties in the bubble approximation thus reduces to the computation of the carrier's spectral function $A(k,\omega)$.

\subsection{Hierarchical equations of motion}
\label{SSec:HEOM}
The HEOM method is a numerically exact density-matrix technique providing access to the dynamics of the system of interest (here, electrons) that is linearly coupled to a collection of harmonic oscillators (here, phonons)~\cite{JChemPhys.153.020901,PhysRevE.75.031107}.
The method has been recently extended to computations of various real-time finite-temperature correlation functions of the operators acting on the system of interest~\cite{PhysRevLett.109.266403,JChemPhys.142.174103,JChemPhys.143.194106,JChemPhys.156.244102,bhattacharyya2023anomalous}.
The method is ultimately based on the formally exact results of the Feynman--Vernon influence functional theory~\cite{AnnPhys.24.118} (though the details do depend on the correlation function).
In Appendix~\ref{App:influence_phases}, we summarize such formally exact results for the current--current correlation function [Eq.~\eqref{Eq:C_jj_t_low_density_limit}]~\cite{JChemPhys.159.094113} and the Green's function [Eqs.~\eqref{Eq:G_gtr_mu_F_neg_lrg} and~\eqref{Eq:G_less_mu_F_neg_lrg}]~\cite{PhysRevB.105.054311} of the Holstein model.
These results can serve as a convenient starting point for analytical studies in various limits; see Sec.~\ref{Sec:Analytical_insights}.
The actual computations are, however, performed numerically, by recasting the formally exact result as a hierarchy of dynamical equations for the correlation function we consider (the hierarchy root) and auxiliary quantities needed to fully take the interactions into account (deeper hierarchy layers).
The hierarchy is, in principle, infinite, and it has to be truncated at a certain maximum depth $D$.

The applications of the HEOM method to the Holstein model featuring a single oscillator per site~\cite{JPhysChemLett.6.3110} have been hindered by the numerical instabilities of the truncated hierarchy~\cite{JChemPhys.150.184109,JChemPhys.153.204109}, which ultimately stem from the finite number of oscillators on a finite lattice.
Within our recently developed momentum-space HEOM~\cite{PhysRevB.105.054311}, we have resolved this issue in a wide range of the model's parameter space by devising a physically motivated hierarchy closing~\cite{JChemPhys.159.094113}.
At the same time, we have lowered the computational requirements with respect to the commonly used real-space HEOM by exploiting the model's translational symmetry.
We summarize the momentum-space HEOM for the current--current correlation function [Eq.~\eqref{Eq:C_jj_t_low_density_limit}]~\cite{JChemPhys.159.094113} and the Green's function [Eqs.~\eqref{Eq:G_gtr_mu_F_neg_lrg} and~\eqref{Eq:G_less_mu_F_neg_lrg}]~\cite{PhysRevB.105.054311} in \textcolor{red}{Sec.~SI of Ref.~\onlinecite{comment050324}}.

Numerical uncertainties in HEOM results can be due to the finite chain length $N$, finite maximum depth $D$, and finite maximum propagation time $t_\mathrm{max}$.
We found~\cite{JChemPhys.159.094113} that finite-size effects can be controlled by following the relative accuracy with which the optical sum rule $\int_{-\infty}^{+\infty}d\omega\:\mathrm{Re}\:\mu(\omega)=-\pi\langle H_\mathrm{e}\rangle_{H,1}$ is satisfied.
We concluded~\cite{JChemPhys.159.094113} that the convergence with respect to $D$ can be enhanced by taking the arithmetic average of HEOM results for two consecutive depths $D-1$ and $D$ (provided that $D$ is sufficiently large, so that the relative accuracies with which the optical sum rule is satisfied at the two depths almost coincide).
The time $t_\mathrm{max}$ should be sufficiently long, so that the integrals $T^{-1}\int_0^{t_\mathrm{max}} ds\:\mathrm{Re}\:C_{jj}(s)$ and $-2\int_0^{t_\mathrm{max}} ds\:s\:\mathrm{Im}\:C_{jj}(s)$, whose $t_\mathrm{max}\to+\infty$ limit defines $\mu_\mathrm{dc}$ [Eq.~\eqref{Eq:mu_dc}], have entered into saturation as a function of $t_\mathrm{max}$. 
In practice, we always choose $N,D,$ and $t_\mathrm{max}$ that are sufficiently large so that: (i) the optical sum rule is satisfied with relative accuracy $\lesssim 10^{-4}$, and (ii) the relative difference between the values of $\mu_\mathrm{dc}$ obtained using the two expressions in Eq.~\eqref{Eq:mu_dc} is $\lesssim 0.1$.
Based on (ii), we estimate that the relative uncertainty of HEOM results for the dc mobility is $\lesssim 10\%$.

For stronger $g$ or at higher $T$, we generally need smaller $N$, shorter $t_\mathrm{max}$, and larger $D$.
However, it is difficult to give an \emph{a priori} estimate of $N,D,$ and $t_\mathrm{max}$ based on the values of model parameters.
As an illustration, we typically use $N\sim 100,D=2-3,$ and $\omega_0t_\mathrm{max}\gtrsim 500$ for small $g$ and $T$, $N\sim 10,D\sim 7,$ and $\omega_0t_\mathrm{max}\simeq 300$ at intermediate $g$ and $T$, and $N\leq 7,D\geq 12,$ and $\omega_0t_\mathrm{max}\lesssim 100$ for large $g$ and $T$.
The HEOM results to be presented are publicly available as a dataset~\cite{veljko.jankovic.2023}.

\subsection{Real-time quantum Monte Carlo}
We also employ path-integral QMC to evaluate the numerically exact current--current correlation function $C_{jj}\qty(t)$, and the same quantity within the bubble approximation $C_{jj}^\mathrm{bbl}\qty(t)$. This method can produce reliable results for imaginary times and for real times that are not too long. For longer real times, the statistical error of the Monte Carlo procedure becomes very large due to the dynamical sign problem and the results could not be obtained. The results obtained using QMC are used to cross-check the results obtained using HEOM, for model parameters and times when both methods give results, as well as to complement HEOM results for some cases in which the results could not be obtained using HEOM.

The path-integral QMC method used in this work is in many aspects the same as the methods that we employed in Refs.~\onlinecite{PhysRevB.107.184315} and \onlinecite{PhysRevB.105.054311}. It is based on a path-integral representation
of the correlation function, where the Suzuki--Trotter expansion is used to decompose the (real- or imaginary-time) evolution operator ($e^{iHt}$ or $e^{-\beta H}$) into evolution operators over small time intervals. Unlike in Ref.~\onlinecite{PhysRevB.107.184315}, where the decomposition is performed to the operators $e^{-\beta H}$, $e^{i H t}$ and $e^{-i H t}$, here we apply it to the operators $e^{-\qty(\beta-it)H}$ and $e^{-i H t}$, which allows as to perform either real- or imaginary-time calculations using the same computational code. As in Ref.~\onlinecite{PhysRevB.107.184315}, in the path-integral representation, we make use of either the momentum or site representation for electronic single-particle states. An appropriate choice of the representation reduces the sign problem and enables the calculations for longer real time. For weaker electron-phonon interaction, the momentum representation of electronic states is more convenient in that respect, while the site representation is more convenient for stronger electron-phonon interaction.

We use QMC to calculate the quantities $C_{jj}\qty(t)$ [as defined in Eq.~\eqref{Eq:def_C_jj_t}], $\mathcal{G}^>(k,t)$ [as defined in Eq.~\eqref{Eq:G_gtr}], $\mathcal{G}^<(k,t)$ [see Eq.~\eqref{Eq:G_less}], and $\langle N_\mathrm{e}\rangle_K$ [see Eq.~\eqref{eq:NeK}]. With these quantities at hand, we can then also evaluate $C_{jj}^\mathrm{bbl}\qty(t)$ using Eq.~\eqref{Eq:C_jj_bubble}.

\subsection{Dynamical mean-field theory}

The DMFT is an approximate, yet nonperturbative method that can treat the models with local interactions~\cite{RevModPhys.68.13}.
The DMFT establishes a mapping between the lattice problem and the impurity problem, supplemented with a self-consistency condition.
For the Holstein model, the polaron impurity model can be efficiently solved in a form of the continued fraction expansion~\cite{PhysRevB.56.4494}.
While this mapping is exact in the infinite-dimensional limit, it remains applicable in the finite-dimensional case as well, yielding approximate results characterized by momentum-independent self-energy.
It was recently shown that this method yields remarkably accurate single-particle properties of the Holstein polaron, regardless of the dimensionality of the system, while demanding minimal computational resources~\cite{PhysRevLett.129.096401}.
Therefore, it can also be used for the calculation of the optical conductivity within the bubble approximation.

\section{Analytical insights into the limits with vanishing vertex corrections}\label{Sec:Analytical_insights}
Using different analytical arguments, this section identifies the limits in which vertex corrections to conductivity vanish.
The main analytical results accompanied with numerical examples are briefly summarized in Secs.~\ref{SSec:analytics_g_to_0}--\ref{SSec:beta_to_0}, while the corresponding technical details are provided in Appendixes~\ref{App:g_to_0}--\ref{App:beta_to_0}.

\subsection{Limit of vanishing electron--phonon interaction ($g\to 0$)}\label{SSec:analytics_g_to_0}

\begin{figure}[htbp!]
    \centering
    \includegraphics[width=\columnwidth]{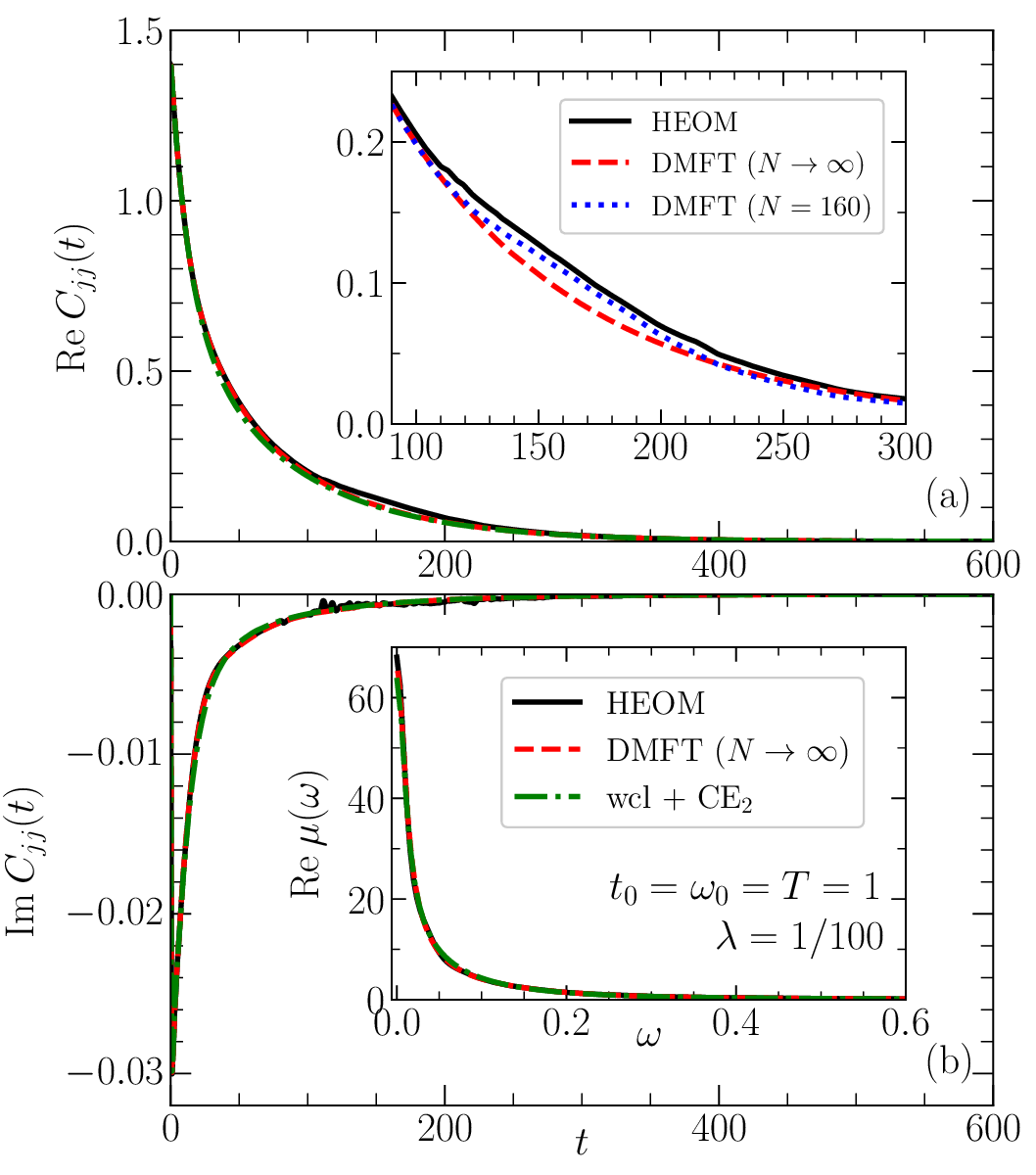}
    \caption{Time dependence of the (a) real and (b) imaginary part of $C_{jj}$ computed using HEOM (solid line), DMFT (dashed line), and Eqs.~\eqref{Eq:C_jj_CE_2} and~\eqref{Eq:Z_e_CE_2} (dash-dotted line, label "wcl+CE$_2$"). HEOM computations use $N=160$ and $D=2$, wcl+CE$_2$ computations use $N=1009$, whereas the DMFT results are in the thermodynamic limit ($N\to\infty$).
    The inset of panel (b) shows the dynamical mobility obtained using the above approaches.
    The inset of panel (a) compares the DMFT result for $N\to\infty$ with the HEOM and finite-chain DMFT results, both of which use $N=160$.
    The model parameters are $t_0=1,\omega_0=1,\lambda=1/100,$ and $T=1$.}
    \label{Fig:slika_wcl}
\end{figure}

In Appendix~\ref{App:g_to_0}, we first demonstrate that the lowest-order terms in the expansions of $C_{jj}(t)$ [Eq.~\eqref{Eq:C_jj_t_low_density_limit}] and $C_{jj}^\mathrm{bbl}(t)$ [Eq.~\eqref{Eq:C_jj_bubble}] in powers of small $g$ are identical.
We then use these terms to partially resum the perturbation series for $C_{jj}^\mathrm{(bbl)}(t)$ in the $g\to 0$ limit by employing the second-order cumulant expansion approach~\cite{arxiv.2212.13846,PhysRevB.105.224304,JChemPhys.142.094106}, which becomes exact in this limit.
The final expressions needed to evaluate the weak-coupling second-order cumulant result are provided in Eqs.~\eqref{Eq:C_jj_CE_2} and~\eqref{Eq:Z_e_CE_2}--\eqref{Eq:def_delta_epsilon_pm_k_q}.

Figures~\ref{Fig:slika_wcl}(a) and~\ref{Fig:slika_wcl}(b) present a numerical example supporting our analytical conclusion that the vertex corrections vanish in the $g\to 0$ limit.
We compare $C_{jj}(t)$ computed using HEOM, DMFT (in the thermodynamic limit), and the weak-coupling second-order cumulant expansion (label "wcl+CE$_2$", obtained on a long, but finite chain).
While the dynamics predicted by the cumulant method almost perfectly agrees with the DMFT result, a small hump in the HEOM result for $\mathrm{Re}\:C_{jj}(t)$ appearing around $t_0t\sim 150$ suggests that it exhibits weak finite-size effects.
This is further corroborated by the inset of Fig.~\ref{Fig:slika_wcl}(a), which shows that HEOM and DMFT results on a finite chain (as implemented in Ref.~\onlinecite{PhysRevLett.129.096401}) exhibit qualitatively (and also quantitatively) similar deviations from the infinite-chain DMFT result for $150<t_0t<200$.  
The HEOM, DMFT ($N\to\infty$), and cumulant dynamical-mobility profiles virtually coincide and assume a Drude-like shape; see the inset of Fig.~\ref{Fig:slika_wcl}(b).

\subsection{Limit of vanishing electronic coupling ($t_0\to 0$)}

One can demonstrate that the vertex corrections vanish in the limit $t_0\to 0$ by establishing the equality of the first non-zero terms in expansions of $C_{jj}(t)$ and $C_{jj}^\mathrm{bbl}(t)$ in powers of small $t_0$.
Since the current operator itself is linear in $t_0$ [Eqs.~\eqref{Eq:def_j} and~\eqref{Eq:j_k}], the lowest-order term in expansions of both $C_{jj}(t)$ and $C_{jj}^\mathrm{bbl}(t)$ as $t_0\to 0$ is of the order of $t_0^2$.
As a starting point, one can again take the formally exact expressions from which the HEOM are derived (Appendix~\ref{App:influence_phases}), in which all operators $e^{-\alpha H_\mathrm{e}}(\alpha=\beta,\pm it)$ are replaced by the unit operator.
The procedure summarized in Sec.~\textcolor{red}{SII of Ref.~\onlinecite{comment050324}} leads to [$n_\mathrm{ph}=(e^{\beta\omega_0}-1)^{-1}$]
\begin{equation}
\label{Eq:atomic_no_vertex_heom}
\begin{split}
    &C_{jj}(t)=C_{jj}^\mathrm{bbl}(t)\approx 2t_0^2\times\\
    &\exp\left\{-2\frac{g^2}{\omega_0^2}\left[(2n_\mathrm{ph}+1)-(n_\mathrm{ph}+1)e^{-i\omega_0 t}-n_\mathrm{ph}e^{i\omega_0 t}\right]\right\}.
\end{split}
\end{equation}
While this proves that the vertex corrections vanish in the limit $t_0\to 0$, the expression in Eq.~\eqref{Eq:atomic_no_vertex_heom} is periodic in real time with period $2\pi/\omega_0$.
Thus, the current--current correlation function does not decay to zero as real time goes to infinity.
Hence, one would obtain infinite dc mobility by integrating Eq.~\eqref{Eq:atomic_no_vertex_heom} over $t$.
This issue has been recognized in the literature~\cite{AnnPhys.8.343,AdvPhys.24.305,AnnPhys.132.163,JPhysChemSolids.28.581}.
To circumvent it, we find it convenient to perform the polaronic (Lang--Firsov) unitary transformation~\cite{SovPhysJETP.16.1301} of the Holstein Hamiltonian and evaluate the current--current correlation function in the $t_0\to 0$ limit in the polaronic frame.
Using the Matsubara Green's function formalism~\cite{Mahanbook}, we eventually obtain an expression for $C_{jj}(t)$ that decays as $t^{-3}$ at long times $t$, which is sufficiently fast to render the time integral of $C_{jj}(t)$, and thus the dc mobility, finite.
While we defer all the details to Appendix~\ref{App:polaronic_matsubara}, here, we only present the final result for the current--current correlation function,
\begin{equation}
\label{Eq:no_vtx_t_0_to_0}
\begin{split}
    &C_{jj}(t)=C_{jj}^\mathrm{bbl}(t)\approx\\
    &2t_0^2\frac{\beta}{t(\beta-it)\sqrt{c_0}}\frac{I_1[-2(\beta-it)\sqrt{c_0}]J_1(2t\sqrt{c_0})}{I_1(-2\beta\sqrt{c_0})}\times\\
    &\exp\left\{-2\frac{g^2}{\omega_0^2}\left[2n_\mathrm{ph}+1-(n_\mathrm{ph}+1)e^{-i\omega_0t}-n_\mathrm{ph}e^{i\omega_0t}\right]\right\},
\end{split}
\end{equation}
where $c_0$ is defined in Eq.~\eqref{Eq:def_c_l}, while $I_1$ ($J_1$) is the (modified) Bessel function of the first kind of order $1$.

\begin{figure}[htbp!]
    \centering
    \includegraphics[width=0.35\textwidth]{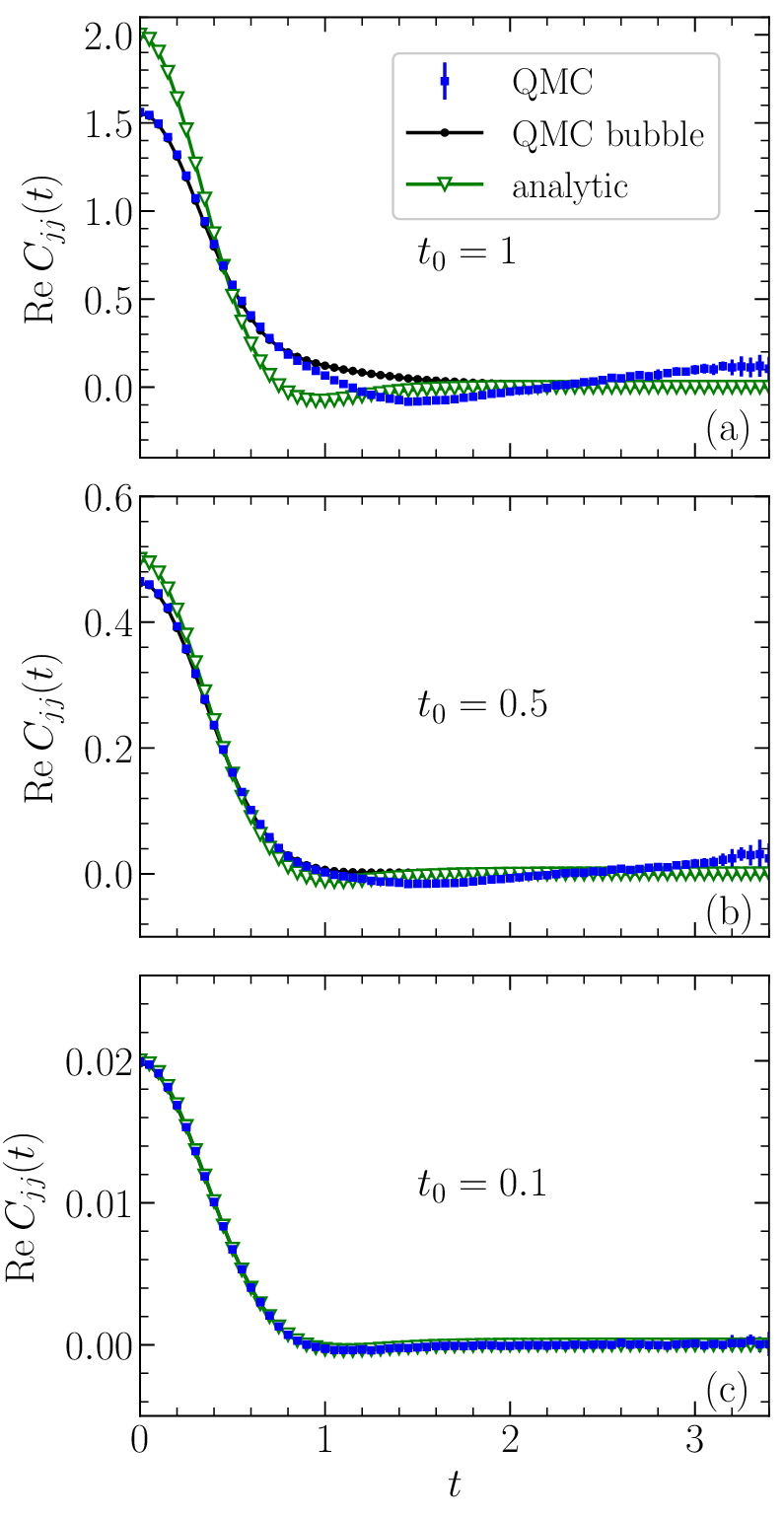}
    \caption{(a)--(c): Real part of the real time current-current correlation function for the Holstein model for different values of the parameter $t_0$. Other parameters are set to $g=1$, $\omega_0=1$, $T=1$.  The results labeled as 'QMC' were obtained using QMC simulations, the results labeled as 'bubble QMC' were obtained from QMC simulations within the bubble approximation, while the results labeled as 'analytic' were obtained using Eq.~\eqref{Eq:no_vtx_t_0_to_0}.}
    \label{Fig:fig-t0-to-0}
\end{figure}

In Figs.~\ref{Fig:fig-t0-to-0}(a)--\ref{Fig:fig-t0-to-0}(c) we present a numerical example that supports our analytical proof that vertex corrections vanish in the limit $t_0\to 0$. We present $\Re C_{jj}\qty(t)$ for $g=1$, $\omega_0=1$, $T=1$, and different values of $t_0$ calculated using QMC, QMC within the bubble approximation, and using the analytical formula given in Eq.~\eqref{Eq:no_vtx_t_0_to_0}. The results clearly demonstrate that the analytical formula and the bubble-approximation results converge towards numerically exact QMC results as $t_0$ decreases towards zero.

\subsection{Limit of infinite temperature ($\beta\to 0$)}\label{SSec:beta_to_0}

\begin{figure}[htbp!]
    \centering
    \includegraphics[width=0.49\textwidth]{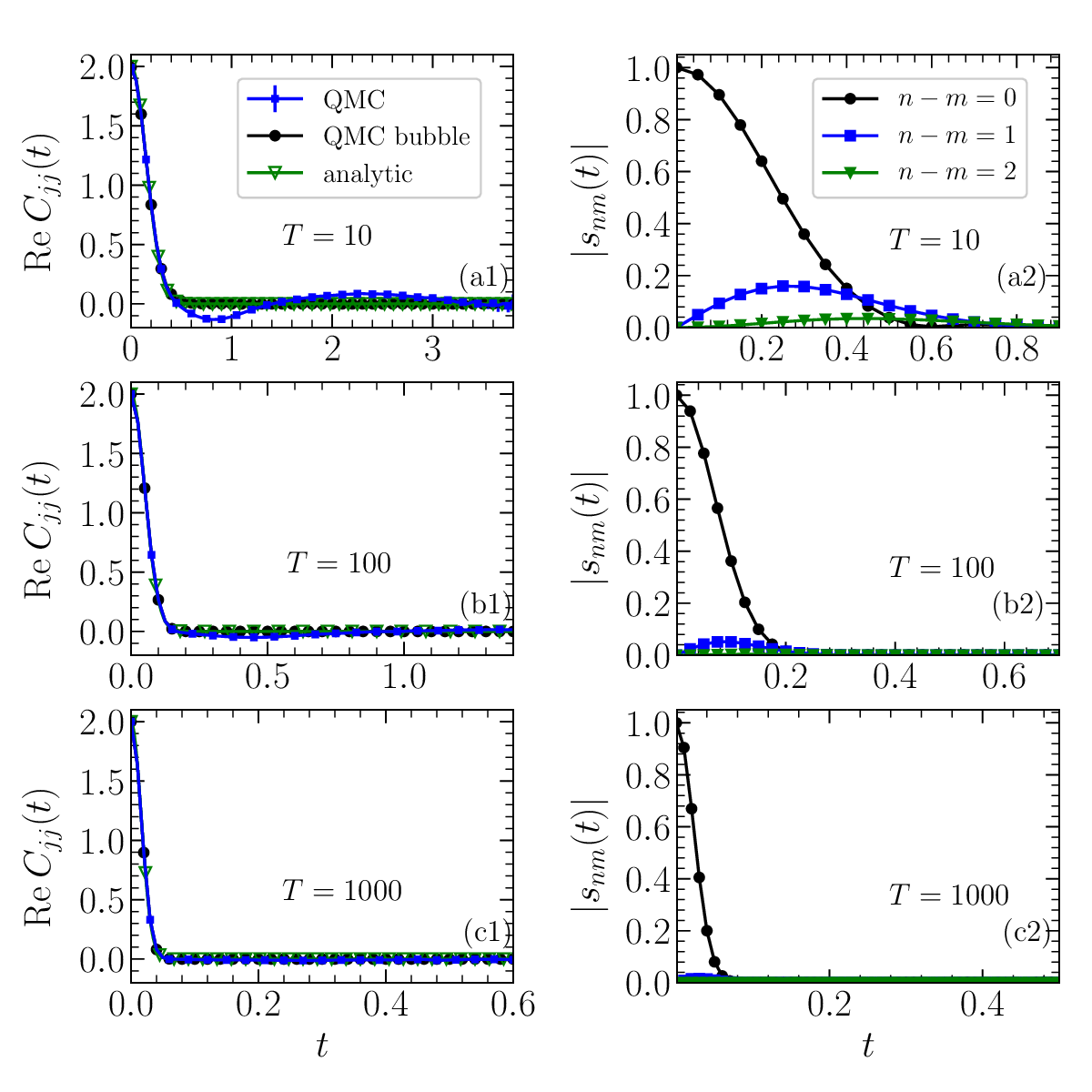}
    \vspace{-0.75cm}
    \caption{(a1)--(c1): Real part of the real time current-current correlation function for the Holstein model for different values of the temperature $T$. Other parameters are set to $g=1$, $\omega_0=1$, $t_0=1$. The results labeled as 'QMC' were obtained using QMC simulations, the results labeled as 'QMC bubble' were obtained from QMC simulations within the bubble approximation, while the results labeled as 'analytic' were obtained using Eq.~\eqref{Eq:C_jj_beta_to_0_no_vtx}.
    (a2)--(c2): Time dependence of the absolute value of the quantity $s_{nm}\qty(t)$ (defined in the text) that describes correlations between annihilation and creation operators at lattice sites $n$ and $m$. The results are presented for the same model parameters as in (a1)--(c1).
    }
    \label{Fig:beta-to-0-new}
\end{figure}

In the limit of infinite temperature, it is permissible to treat phonons as classical harmonic oscillators.
Furthermore, at sufficiently high temperatures, single-particle correlation functions become local (Fig.~\ref{Fig:beta-to-0-new} provides illustrative examples), and their dynamics becomes primarily determined by local (on-site) processes.
In Appendix~\ref{App:beta_to_0}, we derive that the exact and bubble-approximation correlation functions are identical in the $\beta\to 0$ limit:
\begin{equation}
\label{Eq:C_jj_beta_to_0_no_vtx}
    C_{jj}(t)=C_{jj}^\mathrm{bbl}(t)\approx 2t_0^2e^{-\sigma^2t^2-i\sigma^2\beta t}.
\end{equation}
Here,
\begin{equation}
\label{Eq:def_sigma_beta_to_0}
    \sigma^2=g^2\coth(\beta\omega_0/2)\approx 2g^2/(\beta\omega_0)
\end{equation}
is the variance of the thermal fluctuations in the on-site energy $\varepsilon=g(b^\dagger+b)$ evaluated in the equilibrium state $\frac{e^{-\beta H_\mathrm{ph}}}{Z_\mathrm{ph}}$ of free phonons.

We support these analytical results with a numerical example, obtained from QMC simulations, presented in Figs.~\ref{Fig:beta-to-0-new}(a1)--\ref{Fig:beta-to-0-new}(c2).
The results presented in Figs.~\ref{Fig:beta-to-0-new}(a1)--\ref{Fig:beta-to-0-new}(c1) show that the bubble approximation result and the analytical result given by Eq.~\eqref{Eq:C_jj_beta_to_0_no_vtx} converge towards the numerically exact result as $T$ increases.
In Figs.~\ref{Fig:beta-to-0-new}(a2)--\ref{Fig:beta-to-0-new}(c2) we present the absolute value of the quantity $s_{nm}\qty(t)=\expval{c_n\qty(t)c_m^\dagger}_K$, where $c_n$ ($c_m^\dagger$) is the annihilation (creation) operator for an electron at site $n$ ($m$).
This quantity was obtained from Fourier transform to real space of the quantity $\mathcal{G}^>(k,t)$ [see Eq.~\eqref{Eq:G_gtr}] and can be used as a measure of spatial correlations in the system.
It can be seen from Figs.~\ref{Fig:beta-to-0-new}(a2)--\ref{Fig:beta-to-0-new}(c2) that at higher temperatures the spatial correlations for $n-m\ne 0$ become smaller and eventually practically negligible.
This confirms the assumption of locality of the correlations used in our analytical derivation.
One should nevertheless note that unrealistically high temperatures (which would be certainly above the melting point in a real material) are needed for full vanishing of spatial correlations.

\section{Importance of vertex corrections: Analysis of numerical results}\label{Sec:Numerical_results}
Having identified the limiting cases in which the vertex corrections vanish, here we combine numerical results emerging from different methods at our disposal to analyze the importance of vertex corrections in parameter regimes between these limits.
A detailed summary of the parameter regimes examined is provided in \textcolor{red}{Table~S1 of Ref.~\onlinecite{comment050324}}.

\begin{figure*}[htbp!]
    \centering
    \includegraphics[width=\textwidth]{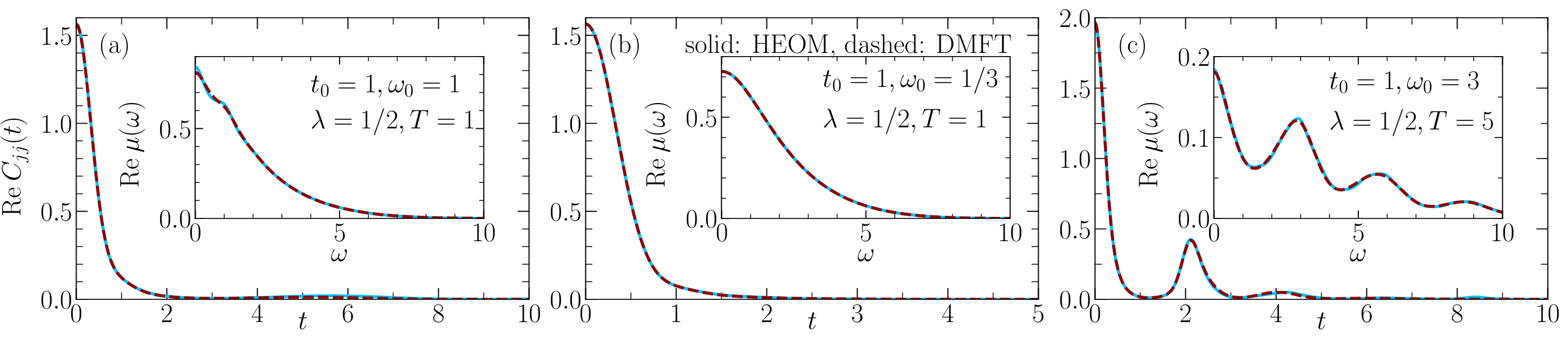}
    \vspace{-0.75cm}
    \caption{Time dependence of $\mathrm{Re}\:C_{jj}$ computed within the bubble approximation using numerically exact HEOM spectral functions (solid blue lines) and approximate DMFT spectral functions (dashed brown lines) for $t_0=1$ and (a) $\omega_0=1,T=1$; (b) $\omega_0=1/3,T=1$; (c) $\omega_0=3,T=5$. The electron--phonon interaction strength in all three panels is $\lambda=1/2$. The insets display the corresponding dynamical-mobility profiles computed within the bubble approximation using HEOM and DMFT spectral functions. HEOM spectral functions are computed using (a) $N=10,D=6$, (b) $N=10,D=8$, and (c) $N=7,D=12$.}
    \label{Fig:heom_bubble_dmft}
\end{figure*}

The numerically exact dynamics on short time scales can be computed using QMC in essentially any parameter regime.
On the other hand, the crossover from the short-time ballistic to the long-time diffusive dynamics, and thus the dynamical-mobility profile down to $\omega=0$, can be captured using the HEOM method, which in practice works best when the hierarchy closing strategy developed in Ref.~\onlinecite{JChemPhys.159.094113} is effective.
This is the case for not too strong interaction ($\lambda\lesssim 1$), at moderate temperatures ($1\lesssim T/t_0\lesssim 10$), and for $\omega_0/t_0\leq 2$.
The results of the two numerically exact methods (HEOM and QMC) will be compared to the results stemming from the bubble approximation, which, in principle, needs numerically exact single-particle properties.
While these are available from appropriate HEOM-method computations~\cite{PhysRevB.105.054311}, we have recently demonstrated that the DMFT, which is formulated directly in the thermodynamic limit, provides close-to-exact single-particle properties of the Holstein model in the whole parameter space at a much smaller computational cost~\cite{PhysRevLett.129.096401}.
The very good agreement between HEOM and DMFT spectral functions translates into the very good agreement between the current--current correlation functions and dynamical-mobility profiles computed using HEOM (within bubble approximation) and DMFT, as shown in Figs.~\ref{Fig:heom_bubble_dmft}(a)--\ref{Fig:heom_bubble_dmft}(c) for different phonon frequencies $\omega_0$.
We have checked that a similar level of agreement persists for all parameters where HEOM bubble computations are performed. 
We thus conclude that the DMFT results can be practically taken as the exact bubble-approximation results, which are available in the whole parameter space.

\subsection{Comparison of typical features of numerically exact and bubble-approximation results in time and frequency domains}

\begin{figure*}[htbp!]
    \centering
    \includegraphics[width=\textwidth]{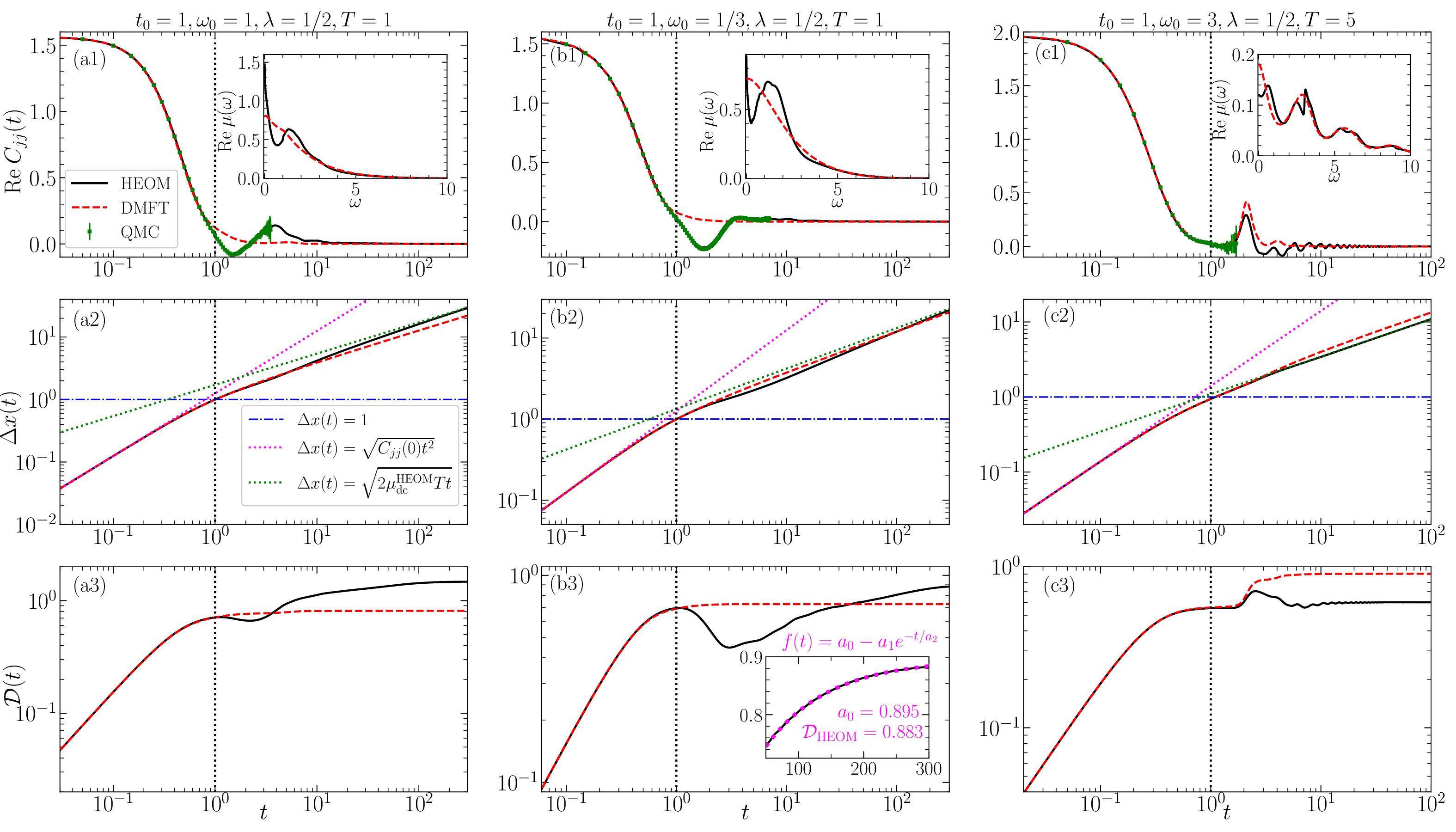}
    \vspace{-0.5cm}
    \caption{Comparison of numerically exact (labels "HEOM" and "QMC") and bubble-approximation (label "DMFT") results for time evolution of (a1)--(c1) $\mathrm{Re}\:C_{jj}$, (a2)--(c2) $\Delta x$, and (a3)--(c3) $\mathcal{D}(t)$. In all panels, $t_0=1,\lambda=1/2$, while the remaining model parameters are: (a1)--(a3) $\omega_0=1$, $T=1$, (b1)--(b3) $\omega_0=1/3$, $T=1$, (c1)--(c3) $\omega_0=3$, $T=5$. Vertical dotted lines indicate time $t=1$. The insets of (a1)--(c1) compare dynamical-mobility profiles in the numerically exact approach and in the bubble approximation. Dotted lines in (a2)--(c2) show the carrier spread in the short-time ballistic ($\Delta x_\mathrm{bal}(t)=\sqrt{C_{jj}(0)}t$) and long-time diffusive ($\Delta x_\mathrm{diff}(t)=\sqrt{2\mu_\mathrm{dc}^\mathrm{HEOM}Tt}$) regimes, while double dash-dotted lines show $\Delta x(t)=1$. The inset of panel (b3) shows the fit of $\mathcal{D}_\mathrm{HEOM}(t)$ to the exponentially decaying function $f(t)=a_0-a_1e^{-t/a_2}$ (magenta dots) for $50\leq t\leq 300$. Fitting parameters are $a_0=0.895,a_1=0.251,a_2=97.1$. HEOM results in (a1)--(a3) are obtained using $N=13,D=6$, while the results displayed in (b1)--(b3) [(c1)--(c3)] are obtained by performing the arithmetic average of HEOM results for $N=10,D=7$ and $N=10,D=8$ [$N=7,D=11$ and $N=7,D=12$]. QMC results are displayed with the associated statistical error bars and are obtained using $N=10$ in (a1), $N=7$ in (b1), and $N=10$ in (c1).}
    \label{Fig:short_t}
\end{figure*}

In Figs.~\ref{Fig:short_t}(a1)--\ref{Fig:short_t}(c3) we compare the numerically exact and bubble-approximation dynamics of $\mathrm{Re}\:C_{jj}$, the carrier spread $\Delta x$, and the diffusion constant $\mathcal{D}$, as well as the corresponding dynamical-mobility profiles.
We perform the comparison in three representative cases spanning the range from slow-phonon [$\omega_0/t_0=1/3$ in (b1)--(b3)] to intermediate-phonon [$\omega_0/t_0=1$ in (a1)--(a3)] and fast-phonon [$\omega_0/t_0=3$ in (c1)--(c3)] regimes.
We choose the intermediate electron--phonon interaction ($\lambda=1/2$), which unveils the most commonly observed differences between the numerically exact and bubble-approximation results.
\textcolor{red}{Figures~S1--S3 of Ref.~\onlinecite{comment050324}} summarize similar comparisons in other parameter regimes examined (see also \textcolor{red}{Table~S1}).

In Appendix~\ref{App:influence_phases}, we prove that the exact and bubble-approximation current--current correlation functions are identical at $t=0$ in all parameter regimes.
Figures~\ref{Fig:short_t}(a1)--\ref{Fig:short_t}(c1) additionally demonstrate that their short-time dynamics are also identical.
The very good agreement between the two dynamics persists beyond the very initial time scales, when the dynamics is ballistic so that $C_{jj,\mathrm{bal}}(t)=C_{jj}(0)$, $\Delta x_\mathrm{bal}(t)=\sqrt{C_{jj}(0)}t$, and $\mathcal{D}_\mathrm{bal}(t)=C_{jj}(0)t$.
Figures~\ref{Fig:short_t}(a2)--\ref{Fig:short_t}(c2) suggest that the numerically exact and bubble-approximation dynamics closely follow one another as long as $\Delta x(t)\lesssim 1$, i.e., the carrier spread is smaller than the lattice constant.
In parameter regimes analyzed in Fig.~\ref{Fig:short_t}, the agreement is good for $t_0t\lesssim 1$, which is the time scale characteristic for the transfer of a free electron between neighboring sites.
This translates into the very good agreement between the two dynamical-mobility profiles in the high-frequency region $\omega/t_0\gtrsim 2\pi$, see the insets of Figs.~\ref{Fig:short_t}(a1)--\ref{Fig:short_t}(c1).

On intermediate time scales, the numerically exact results in Figs.~\ref{Fig:short_t}(a) and~\ref{Fig:short_t}(b) predict a time-limited slow-down of the carrier [negative values of $\mathrm{Re}\:C_{jj}(t)$, decrease of $\mathcal{D}(t)$] that is followed by a steady increase in the diffusion constant until it saturates to its long-time limit.
On the other hand, in the regimes analyzed here, the results in the bubble approximation do not display any transient slow-down of the carrier, and the diffusion constant is a monotonically increasing function of time.
As a consequence, the dynamical-mobility profile in the bubble approximation has only the Drude-like peak centered at $\omega=0$ (qualitatively similar as in the $g\to 0$ limit analyzed in Sec.~\ref{SSec:analytics_g_to_0}), while the numerically exact dynamical-mobility profile additionally develops a finite-frequency peak.
In the regimes analyzed here, the numerically exact dynamical-mobility profile still has a local maximum at $\omega=0$.
Namely, rewriting Eq.~\eqref{Eq:def_re_mu_ac_omega} as $\mathrm{Re}\:\mu(\omega)=2\frac{\tanh(\beta\omega/2)}{\omega}\int_0^{+\infty}dt\:\cos(\omega t)\mathrm{Re}\:C_{jj}(t)$, one realizes that $\omega=0$ is a stationary point of the dynamical-mobility profile [$\mathrm{Re}\:\mu(\omega)\propto\omega^2$ as $\omega\to 0$].
The convexity of $\mathrm{Re}\:\mu(\omega)$ around $\omega=0$ then follows from the sign of the corresponding second derivative that reads as~\cite{PhysRevResearch.2.013001}
\begin{equation}
\label{Eq:character_omega_0}
\begin{split}
    &\left(\frac{d^2}{d\omega^2}\mathrm{Re}\:\mu(\omega)\right)_{\omega=0}=\\&-\beta\left\{\frac{\mathcal{D}_\infty\beta^2}{6}+2\int_0^{+\infty}dt\:t\left[\mathcal{D}_\infty-\mathcal{D}(t)\right]\right\}.
\end{split}
\end{equation}
In Figs.~\ref{Fig:short_t}(a) and~\ref{Fig:short_t}(b), the function $\mathcal{D}_\infty-\mathcal{D}(t)$ is non-negative for $t\geq 0$, and $\omega=0$ is a local maximum.
In general, the sign of $\mathcal{D}_\infty-\mathcal{D}(t)$ can change with $t$, and direct analytical arguments based on Eq.~\eqref{Eq:character_omega_0} cannot be developed.
It is then notable that our HEOM results suggest that $\omega=0$ remains a local maximum of $\mathrm{Re}\:\mu(\omega)$ in all parameter regimes amenable to HEOM computations; see \textcolor{red}{Figs.~S1--S3 of Ref.~\onlinecite{comment050324}}.

While the long-time saturation of $\mathcal{D}_\mathrm{HEOM}(t)$ is apparent in Fig.~\ref{Fig:short_t}(a3), Fig.~\ref{Fig:short_t}(b3) might suggest that the corresponding $t_\mathrm{max}$ is not sufficiently long to guarantee that the relative uncertainty of $\mathcal{D}_\mathrm{HEOM}$ (or $\mu^\mathrm{HEOM}_\mathrm{dc}=\mathcal{D}_\mathrm{HEOM}/T$) is below or of the order of the target ten-percent accuracy (Sec.~\ref{SSec:HEOM}).
To exclude this possibility, we fit $\mathcal{D}_\mathrm{HEOM}(t)$ in Fig.~\ref{Fig:short_t}(b3) for $t\geq 50$ (when all the transients have certainly vanished) to the exponentially saturating function $f(t)=a_0-a_1\:e^{-t/a_2}$.
The high quality of the fit is apparent from the inset of Fig.~\ref{Fig:short_t}(b3), and the relative difference between $\mathcal{D}_\mathrm{HEOM}$ and $a_0$ is well below 10\%.

Differently from the situation in Figs.~\ref{Fig:short_t}(a) and~\ref{Fig:short_t}(b), in Fig.~\ref{Fig:short_t}(c), the bubble-approximation dynamical-mobility profile qualitatively resembles its numerically exact counterpart.
Both profiles display relatively broad peaks at integer multiples of $\omega_0$ that originate from peaks in $\mathrm{Re}\:C_{jj}(t)$ at integer multiples of $2\pi/\omega_0$.
The bubble approximation predicts peaks without internal structure, whereas numerically exact results predict structured peaks.
Such peaks may be ascribed to a more complicated dynamics of $\mathrm{Re}\:C_{jj}$, which becomes negative after the first peak.
A word of caution is in order here as we have established~\cite{JChemPhys.159.094113} that our HEOM results for $\omega_0/t_0=3$ may not be entirely reliable due to possible problems with the HEOM closing strategy for $\omega_0/t_0\geq 2$.

\begin{figure}[htbp!]
    \centering
    \includegraphics[width=\columnwidth]{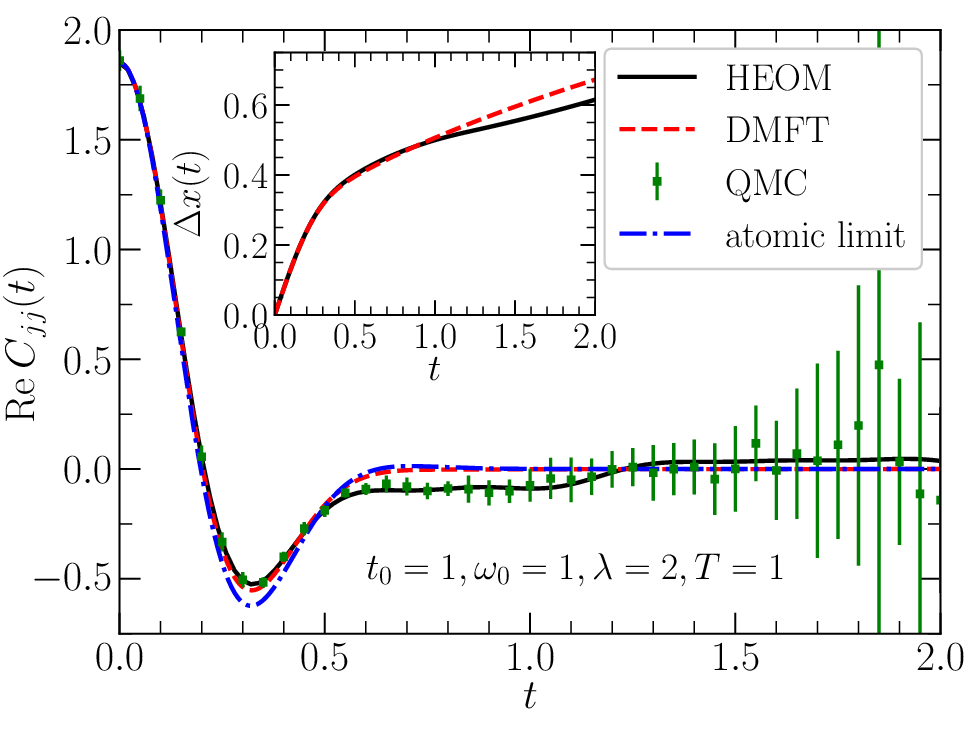}
    \vspace{-.75cm}
    \caption{Comparison of numerically exact (labels "HEOM" and "QMC") and bubble-approximation (label "DMFT") dynamics of $\mathrm{Re}\:C_{jj}$ for $t_0=1,\omega_0=1,\lambda=2,$ and $T=1$. The results labeled "atomic limit" are obtained using Eq.~\eqref{Eq:atomic-limit-like}. The inset shows time-dependent carrier spread $\Delta x(t)$ computed numerically exactly and within the bubble approximation. HEOM computations use $N=10,D=8$. QMC results are displayed with the associated statistical error bars and are obtained for $N=10$.}
    \label{Fig:illustration_261223}
\end{figure}

Finally, in contrast to the regimes studied in Figs.~\ref{Fig:short_t}(a) and~\ref{Fig:short_t}(b), there are situations in which the bubble approximation does partially capture the time-limited slow-down of the carrier.
This typically happens for strong interaction and at not too high temperatures.
One example ($\omega_0/t_0=1,\lambda=2,T/t_0=1$) is analyzed in Fig.~\ref{Fig:illustration_261223}, whose inset shows the short-time dynamics of the carrier spread.
While in Figs.~\ref{Fig:short_t}(a) and~\ref{Fig:short_t}(b) the carrier slows down having covered more than a lattice constant, the slow-down in Fig.~\ref{Fig:illustration_261223} happens over the time interval in which $\Delta x$ remains well below a single lattice constant.
This suggests that the dynamics shown in Fig.~\ref{Fig:illustration_261223} predominantly reflects on-site phonon-assisted processes.
It is thus not surprising that the short-time bubble-approximation dynamics can be qualitatively (and to a large extent quantitatively) reproduced by the atomic-limit formula [Eq.~\eqref{Eq:no_vtx_t_0_to_0}] corrected so that it reproduces the value of $C_{jj}(t=0)$:
\begin{equation}
\label{Eq:atomic-limit-like}
\begin{split}
    &C_{jj}(t)=C_{jj}(0)\times\\&\exp\left\{-2\frac{g^2}{\omega_0^2}\left[(2n_\mathrm{ph}+1)-(n_\mathrm{ph}+1)e^{-i\omega_0 t}-n_\mathrm{ph}e^{i\omega_0 t}\right]\right\},
\end{split}
\end{equation}
compare the lines labeled "DMFT" and "atomic limit" in Fig.~\ref{Fig:illustration_261223}.
[We have checked that, on time scales analyzed in Fig.~\ref{Fig:illustration_261223}, the attenuating time-dependent prefactor entering Eq.~\eqref{Eq:no_vtx_t_0_to_0} does not introduce any quantitative changes to the result of Eq.~\eqref{Eq:atomic-limit-like}.]  
The numerically exact dynamics shows that the slow-down is prolonged with respect to the bubble-approximation results, meaning that the latter captures the temporal slow-down only partially.

\subsection{Vertex corrections to the dc mobility}

\begin{figure*}[htbp!]
    \centering
    \includegraphics[width=\textwidth]{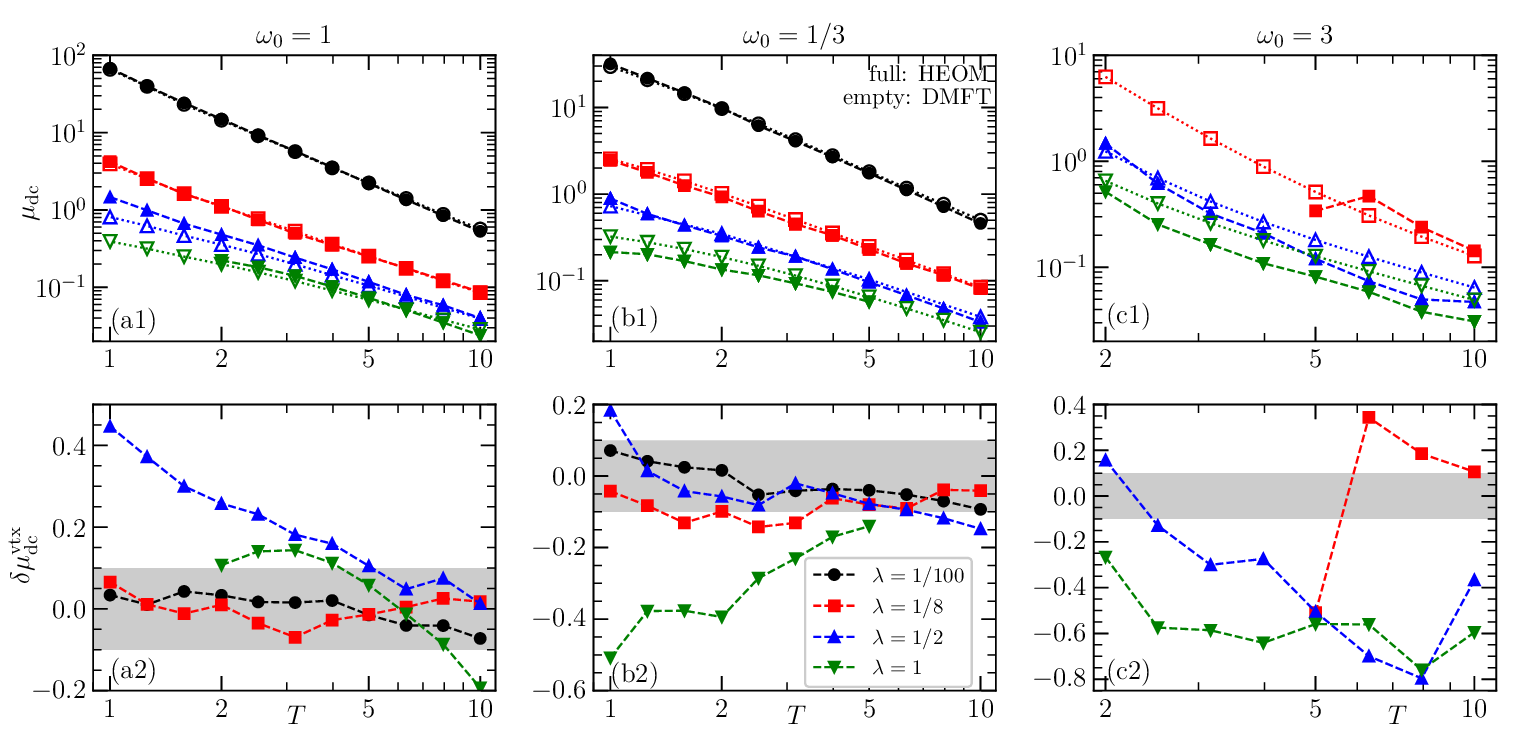}
    \vspace{-0.75cm}
    \caption{(a1)--(c1): Temperature dependence of the dc mobility computed using the HEOM (full symbols connected by dashed lines) and DMFT (empty symbols connected by dotted lines) for different strengths of the electron--phonon interaction $\lambda$ and fixed phonon frequency $\omega_0$.
    (a2)--(c2): Temperature dependence of the quantity $\delta\mu_\mathrm{dc}^\mathrm{vtx}$ [Eq.~\eqref{Eq:def_delta_mu_dc_vtx}], which quantifies the importance of vertex corrections to the dc mobility, for different values of $\lambda$ and fixed $\omega_0$.
    Gray regions in (a2)--(c2) delimit the range $\delta\mu_\mathrm{dc}^\mathrm{vtx}\in[-0.1,0.1]$ in which the vertex corrections to the dc mobility can be considered as vanishing.
    $\omega_0$ is equal to 1 in (a1) and (a2), 1/3 in (b1) and (b2), and 3 in (c1) and (c2), while $t_0=1$ in all panels.}
    \label{Fig:slika_all_omega}
\end{figure*}

The importance of vertex corrections to the dc mobility will be quantified by the relative deviation of the dc mobility in the bubble approximation from the numerically exact result, i.e.,
\begin{equation}
\label{Eq:def_delta_mu_dc_vtx}
    \delta\mu_\mathrm{dc}^\mathrm{vtx}=\frac{\mu_\mathrm{dc}^\mathrm{HEOM}-\mu_\mathrm{dc}^\mathrm{DMFT}}{\mu_\mathrm{dc}^\mathrm{HEOM}}.
\end{equation}
The results in the bubble approximation can be considered to carry no intrinsic numerical error because they follow from the DMFT equations formulated directly in the thermodynamic limit, see also Fig.~\ref{Fig:heom_bubble_dmft}.
On the other hand, the relative uncertainty that should accompany HEOM results for the dc mobility does not surpass 10\%, as discussed in Sec.~\ref{SSec:HEOM} and Ref.~\onlinecite{JChemPhys.159.094113}.
In other words, whenever $\left|\delta\mu_\mathrm{dc}^\mathrm{vtx}\right|\lesssim 0.1$, one can regard the vertex corrections to the dc mobility as unimportant.

Figures~\ref{Fig:slika_all_omega}(a1)--\ref{Fig:slika_all_omega}(c2) provide an overall picture of the importance of the vertex corrections to the dc mobility for different values of $\omega_0/t_0,\lambda,$ and $T/t_0$.
Interestingly, in most of the parameter regimes examined for $\omega_0/t_0=1$ and $1/3$, $\delta\mu_\mathrm{dc}^\mathrm{vtx}$ falls in the gray-shaded regions delimiting the aforementioned range $\left|\delta\mu_\mathrm{dc}^\mathrm{vtx}\right|\lesssim 0.1$, in which the vertex corrections to the dc mobility can be regarded as vanishing.
Our scarce results for $\omega_0/t_0=3$ might suggest that the vertex corrections to the dc mobility are more important than in the other two cases analyzed in Fig.~\ref{Fig:slika_all_omega}.
However, possible problems with HEOM results for $\omega_0/t_0\geq 2$~\cite{JChemPhys.159.094113} prevent us from giving a definite statement on the importance of the vertex corrections to the dc mobility in the case of fast phonons.

In the parameter regimes analyzed in Figs.~\ref{Fig:short_t}(a) and~\ref{Fig:short_t}(b), the vertex corrections to the dc mobility can be deemed important as $\delta\mu_\mathrm{dc}^\mathrm{vtx}$ is around 0.5 and 0.2, respectively.
On the other hand, while $\delta\mu_\mathrm{dc}^\mathrm{vtx}\approx-0.4$ points towards significant vertex corrections to the dc mobility for parameters in Fig.~\ref{Fig:short_t}(c), the overall shape of the dynamical-mobility profile suggests that their importance may be much smaller for the ac mobility.
The vertex corrections in Figs.~\ref{Fig:short_t}(a) and~\ref{Fig:short_t}(b) are positive, i.e., $\mu_\mathrm{dc}^\mathrm{HEOM}>\mu_\mathrm{dc}^\mathrm{DMFT}$.
The dominant contribution to $\mu_\mathrm{dc}^\mathrm{DMFT}$ comes from the dynamics on short time scales $t_0t\lesssim 1$, on which the approximate and numerically exact results agree quite well [see also Eq.~\eqref{Eq:mu_dc}].
This is most conveniently seen from Figs.~\ref{Fig:short_t}(a3)--\ref{Fig:short_t}(c3) showing the dynamics of $\mathcal{D}$.
The subsequent slow-down of the carrier is fully compensated by the speed-up of the carrier [$\mathrm{Re}\:C_{jj}(t)>0$, $\mathcal{D}(t)$ increases with $t$] on somewhat longer time scales.
Ultimately, the effects of the transient slow-down are overpowered by the speed-up, so that the numerically exact dc mobility becomes larger than the approximate one.

Our results for $\lambda=1/2$ in Fig.~\ref{Fig:slika_all_omega}(a2) and for $\lambda=1$ in Fig.~\ref{Fig:slika_all_omega}(b2) suggest that the vertex corrections to the dc mobility decrease with temperature for the parameters studied.
However, even at the highest temperatures accessible in these two cases, the numerically exact results (and the bubble-approximation results, too) are not close to the results in the infinite-temperature limit analyzed in Sec.~\ref{SSec:beta_to_0}.
Also, one should keep in mind that even when the vertex corrections to the dc mobility can be considered as vanishing [e.g., for $\lambda=1/8$ in Figs.~\ref{Fig:slika_all_omega}(a2) and~\ref{Fig:slika_all_omega}(b2)], there may be important differences between the numerically exact and bubble-approximation dynamical-mobility profiles (see also \textcolor{red}{Figs.~S1 and~S2 of Ref.~\onlinecite{comment050324}}).

Considering all these things, in the following three sections we discuss in more detail specific results related to the importance of vertex corrections for the three values of $\omega_0/t_0$ studied.

\subsection{Intermediate-frequency phonons ($\omega_0/t_0=1$)}
For $\lambda=1/100$ and at all temperatures examined, our numerical results show vanishing vertex corrections to the dc mobility, as demonstrated both numerically in Fig.~\ref{Fig:slika_all_omega}(a2) and analytically in Sec.~\ref{SSec:analytics_g_to_0}.
The small nonzero values of $\delta\mu_\mathrm{dc}^\mathrm{vtx}$ can be attributed to the weakly pronounced finite-size effects in the HEOM results, see also Fig.~\ref{Fig:slika_wcl}.
The vertex corrections to dc mobility are also small for $\lambda=1/8$. 
However, as the temperature is increased, the dynamical-mobility profile develops a finite-frequency peak qualitatively similar to that discussed in Fig.~\ref{Fig:short_t}(a), see \textcolor{red}{Fig.~S1 of Ref.~\onlinecite{comment050324}}.

\begin{figure}[htbp!]
 \centering
 \includegraphics[width = \columnwidth]{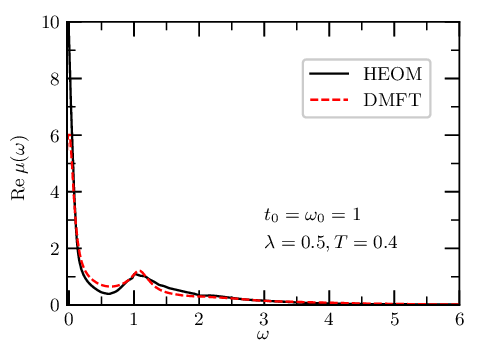}
 \caption{Dynamical-mobility profile computed using HEOM and DMFT for $t_0=\omega_0=1,\lambda=1/2$, and $T=0.4$.
 HEOM computations use $N=29,D=4$.}
 \label{Fig:fig-T-0.4}
\end{figure}

Although the vertex corrections to the dc mobility for $\lambda=1/2$ become insignificant at the highest temperatures considered, see Fig.~\ref{Fig:slika_all_omega}(a2), the differences between the numerically exact and bubble-approximation dynamics of $\mathrm{Re}\:C_{jj}$ and dynamical-mobility profiles persist, as shown in Fig.~\ref{Fig:short_t}(a) and \textcolor{red}{Fig.~S1 of Ref.~\onlinecite{comment050324}}.
\textcolor{red}{Figure~S1} additionally suggests that the carrier slow-down becomes less pronounced and shifts towards later times as the temperature is decreased, meaning that the finite-frequency peak in the numerically exact $\mathrm{Re}\:\mu(\omega)$ then shifts towards lower frequencies.
Also, in the inset of Fig.~\ref{Fig:heom_bubble_dmft}(a), we observe a nascent atomic-limit-like peak in the bubble-approximation optical response at $\omega_0$, indicating the presence of a more pronounced peak at $\omega_0$ at even lower temperatures ($T/t_0<1$). 
These expectations are confirmed in Fig.~\ref{Fig:fig-T-0.4}, which compares the numerically exact and bubble-approximation dynamical-mobility profiles at $T/t_0=0.4$.
Both HEOM and DMFT results show that the finite-frequency peak is centered exactly around $\omega_0$.
This peak can be associated with the transitions between the quasiparticle and the satellite peak in the single-particle spectral function.
While Fig.~\ref{Fig:fig-T-0.4} indicates that the vertex corrections to the dc mobility remain important at lower temperatures, we note that reaching lower temperatures is problematic for the HEOM method here, mainly because of the system size necessary to minimize finite-size effects.

Figure~\ref{Fig:slika_all_omega}(a2) suggests that the vertex corrections to the dc mobility for $\lambda=1$ are overall smaller than for $\lambda=1/2$.
The numerically exact dynamics of $\mathrm{Re}\:C_{jj}$ displays similar features to those in Figs.~\ref{Fig:short_t}(a1) ($\lambda=1/2$) and~\ref{Fig:illustration_261223} ($\lambda=2$), see also \textcolor{red}{Fig.~S1 of Ref.~\onlinecite{comment050324}}.
Setting $\lambda=1$ and increasing $T$, the overall decrease of $\mathrm{Re}\:C_{jj}$ towards zero becomes generally faster, so that the difference between $\mu_\mathrm{dc}$ and $\mu_\mathrm{dc}^\mathrm{DMFT}$ stemming from the carrier slow-down becomes more pronounced than that caused by the subsequent speed-up.
This argument can explain positive (negative) vertex corrections at the lower (upper) end of the temperature range considered for $\lambda=1$ in Fig.~\ref{Fig:slika_all_omega}(a2).
In a similar vein, we argue that the vertex corrections to the dc mobility are not substantial for $\lambda=2$.
Namely, at a fixed temperature $T$, the overall decrease of $\mathrm{Re}\:C_{jj}$ towards zero is faster for stronger interaction.
Therefore, for sufficiently strong $g$,
we may expect that the features specific to the numerically exact result will not appreciably affect $\mu_\mathrm{dc}$, which is then primarily determined by $\mathrm{Re}\:C_{jj}(t)$ up to times at which the corresponding bubble-approximation and numerically exact results agree well, see also Fig.~\ref{Fig:illustration_261223}.

\subsection{Slow phonons ($\omega_0/t_0=1/3$)}
\begin{figure}[htbp!]
    \centering
    \includegraphics[width=\columnwidth]{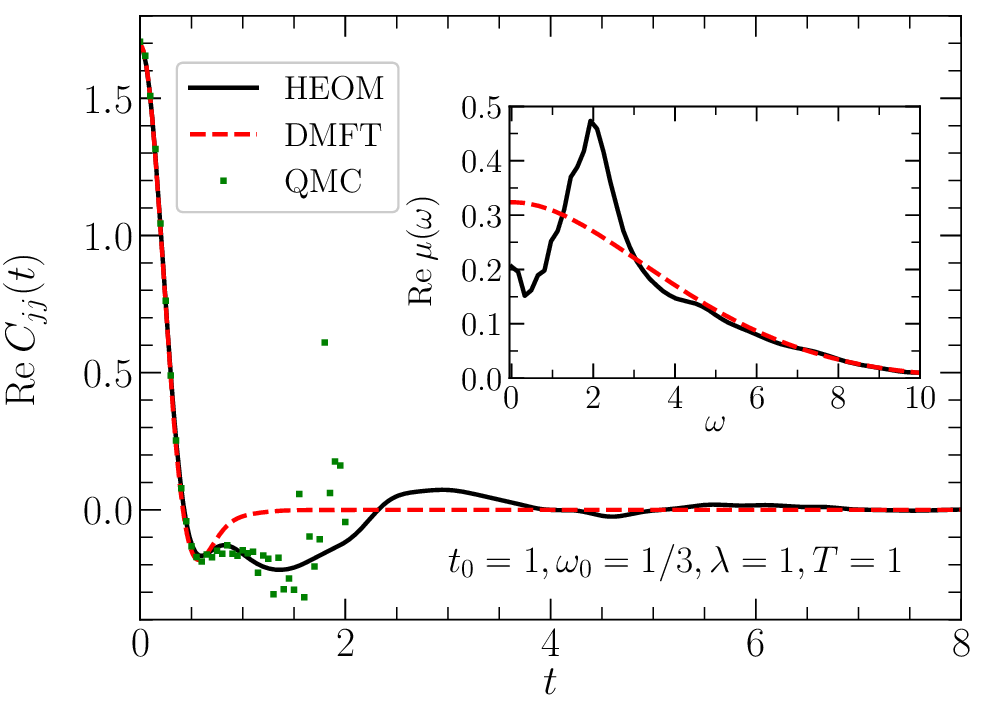}
    \vspace{-0.75cm}
    \caption{Time dependence of $\mathrm{Re}\:C_{jj}$ computed within the HEOM, DMFT, and QMC for $t_0=1,\omega_0=1/3,\lambda=1,$ and $T=1$. The QMC results are displayed without the associated statistical error bars for visual clarity. The inset compares the corresponding dynamical-mobility profiles computed within HEOM and DMFT. HEOM computations use $N=7,D=11$ and 12 (arithmetic average). QMC simulations use $N=7$.}
    \label{Fig:fig-schubert}
\end{figure}

For $\lambda=1/100$, the vertex corrections are negligible.
For $\lambda=1/8$, the vertex corrections are very small at $T/t_0=1$, but with increasing
temperature, a characteristic two-peak structure emerges in the numerically exact dynamical-mobility profile, see \textcolor{red}{Fig.~S2 of Ref.~\onlinecite{comment050324}}.
Intermediate interactions $\lambda=1/2$ [Fig.~\ref{Fig:short_t}(b1)] and $\lambda=1$ (Fig.~\ref{Fig:fig-schubert}) bring about the appearance of the two-peak structure also at $T/t_0=1$.
For $\lambda=1$, the finite-frequency peak is centered around $\omega=2t_0$, in agreement with earlier numerically exact studies in the slow-phonon regime~\cite{PhysRevB.72.104304}.
While the bubble approximation partially captures the time-limited carrier slow-down at $T/t_0=1$, compare Fig.~\ref{Fig:fig-schubert} with Fig.~\ref{Fig:illustration_261223}, it does not capture the finite-frequency absorption feature, see the inset of Fig.~\ref{Fig:fig-schubert} and Sec. IV.~C of Ref.~\onlinecite{PhysRevB.74.075101}.
To elucidate the origin of this feature, the assumption of strictly on-site phonon dynamics underlying the atomic-limit-like Eq.~\eqref{Eq:atomic-limit-like} (which reproduces the bubble-approximation result reasonably well, similarly to Fig.~\ref{Fig:illustration_261223}) has to be relaxed, as has been done by Schubert et al.~\cite{PhysRevB.72.104304}.
Their analysis attributes the $2t_0$ absorption feature to the optical transition between the symmetric and antisymmetric states of the electron residing on two neighboring sites, over which phonon dynamics is considered.
\textcolor{red}{Figure~S2 of Ref.~\onlinecite{comment050324}} shows that the finite-frequency peak remains around $2t_0$ also at higher temperatures, in agreement with earlier results~\cite{PhysRevB.72.104304}.

Despite important differences in the overall dynamics of $C_{jj}$, see Fig.~\ref{Fig:fig-schubert}, the results summarized in Fig.~\ref{Fig:slika_all_omega} show that the bubble approximation correctly predicts the order of magnitude of the dc mobility.
However, as we approach the adiabatic limit $\omega_0\to 0$, the exact result for the dc mobility tends to zero (phonon motion is effectively frozen and an electron experiences a random potential in one dimension where it cannot move over a large distance due to the effect of Anderson localization), while the bubble-approximation result keeps a finite value~\cite{Bruus-Flensberg-book}.
Therefore, in this limit, vertex corrections are most pronounced.
They give a negative contribution to dc mobility, which completely cancels out the bubble-approximation result (so that $\delta\mu_\mathrm{dc}^\mathrm{vtx}\to-\infty$).
The fact that the results presented in Fig.~\ref{Fig:slika_all_omega} show positive or somewhat negative vertex corrections implies that the system is still relatively far from the adiabatic limit in these cases.
We illustrate this in Fig.~\textcolor{red}{S4} \textcolor{red}{Ref.~\onlinecite{comment050324}}, where we compare $C_{jj}\qty(t)$ and $\mathcal{D}\qty(t)$ for certain parameter values with the adiabatic-limit result for these quantities.
The figure confirms that these quantities are far from their adiabatic-limit values.
Adiabatic-limit results are obtained using a very computationally efficient Monte Carlo procedure that exploits the fact that phonon momentum is negligible in the adiabatic limit.
This procedure is described in detail in \textcolor{red}{Sec.~SV of Ref.~\onlinecite{comment050324}}.

\begin{figure}[htbp!]
    \centering
    \includegraphics[width=\columnwidth]{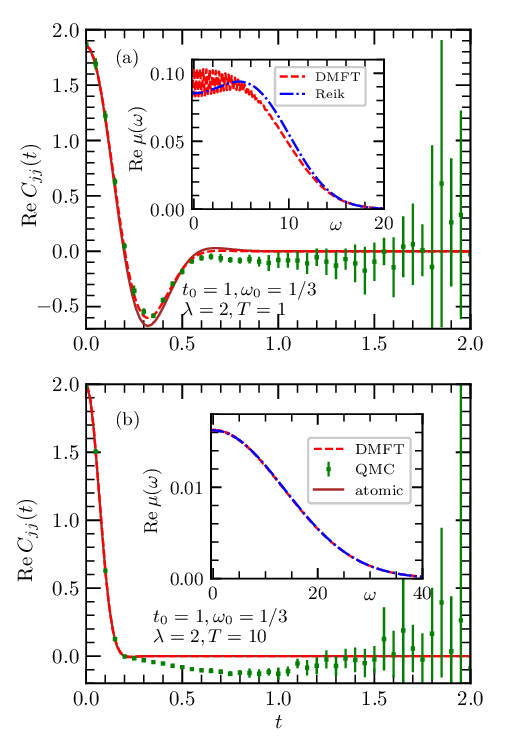}
    \vspace{-1cm}
    \caption{Time dependence of $\mathrm{Re}\:C_{jj}$ computed within DMFT and QMC for $t_0=1,\omega_0=1/3,\lambda=2,$ and (a) $T=1$, (b) $T=10$.
    Solid brown lines represent the results of the atomic-limit Eq.~\eqref{Eq:atomic-limit-like}.
    The insets compare the corresponding dynamical-mobility profiles with those evaluated using the Reik formula [Eq.~\eqref{Eq:reik_symmetric}].
    QMC simulations use $N=7$.}
    \label{Fig:polaron-peak}
\end{figure}

In the strong-interaction regime $\lambda=2$, and at $T/t_0=1$, the DMFT captures the time-limited slow-down observed in QMC data quite well, see Fig.~\ref{Fig:polaron-peak}(a) and compare to Fig.~\ref{Fig:fig-schubert}.
The short-time dynamics of $\mathrm{Re}\:C_{jj}$ is also very well reproduced by the atomic-limit-like Eq.~\eqref{Eq:atomic-limit-like}.
The DMFT dynamical-mobility profile displays atomic-limit-like peaks at integer multiples of $\omega_0$.
As the temperature is increased from $T/t_0=1$ to $10$, the peaks become smoothed out, and their envelope changes from a function displaying a finite-frequency local maximum and a zero-frequency local minimum to a function displaying only a zero-frequency local maximum; see the insets of Figs.~\ref{Fig:polaron-peak}(a) and~\ref{Fig:polaron-peak}(b).
This finite-frequency peak of the envelope, appearing at sufficiently low temperatures, is the well-known polaron peak because the shape of the envelope of the dynamical-mobility profile compares reasonably well with the high-temperature (small-$\omega_0$) limit of the Reik formula [Eq.~(29) of Ref.~\onlinecite{JPhysChemSolids.28.581} in which $\sinh x\approx\tanh x\approx x$, with $x=\mathrm{const}\times\beta\omega_0\ll 1$], which reads as
\begin{equation}
\label{Eq:reik_symmetric}
\begin{split}
    \mathrm{Re}\:\mu(\omega)=\frac{\sqrt{\pi}t_0^2}{\sigma\omega}\left[\exp\left(-\frac{(\omega-2\varepsilon_\mathrm{pol})^2}{4\sigma^2}\right)- \right. \\ \left.
    \exp\left(-\frac{(\omega+2\varepsilon_\mathrm{pol})^2}{4\sigma^2}\right)\right].
\end{split}
\end{equation}
Here, $\varepsilon_\mathrm{pol}=g^2/\omega_0$ is the polaron binding energy, and $\sigma$ is defined in Eq.~\eqref{Eq:def_sigma_beta_to_0}.
Even at the highest temperatures studied, QMC results predict a protracted temporally limited slow-down of the electron, pointing towards possibly nontrivial negative vertex corrections to the dc mobility (because the intensity of the possible atomic-limit-like features appearing at integer multiples of $2\pi/\omega_0$ decreases quickly with time, which is compatible with the rather wide DMFT optical response).

\subsection{Fast phonons ($\omega_0/t_0=3$)}

\begin{figure}[htbp!]
    \centering
    \includegraphics[width=\columnwidth]{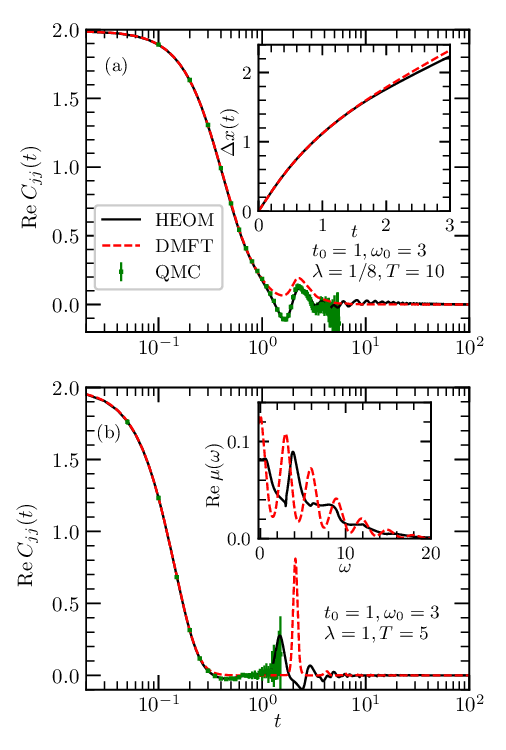}
    \vspace{-1cm}
    \caption{Time dependence of $\mathrm{Re}\:C_{jj}$ computed using HEOM (solid black lines), DMFT (dashed red lines), and QMC (full circles) for $t_0=1,\omega_0=3$, and (a) $\lambda=1/8,T=10$, and (b) $\lambda=1,T=5$. QMC results are accompanied with their statistical error bars. The insets show (a) the carrier spread, and (b) the dynamical-mobility profile computed using HEOM and DMFT. HEOM computations use (a) $N=10,D=8$, (b) $N=5$, $D=20$ and 21 (arithmetic average). QMC simulations use (a) $N=10$, (b) $N=7$.}
    \label{Fig:fig_omega_3_all}
\end{figure}

HEOM results are available only in a limited portion of the parameter space, characterized by sufficiently, but not excessively high temperatures and interaction strengths.

We first compare numerically exact and bubble-approximation results in the weak-coupling regime $\lambda=1/8$ and $T/t_0=10$, see Fig.~\ref{Fig:fig_omega_3_all}(a).
While the bubble-approximation result predicts a monotonically increasing diffusion constant [$\mathrm{Re}\:C_{jj}(t)>0$], the numerically exact result predicts a time-limited slow-down of the carrier motion after it has covered a single lattice constant (for $1\leq t_0t\leq 2$, when $\Delta x\gtrsim 1$), similarly to what we observed in Figs.~\ref{Fig:short_t}(a) and~\ref{Fig:short_t}(b) for smaller values of $\omega_0/t_0$.
Interestingly, both numerically exact and approximate results predict a local maximum in $\mathrm{Re}\:C_{jj}(t)$ around $2\pi/\omega_0$, which is typical for the atomic limit.
With this in mind, the reasonable overall agreement between the numerically exact and approximate dynamical-mobility profiles, similar to our observations in Fig.~\ref{Fig:short_t}(c), is not surprising, see \textcolor{red}{Fig.~S3 of Ref.~\onlinecite{comment050324}}.
The vertex corrections to the dc mobility may be considered as vanishing, see Fig.~\ref{Fig:slika_all_omega}(c), as expected in the limit of weak electron--phonon coupling.
The potential problems with our HEOM closing are exacerbated at a lower temperature ($T/t_0=5$), where our results predict significant vertex corrections, contrary to expectations for small $\lambda$.

For $\lambda=1$ and $T/t_0=5$, see Fig.~\ref{Fig:fig_omega_3_all}(b), the bubble approximation again predicts strictly non-negative $\mathrm{Re}\:C_{jj}(t)$ with peaks at integer multiples of $2\pi/\omega_0$.
On the other hand, numerically exact results show that the peaks' centers are generally shifted towards somewhat earlier times, while $\mathrm{Re}\:C_{jj}(t)$ may assume negative values.
This is clearly observed for the first peak, whose intensity within the numerically exact framework is smaller than the intensity of its bubble-approximation counterpart.
As a consequence, the numerically exact dynamical-mobility spectrum is qualitatively different from its bubble-approximation counterpart, which has equidistant peaks at integer multiples of $\omega_0$.

\subsection{Insights from the imaginary-time domain}

\begin{figure}[htbp!]
    \centering
    \includegraphics[width=\columnwidth]{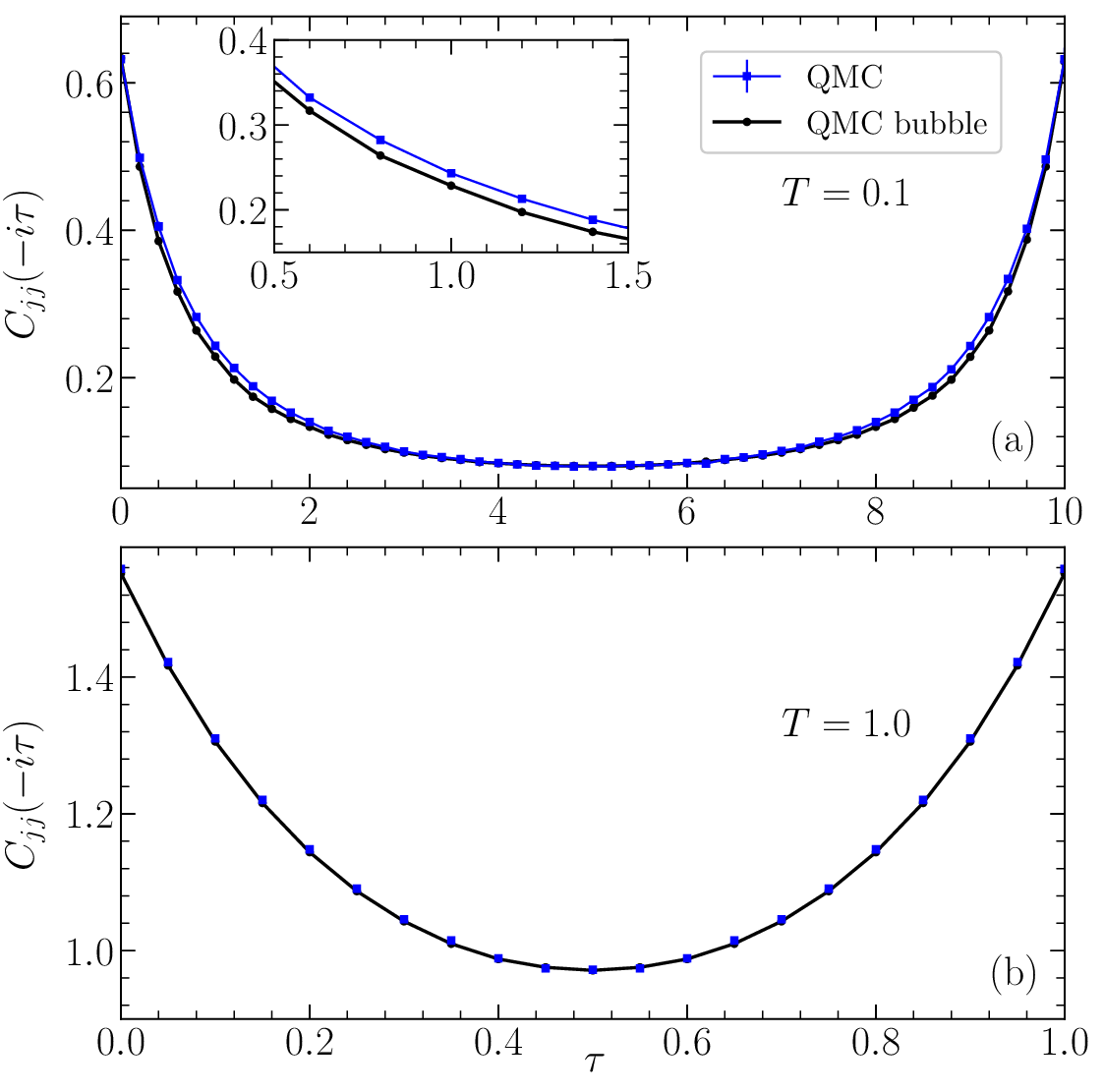}
    \caption{
    Imaginary-time current--current correlation function for $t_0=1,\omega_0=1/3,$ and $\lambda=1/2$ at temperatures (a) $T=0.1$ and (b) $T=1$. The QMC results obtained from a full calculation and by making use of the bubble approximation are presented. The inset of the top panel shows a zoom to the region where the difference between numerically exact and bubble approximation results is most pronounced.
    }
    \label{Fig:imag_time}
\end{figure}

It is known that the extraction of transport properties from imaginary-axis data for the current--current correlation function can often be unreliable as it relies on procedures for numerical analytical continuation.
In \textcolor{red}{Fig.~S5 of Ref.~\onlinecite{comment050324}}, we compare some of our results for dc mobility, which entirely follow from real-axis data, to the corresponding results of Ref.~\onlinecite{PhysRevLett.114.146401}, which follow from imaginary-axis QMC data.
Nevertheless, one may hope to gain insights into the importance of vertex corrections by comparing the imaginary-time current--current correlation functions computed numerically exactly and within the bubble approximation.
We have done this by computing, on the one hand, the quantity $C_{jj}\qty(-i\tau)$ [see Eq.~\eqref{Eq:def_C_jj_t}] for $0\le \tau \le \beta$, and, on the other hand, the quantities $\mathcal{G}^>(k,-i\tau)$ [see Eq.~\eqref{Eq:G_gtr}] and $\mathcal{G}^<(k,-i\tau)$ [see Eq.~\eqref{Eq:G_less}], which are needed to obtain $C_{jj}^\mathrm{bbl}(-i\tau)$ in accordance with Eq.~\eqref{Eq:C_jj_bubble}.
The computations were performed using QMC.
It is easier to obtain these quantities in imaginary time than in real time because the dynamical sign problem occurs only for real times.

For the investigated values of parameters when $T/t_0\ge 1$ we find that the differences between $C_{jj}\qty(-i\tau)$ and $C_{jj}^\mathrm{bbl}(-i\tau)$ are very small.
They typically differ by less than 1\% and this difference is typically below or comparable to the statistical error of QMC results.
One such example is presented in Fig.~\ref{Fig:imag_time}(b).
These results should be contrasted with our conclusions reached by analyzing real-time results in the very same parameter regime; see Figs.~\ref{Fig:short_t}(b) and~\ref{Fig:slika_all_omega}(b2).
We can thus conclude that good agreement between numerically exact and bubble approximation results in imaginary time does not imply that the same level of agreement will be present in real time.

When the temperature is lowered to $T/t_0=0.1$, the differences between numerically exact and bubble approximation results become noticeable, see Fig.~\ref{Fig:imag_time}(a).
The inset of that panel shows a zoom to the region where this difference is most significant.
In this region, the difference is above 5\%, while the statistical error of QMC results is on the order of 0.5\%, hence we are confident that the difference observed is above the level of statistical noise.
While these data point towards non-negligible vertex corrections, a definitive conclusion could be reached only if we had access to the corresponding real-time data.

\section{Summary and outlook}\label{Sec:Conclusions}
In summary, we presented a detailed analysis of the importance of vertex corrections for charge transport in the one-dimensional Holstein model.
Our analytical results demonstrate that the vertex corrections vanish in the weak-interaction, atomic, and infinite-temperature limits, which is supported by numerical results.
The numerically exact HEOM calculations were performed for three phonon frequencies: intermediate $\omega_0/t_0=1$, low $\omega_0/t_0=1/3$, and high $\omega_0/t_0=3$.

For $\omega_0/t_0=1$, as the electron--phonon coupling is increased, a characteristic two-peak profile of the numerically exact optical conductivity emerges instead of a single Drude peak.
At low temperatures, the other peak is centered precisely at $\omega=\omega_0$ and is also captured within the bubble approximation because it corresponds to the transitions between the quasiparticle and the first satellite peak in the single-particle spectral density.
At higher temperatures, the peak shifts to $\omega>\omega_0$.
Then, the two-peak structure in optical conductivity, which cannot be reproduced in the bubble approximation, is traced back to a downturn in the time-dependent diffusion constant $\mathcal{D}(t)$ at intermediate time scales.
Interestingly, for strong interactions, such a slow-down of charge carrier is observed also in the bubble approximation, but at shorter times.
In this case, it is primarily governed by the on-site phonon-assisted processes and not by the vertex corrections.
This observation is compatible with the absence of vertex corrections in the atomic limit.
Both at low and high temperatures, we find that the vertex corrections either do not affect or increase the dc mobility by a few tens of percent in comparison to the bubble term.

Interestingly, for $\omega_0/t_0=1/3$ and for moderate electron--phonon interaction, the optical-conductivity profile looks qualitatively similar to that for $\omega_0/t_0=1$.
We find that the peak at zero frequency persists, giving the dc mobility that is somewhat smaller, but still comparable to the one obtained in the bubble approximation.
The height of this peak is expected to diminish as the frequency is lowered further, approaching the adiabatic limit.
However, this calculation is not feasible within our implementation of the HEOM method, and is generally challenging for numerically exact methods on the market.

The HEOM solution in the bubble approximation almost coincides with the DMFT solution for optical conductivity for all available parameter values.
Hence, our results demonstrate that the vertex corrections can be substantial also in the cases in which the single-particle correlations are almost local, in agreement with previous findings on the Hubbard model~\cite{PhysRevLett.123.036601}.
Our results also illustrate the challenges in numerical analytical continuation of the imaginary axis data: in all parameter regimes, the difference between the full and bubble correlation functions in imaginary time is minuscule, and yet the difference in optical conductivity can be large.

To our knowledge, in this work we have presented the most comprehensive study so far of vertex corrections to conductivity on a specific model of the electron--phonon interaction.
However, there are several challenges left for future work.
It would be interesting to calculate the charge transport in higher dimensions and make a comparison to the one-dimensional case.
Also, microscopic calculations for lower phonon frequencies and for models with nonlocal electron--phonon interactions are highly desirable, especially in connection with real-world materials.

\acknowledgments
This research was supported by the Science Fund of the Republic of Serbia, Grant No. 5468, Polaron Mobility in Model Systems and Real Materials--PolMoReMa.
The authors acknowledge funding provided by the Institute of Physics Belgrade, through the grant from the Ministry of Science, Technological Development, and Innovation of the Republic of Serbia.
Numerical computations were performed on the PARADOX-IV supercomputing facility at the Scientific Computing Laboratory, National Center of Excellence for the Study of Complex Systems, Institute of Physics Belgrade.
V.J. and N.V. also acknowledge computational time on the ARIS supercomputing facility (GRNET, Athens, Greece) that was granted by the NI4OS-Europe network under the CoNTraSt project (Open Call 2022, Project No. ni4os002).

\section*{Author contributions}
V.J. and N.V. performed analytical work.
V.J. performed HEOM calculations.
P.M. performed DMFT calculations with guidance of D.T.
QMC calculations were performed by N.V.
The first version of the manuscript was prepared by V.J.
All authors discussed the results and contributed to the final manuscript.

\begin{widetext}
\appendix
\section{Formally exact expressions for the electronic dynamics}
\label{App:influence_phases}
It is convenient to switch from the electronic creation and annihilation operators, which act in the Fock space for the electrons, to their counterparts acting in the subspaces containing at most one electron.
The corresponding replacements $c_k^\dagger\to|k\rangle\langle\mathrm{vac}|$ and $c_k\to|\mathrm{vac}\rangle\langle k|$, where the state $|\mathrm{vac}\rangle$ contains no electrons, are appropriate in the limit of low carrier concentration in which we are interested.

We first summarize the formally exact result for the dynamics of the current--current correlation function in Eq.~\eqref{Eq:C_jj_t_low_density_limit}, which can be expressed as
\begin{equation}
\label{Eq:C_jj_full_iota}
    C_{jj}(t)=\mathrm{Tr}_\mathrm{1e}\{j\iota(t)\}=\sum_k j_k\langle k|\iota^{(I)}(t)|k\rangle,
\end{equation}
where
\begin{equation}
    \iota(t)=\frac{1}{Z}\mathrm{Tr}_\mathrm{ph}\left\{e^{-iHt}je^{-\beta H}e^{iHt}\right\}.
\end{equation}
In Ref.~\onlinecite{JChemPhys.159.094113}, we derived that the interaction-picture counterpart of the purely electronic operator $\iota(t)$ reads as
\begin{equation}
\label{Eq:iota_I_t}
    \iota^{(I)}(t)=\mathcal{T}e^{-[\Phi_1(t)+\Phi_2(\beta)+\Phi_3(t,\beta)]}j\frac{e^{-\beta H_\mathrm{e}}}{Z_\mathrm{e}}.
\end{equation}
The corresponding influence phases are given as
\begin{equation}
    \begin{split}
    &\Phi_1(t)=\sum_{qm}\int_0^t ds_2\int_0^{s_2}ds_1\:V_q^{(I)}(s_2)^\times\:e^{-\mu_{m}(s_2-s_1)}\\&\times\left[\frac{c_{m}+c_{\overline{m}}}{2}\:V_{-q}^{(I)}(s_1)^\times+\frac{c_{m}-c_{\overline{m}}}{2}\:V_{-q}^{(I)}(s_1)^\circ\right],
\end{split}\label{Eq:def_Phi_1}
\end{equation}
\begin{equation}
    \begin{split}
\Phi_2(\beta)=&-\sum_{qm}\int_0^\beta d\tau_2\int_0^{\tau_2}d\tau_1\:^C\overline{V}_{-q}(\tau_1)\\
&\times e^{i\mu_{m}(\tau_2-\tau_1)}c_{m}\:^C\overline{V}_q(\tau_2),
\end{split}\label{Eq:def_Phi_2}
\end{equation}
\begin{equation}
    \begin{split}
    \Phi_3(t,\beta)=&-i\sum_{qm}\int_0^t ds\int_0^\beta d\tau\:V_q^{(I)}(s)^\times\\
    &\times e^{-\mu_{m}s}e^{i\mu_{m}(\beta-\tau)}c_{m}\:^C\overline{V}_{-q}(\tau),
\end{split}\label{Eq:def_Phi_3}
\end{equation}
where the purely electronic operator $V_q$ reads as
\begin{equation}
\label{Eq:def_V_q}
    V_q=\sum_k|k+q\rangle\langle k|,
\end{equation}
while the coefficients $c_m$ and $\mu_m$ ($m=0,1$) are
\begin{equation}
    c_{0}=\left(\frac{g}{\sqrt{N}}\right)^2(1+n_\mathrm{ph}),\quad\mu_{0}=+i\omega_0,\quad n_\mathrm{ph}=\frac{1}{e^{\beta\omega_0}-1},
    \label{Eq:def_c_q0}
\end{equation}
\begin{equation}
    c_{1}=\left(\frac{g}{\sqrt{N}}\right)^2n_\mathrm{ph},\quad\mu_{1}=-i\omega_0.\label{Eq:def_c_q1}
\end{equation}
For later use, we introduce the index $\overline{m}$ defined by $\mu_{\overline{m}}=\mu_m^*$.
In other words, $\overline{0}=1$ and \emph{vice versa}.
The hyperoperators entering Eqs.~\eqref{Eq:def_Phi_1}--\eqref{Eq:def_Phi_3} are defined by their action on an arbitrary operator $O$ which is as follows:
\begin{subequations}
\begin{align}
V^C\:O=VO,\\
^CV\:O=OV,\\
V^\times\:O=VO-OV,\\
V^\circ\:O=VO+OV.
\end{align}
\end{subequations}
The operators $V_q$ in the real-time and imaginary-time interaction picture are defined as
\begin{equation}
    V_q^{(I)}(t)=e^{iH_\mathrm{e}t}V_qe^{-iH_\mathrm{e}t},\quad \overline{V}_q(\tau)=e^{H_\mathrm{e}\tau}V_qe^{-H_\mathrm{e}\tau}.
\end{equation}
The time-ordering sign $\mathcal{T}$ in Eq.~\eqref{Eq:iota_I_t} orders the hyperoperators so that one first applies imaginary time-dependent hyperoperators that are mutually antichronologically ordered (descending imaginary-time instants) and subsequently applies real time-dependent hyperoperators that are mutually chronologically ordered (ascending real-time instants).
The so-called electronic partition sum entering Eq.~\eqref{Eq:iota_I_t} is defined as
\begin{equation}
\label{Eq:def_Z_e}
    Z_\mathrm{e}=\frac{Z}{Z_\mathrm{ph}}=\mathrm{Tr}_\mathrm{1e}\left\{\mathcal{T}e^{-\Phi_2(\beta)}e^{-\beta H_\mathrm{e}}\right\},
\end{equation}
where $Z_\mathrm{ph}=\mathrm{Tr}_\mathrm{ph}e^{-\beta H_\mathrm{ph}}$ is the free-phonon partition sum.

Concerning single-particle quantities, it is convenient to redefine them as follows 
\begin{equation}
\label{Eq:G_gtr_app}
\begin{split}
    G^>(k,t)&=\mathcal{G}^>(k,t)e^{-i\mu_\mathrm{F}t}\\
    &=-i\frac{\mathrm{Tr}_{0\mathrm{e}}\{e^{iHt}|\mathrm{vac}\rangle\langle k|e^{-iHt}|k\rangle\langle\mathrm{vac}| e^{-\beta H}\}}{Z_\mathrm{ph}}\\
    &=-i\frac{\mathrm{Tr}_\mathrm{ph}\left\{e^{iH_\mathrm{ph}t}\left\langle k\left|e^{-iHt}\right|k\right\rangle e^{-\beta H_\mathrm{ph}}\right\}}{Z_\mathrm{ph}}\\
    &=-ie^{-i\varepsilon_k t}\left\langle k\left|\mathcal{T}e^{-\varphi_1(t)}\mathbbm{1}_\mathrm{e}\right|k\right\rangle,
\end{split}
\end{equation}
\begin{equation}
\label{Eq:G_less_app}
\begin{split}
    G^<(k,t)&=\frac{\mathcal{G}^<(k,t)e^{-i\mu_\mathrm{F}t}}{\langle N_\mathrm{e}\rangle_K}\\
    &=i\frac{\mathrm{Tr}_{1\mathrm{e}}\{|k\rangle\langle\mathrm{vac}| e^{iHt}|\mathrm{vac}\rangle\langle k|e^{-iHt}e^{-\beta H}\}}{Z}\\
    &=i\frac{\mathrm{Tr}_\mathrm{ph}\left\{e^{iH_\mathrm{ph}t}\left\langle k\left|e^{-iHt}e^{-\beta H}\right|k\right\rangle\right\}}{Z}\\
    &=ie^{-i\varepsilon_k t}\left\langle k\left|\mathcal{T}e^{-[\varphi_1(t)+\varphi_2(\beta)+\varphi_3(t,\beta)]}\frac{e^{-\beta H_\mathrm{e}}}{Z_\mathrm{e}}\right|k\right\rangle.
\end{split}
\end{equation}
The redefinition embodied in the first equalities of Eqs.~\eqref{Eq:G_gtr_app} and~\eqref{Eq:G_less_app} is fully compatible with the frequency shift used to transform Eqs.~\eqref{Eq:G_gtr} and~\eqref{Eq:G_less} into Eqs.~\eqref{Eq:G_gtr_mu_F_neg_lrg} and~\eqref{Eq:G_less_mu_F_neg_lrg}, respectively.
It also introduces a different normalization for $G^<$, $-i\sum_k G^<(k,t=0)=1$, which explicitly shows that we consider a single electron, while the current--current correlation function in the bubble approximation is expressed as [cf. Eq.~\eqref{Eq:C_jj_bubble}]
\begin{equation}
\label{Eq:def_C_jj_bbl_appendix_a}
    C_{jj}^\mathrm{bbl}(t)=-\sum_k j_k^2 G^>(k,t) G^<(k,t)^*.
\end{equation}
In Ref.~\onlinecite{PhysRevB.105.054311}, we derived that the influence phases for single-particle quantities read as
\begin{equation}
    \begin{split}
    \varphi_1(t)=&\sum_{qm}\int_0^t ds_2\int_0^{s_2}ds_1 V_q^{(I)}(s_2)^C\\&\times e^{-\mu_{m}(s_2-s_1)}c_{m} V_{-q}^{(I)}(s_1)^C,
\end{split}\label{Eq:def_varphi_1}
\end{equation}
\begin{equation}
\label{Eq:def_varphi_2}
    \begin{split}
          \varphi_2(\beta)=\Phi_2(\beta),
    \end{split}
\end{equation}
\begin{equation}
    \begin{split}
        \varphi_3(t,\beta)=&-i\sum_{qm}\int_0^t ds\int_0^{\beta}d\tau\:V_q^{(I)}(s)^C\\
  &\times e^{-\mu_{m}s}e^{i\mu_{m}(\beta-\tau)}c_{m}
  \:^C\overline{V}_{-q}(\tau).
    \end{split}\label{Eq:def_varphi_3}
\end{equation}

As the first illustration of the utility of these formally exact results,
we prove the equality $C_{jj}(t=0)=C_{jj}^\mathrm{bbl}(t=0)$.
Because of $\Phi_1(t=0)=\Phi_3(t=0,\beta)=0$, see Eqs.~\eqref{Eq:def_Phi_1} and~\eqref{Eq:def_Phi_3}, Eqs.~\eqref{Eq:C_jj_full_iota} and~\eqref{Eq:iota_I_t} imply that
\begin{equation}
\label{Eq:C_jj_t_0}
\begin{split}
    C_{jj}(t=0)&=\sum_k j_k\left\langle k\left|\mathcal{T}e^{-\Phi_2(\beta)}j\frac{e^{-\beta H_\mathrm{e}}}{Z_\mathrm{e}}\right|k\right\rangle\\
    &=\sum_k j_k\left\langle k\left|j\mathcal{T}e^{-\Phi_2(\beta)}\frac{e^{-\beta H_\mathrm{e}}}{Z_\mathrm{e}}\right|k\right\rangle\\
    &=\sum_k j_k^2\left\langle k\left|\mathcal{T}e^{-\Phi_2(\beta)}\frac{e^{-\beta H_\mathrm{e}}}{Z_\mathrm{e}}\right|k\right\rangle.
\end{split}
\end{equation}
Since the hyperoperators in $\Phi_2(\beta)$ act on the operator $je^{-\beta H_\mathrm{e}}/Z_\mathrm{e}$ from the right-hand side, see Eq.~\eqref{Eq:def_Phi_2}, the current operator can be moved in front of $\mathcal{T}e^{-\Phi_2(\beta)}$.
Because of $\varphi_1(t=0)=\varphi_3(t=0,\beta)=0$, see Eqs.~\eqref{Eq:def_varphi_1} and~\eqref{Eq:def_varphi_3}, Eqs.~\eqref{Eq:def_C_jj_bbl_appendix_a},~\eqref{Eq:G_gtr_app}, and~\eqref{Eq:G_less_app} imply that
\begin{equation}
\label{Eq:C_jj_bbl_t_0}
    C_{jj}^\mathrm{bbl}(t=0)=\sum_k j_k^2\left\langle k\left|\mathcal{T}e^{-\varphi_2(\beta)}\frac{e^{-\beta H_\mathrm{e}}}{Z_\mathrm{e}}\right|k\right\rangle.
\end{equation}
The right-hand sides of Eqs.~\eqref{Eq:C_jj_t_0} and~\eqref{Eq:C_jj_bbl_t_0} are identical because of Eq.~\eqref{Eq:def_varphi_2}.

\section{Evaluation of $C_{jj}(t)$ and $C_{jj}^\mathrm{bbl}(t)$ in the $g\to 0$ limit}\label{App:g_to_0}
\subsection{Equality of the lowest-order terms}

Let us start from the lowest-order term in the bubble result [Eq.~\eqref{Eq:def_C_jj_bbl_appendix_a}], which we derive by separately considering $G^>(k,t)$ [Eq.~\eqref{Eq:G_gtr_app}] and $G^<(k,t)$ [Eq.~\eqref{Eq:G_less_app}] in the lowest-order approximation.
Up to the second order in $g$, we have
\begin{equation}
\label{Eq:G_gtr_start}
    [G^>(k,t)]_2=-ie^{-i\varepsilon_kt}\left[1-\langle k|\varphi_1(t)\mathbbm{1}_\mathrm{e}|k\rangle\right],
\end{equation}
\begin{equation}
\label{Eq:G_less_start}
\begin{split}
    [G^<(k,t)]_2=ie^{-i\varepsilon_kt}\frac{e^{-\beta\varepsilon_k}}{Z_\mathrm{e}}\left[
    1-
    \langle k|\varphi_1(t)\mathbbm{1}_\mathrm{e}|k\rangle- \right. \\ \left.
    \langle k|\varphi_2(\beta)\mathbbm{1}_\mathrm{e}|k\rangle-
    \langle k|\varphi_3(t,\beta)e^{-\beta(H_\mathrm{e}-\varepsilon_k)}|k\rangle
    \right],
\end{split}
\end{equation}
where we have used the relation
\begin{equation}
\label{Eq:commute_varphi_2}
    \langle k|\varphi_2(\beta)e^{-\beta(H_\mathrm{e}-\varepsilon_k)}|k\rangle=
    \langle k|\varphi_2(\beta)\mathbbm{1}_\mathrm{e}|k\rangle
\end{equation}
and a similar relation for $\langle k|\varphi_1(t)e^{-\beta(H_\mathrm{e}-\varepsilon_k)}|k\rangle$.
The time-ordering sign can be safely omitted because the hyperoperators entering Eqs.~\eqref{Eq:G_gtr_start} and~\eqref{Eq:G_less_start} are already properly ordered.
Therefore, up to the second order in $g$, the current--current correlation function in the bubble approximation reads as
\begin{equation}
\label{Eq:C_jj_bbl_g2}
\begin{split}
    [C_{jj}^\mathrm{bbl}(t)]_2=\sum_k j_k^2\frac{e^{-\beta\varepsilon_k}}{Z_\mathrm{e}}\left[
    1-\langle k|\varphi_2(\beta)\mathbbm{1}_\mathrm{e}|k\rangle- \right. \\ \left.
    2\mathrm{Re}\:\langle k|\varphi_1(t)\mathbbm{1}_\mathrm{e}|k\rangle-
    \langle k|\varphi_3(t,\beta)e^{-\beta(H_\mathrm{e}-\varepsilon_k)}|k\rangle^*
    \right]
\end{split}
\end{equation}
where we have observed that the matrix element $\langle k|\varphi_2(\beta)\mathbbm{1}_\mathrm{e}|k\rangle$ is purely real.

We proceed to find the expression for the full current--current correlation function [Eq.~\eqref{Eq:C_jj_full_iota}] up to the second order in $g$.
Because of Eqs.~\eqref{Eq:def_varphi_2} and~\eqref{Eq:commute_varphi_2}, together with $[j,H_\mathrm{e}]=0$, we can start from
\begin{equation}
\label{Eq:C_jj_full_g2}
\begin{split}
    [C_{jj}(t)]_2=\sum_k j_k^2\frac{e^{-\beta H_\mathrm{e}}}{Z_\mathrm{e}}\left[1-\left\langle k\left|\varphi_2(\beta)\mathbbm{1}_\mathrm{e}\right|k\right\rangle- \right. \\ \left.
    \left\langle k\left|\Phi_1(t)j_k^{-1}je^{-\beta(H_\mathrm{e}-\varepsilon_k)}\right|k\right\rangle- \right. \\ \left.
    \left\langle k\left|\Phi_3(t,\beta)j_k^{-1}je^{-\beta(H_\mathrm{e}-\varepsilon_k)}\right|k\right\rangle
    \right].
\end{split}
\end{equation}

In the following two paragraphs, we demonstrate that
\begin{equation}
\label{Eq:equality_varphi1_Phi1}
    \Delta_1(k,t)=\left\langle k\left|\Phi_1(t)j_k^{-1}je^{-\beta(H_\mathrm{e}-\varepsilon_k)}\right|k\right\rangle-2\mathrm{Re}\:\langle k|\varphi_1(t)\mathbbm{1}_\mathrm{e}|k\rangle=0
\end{equation}
and
\begin{equation}
\label{Eq:equality_varphi3_Phi3}
    \begin{split}
        \Delta_3(k,t)=\left\langle k\left|\Phi_3(t,\beta)j_k^{-1}je^{-\beta(H_\mathrm{e}-\varepsilon_k)}\right|k\right\rangle-
        \langle k|\varphi_3(t,\beta)e^{-\beta(H_\mathrm{e}-\varepsilon_k)}|k\rangle^*=0.
    \end{split}
\end{equation}
Collectively, Eqs.~\eqref{Eq:C_jj_bbl_g2},~\eqref{Eq:C_jj_full_g2},~\eqref{Eq:equality_varphi1_Phi1}, and~\eqref{Eq:equality_varphi3_Phi3} show that, in the lowest order in the electron--phonon coupling $g$, the expressions for the full and bubble current--current correlation function coincide.
Since these terms are the most important ones in the limit $g\to 0$, we can conclude that there are no vertex corrections in the limit of vanishing $g$.

We first prove that $\Delta_1(k,t)=0$. We start from
\begin{equation}
\label{Eq:app_a_1}
\begin{split}
    \left\langle k\left|\Phi_1(t)j_k^{-1}je^{-\beta(H_\mathrm{e}-\varepsilon_k)}\right|k\right\rangle=
    \sum_{qm}\int_0^t ds_2\int_0^{s_2}ds_1\:e^{-\mu_{m}(s_2-s_1)}\left[
    c_{m}\langle k|V_{q}^{(I)}(s_2)V_{-q}^{(I)}(s_1)|k\rangle+
    c_{\overline{m}}\langle k|V_{-q}^{(I)}(s_1)V_q^{(I)}(s_2)|k\rangle- \right. \\ \left.
    c_{m}\left\langle k\left|V_{-q}^{(I)}(s_1)j_k^{-1}je^{-\beta(H_\mathrm{e}-\varepsilon_k)}V_q^{(I)}(s_2)\right| k\right\rangle-
    c_{\overline{m}}\left\langle k\left|V_{q}^{(I)}(s_2)j_k^{-1}je^{-\beta(H_\mathrm{e}-\varepsilon_k)}V_{-q}^{(I)}(s_1)\right| k\right\rangle
    \right].
\end{split}
\end{equation}
Since $V_q^\dagger=V_{-q}$ [see Eq.~\eqref{Eq:def_V_q}], 
one observes that the first two summands within the square brackets in Eq.~\eqref{Eq:app_a_1} are complex conjugates of one another, and the same applies to the last two summands.
On the other hand,
\begin{equation}
    \langle k|\varphi_1(t)\mathbbm{1}_\mathrm{e}|k\rangle=\sum_{qm}\int_0^t ds_2\int_0^{s_2}ds_1\:c_{m}e^{-\mu_{m}(s_2-s_1)}\langle k|V_q^{(I)}(s_2) V_{-q}^{(I)}(s_1)|k\rangle,
\end{equation}
so that the following equation holds
\begin{equation}
\label{Eq:app_a_3}
\begin{split}
    \Delta_1(k,t)&=\left\langle k\left|\Phi_1(t)j_k^{-1}je^{-\beta(H_\mathrm{e}-\varepsilon_k)}\right|k\right\rangle-2\mathrm{Re}\:\langle k|\varphi_1(t)\mathbbm{1}_\mathrm{e}|k\rangle=\\
    &-2\mathrm{Re}\:\sum_{qm}\int_0^t ds_2\int_0^{s_2}ds_1\:c_{\overline{m}}e^{-\mu_{m}(s_2-s_1)}\left\langle k\left|V_{q}^{(I)}(s_2)j_k^{-1}je^{-\beta(H_\mathrm{e}-\varepsilon_k)}V_{-q}^{(I)}(s_1)\right| k\right\rangle.
\end{split}
\end{equation}
An explicit calculation of the matrix element in Eq.~\eqref{Eq:app_a_3} gives
\begin{equation}
    \Delta_1(k,t)=-2\mathrm{Re}\:\sum_{qm}\int_0^t ds_2\int_0^{s_2}ds_1\:c_{\overline{m}}e^{i(\varepsilon_k-\varepsilon_{k-q}+i\mu_{m})(s_2-s_1)}\:\frac{j_{k-q}}{j_k}e^{-\beta(\varepsilon_{k-q}-\varepsilon_k)}.
\end{equation}
The following sum over $q$ should be performed (with $q'=k-q$)
\begin{equation}
\label{Eq:app_a_5}
    \sum_q j_{k-q}e^{-(\beta+it)\varepsilon_{k-q}}=\sum_{q'}j_{q'}e^{-(\beta+it)\varepsilon_{q'}}.
\end{equation}
The second sum in Eq.~\eqref{Eq:app_a_5} is equal to zero because $j_{-q}=-j_q$, while $\varepsilon_{-q}=\varepsilon_q$.
This proves Eq.~\eqref{Eq:equality_varphi1_Phi1}.

Let us now prove that $\Delta_3(k,t)=0$. We start from
\begin{equation}
\label{Eq:app_a_6}
\begin{split}
    \left\langle k\left|\Phi_3(t,\beta)j_k^{-1}je^{-\beta(H_\mathrm{e}-\varepsilon_k)}\right|k\right\rangle=&i\sum_{qm}\int_0^t ds\int_0^{\beta}d\tau\:c_{m}e^{-\mu_{m}s}e^{i\mu_{m}(\beta-\tau)}\times\\
    &\left[\langle k|\overline{V}_{-q}(\tau)V_{q}^{(I)}(s)|k\rangle-\left\langle k\left|V_{q}^{(I)}(s)j_k^{-1}je^{-\beta(H_\mathrm{e}-\varepsilon_k)}\overline{V}_{-q}(\tau)\right|k\right\rangle\right].
\end{split}
\end{equation}
On the other hand, using Eq.~\eqref{Eq:def_varphi_3}, we find that
\begin{equation}
\label{Eq:app_a_7}
\begin{split}
    \langle k|\varphi_3(t,\beta)e^{-\beta(H_\mathrm{e}-\varepsilon_k)}|k\rangle^*&=i\sum_{qm}\int_0^t ds\int_0^\beta d\tau\:c_{m}e^{-\mu_{\overline{m}}s}e^{-i\mu_{\overline{m}}(\beta-\tau)}
    \left\langle k\left|\overline{V}_{q}(-\tau)e^{-\beta(H_\mathrm{e}-\varepsilon_k)}V_{-q}^{(I)}(s)\right|k\right\rangle\\
    &=i\sum_{qm}\int_0^t ds\int_0^\beta d\tau\:c_{\overline{m}}e^{-\mu_{m}s}e^{-i\mu_{m}(\beta-\tau)}
    \left\langle k\left|\overline{V}_{-q}(\beta-\tau)V_{q}^{(I)}(s)\right|k\right\rangle\\
    &=i\sum_{qm}\int_0^t ds\int_0^\beta d\tau\:c_{\overline{m}}e^{-\mu_{m}s}e^{-i\mu_{m}\tau}\left\langle k\left|\overline{V}_{-q}(\tau)V_{q}^{(I)}(s)\right|k\right\rangle\\
    &=i\sum_{qm}\int_0^t ds\int_0^\beta d\tau\:c_{m}e^{-\mu_{m}s}e^{i\mu_{m}(\beta-\tau)}\left\langle k\left|\overline{V}_{-q}(\tau)V_{q}^{(I)}(s)\right|k\right\rangle
\end{split}
\end{equation}
Writing the first equality in Eq.~\eqref{Eq:app_a_7}, we used $\mu_{m}^*=\mu_{\overline{m}}$, as well as $\overline{V}_{-q}(\tau)=\overline{V}_{q}(-\tau)$.
In going from the first to the second equality in Eq.~\eqref{Eq:app_a_7}, we performed the dummy-index change $q\to -q,m\to\overline{m}$, and observed that $\overline{V}_{-q}(-\tau)e^{-\beta H_\mathrm{e}}=e^{-\beta H_\mathrm{e}}\overline{V}_{-q}(\beta-\tau)$.
The third equality is obtained from the second by the integral variable change $\beta-\tau\to\tau$.
The last equality in Eq.~\eqref{Eq:app_a_7} follows from the identity $c_{\overline{m}}e^{-i\mu_{m}\tau}=c_{m}e^{i\mu_{m}(\beta-\tau)}$, which can be checked by direct inspection. Equations~\eqref{Eq:app_a_6} and~\eqref{Eq:app_a_7} imply that
\begin{equation}
\label{Eq:app_a_8}
\begin{split}
    \Delta_3(k,t)&=\left\langle k\left|\Phi_3(t,\beta)j_k^{-1}je^{-\beta(H_\mathrm{e}-\varepsilon_k)}\right|k\right\rangle-\langle k|\varphi_3(t,\beta)e^{-\beta(H_\mathrm{e}-\varepsilon_k)}|k\rangle^*\\&=
    -i\sum_{qm}\int_0^t ds\int_0^{\beta}d\tau\:c_{m}e^{-\mu_{m}s}e^{i\mu_{m}(\beta-\tau)}\left\langle k\left|V_{q}^{(I)}(s)j_k^{-1}je^{-\beta(H_\mathrm{e}-\varepsilon_k)}\overline{V}_{-q}(\tau)\right|k\right\rangle.
\end{split}
\end{equation}
An explicit calculation of the matrix element in Eq.~\eqref{Eq:app_a_8} leads to
\begin{equation}
   \Delta_3(k,t)=-i\sum_{qm}\int_0^t ds\int_0^{\beta}d\tau\:c_{m}e^{i(\varepsilon_k-\varepsilon_{k-q}+i\mu_{m})s}e^{i\mu_{m}\beta}e^{(\varepsilon_{k-q}-\varepsilon_k-i\mu_{m})\tau}\frac{j_{k-q}}{j_k}e^{-\beta(\varepsilon_{k-q}-\varepsilon_k)}.
\end{equation}
The same reasoning as in Eq.~\eqref{Eq:app_a_5} proves that $\Delta_3(k,t)=0$.

\subsection{Second-order cumulant expansion}
Still, Eqs.~\eqref{Eq:C_jj_bbl_g2} or~\eqref{Eq:C_jj_full_g2} do not suffice to obtain an expression for $C_{jj}(t)$ in the weak-coupling limit.
To that end, the perturbation series for $C_{jj}(t)$ in powers of $g$ has to be (at least partially) resummed.
We~\cite{arxiv.2212.13846} and other groups~\cite{PhysRevB.105.224304} have recently promoted the second-order cumulant expansion as a computationally viable and accurate approach to resum the perturbation series for time-dependent quantities in the limit $g\to 0$.
The second-order cumulant expansion starts from Eq.~\eqref{Eq:C_jj_bbl_g2} and produces the following expression for $C_{jj}(t)$:
\begin{equation}
\label{Eq:C_jj_CE_2}
\begin{split}
    &C_{jj}(t)=C_{jj}^\mathrm{bbl}(t)\approx\sum_k j_k^2\frac{e^{-\beta\varepsilon_k}}{Z_\mathrm{e}}e^{-\left\langle k\left|\varphi_2(\beta)\mathbbm{1}_\mathrm{e}\right|k\right\rangle}\times\\
    &e^{-2\mathrm{Re}\:\langle k|\varphi_1(t)\mathbbm{1}_\mathrm{e}|k\rangle-
    \langle k|\varphi_3(t,\beta)e^{-\beta(H_\mathrm{e}-\varepsilon_k)}|k\rangle^*}.
\end{split}
\end{equation}

Up to now, we have not discussed the electronic partition sum $Z_\mathrm{e}$, which has its own perturbation expansion in $g$ that up to second order reads
\begin{equation}
\label{Eq:Z_e_raw}
    Z_\mathrm{e}=\sum_k e^{-\beta\varepsilon_k}\left[1-\langle k|\varphi_2(\beta)\mathbbm{1}_\mathrm{e}|k\rangle\right].
\end{equation}
Performing the second-order cumulant resummation in Eq.~\eqref{Eq:Z_e_raw}, we obtain the following expression for $Z_\mathrm{e}$:
\begin{equation}
\label{Eq:Z_e_CE_2}
    Z_\mathrm{e}=\sum_k e^{-\beta\varepsilon_k}e^{-\left\langle k\left|\varphi_2(\beta)\mathbbm{1}_\mathrm{e}\right|k\right\rangle}.
\end{equation}
We note that Eq.~\eqref{Eq:C_jj_CE_2} suggests that the unnormalized equilibrium occupation of state $|k\rangle$ is proportional to $e^{-\beta\varepsilon_k}e^{-\left\langle k\left|\varphi_2(\beta)\mathbbm{1}_\mathrm{e}\right|k\right\rangle}$, so that $Z_\mathrm{e}$ given in Eq.~\eqref{Eq:Z_e_CE_2} ensures the correct normalization of equilibrium occupations.

We finally list the explicit expressions for the matrix elements needed to evaluate Eqs.~\eqref{Eq:C_jj_CE_2} and~\eqref{Eq:Z_e_CE_2}:
\begin{equation}
\label{Eq:me_varphi_1}
    \langle k|\varphi_1(t)\mathbbm{1}_\mathrm{e}|k\rangle=\frac{g^2}{N}\sum_q\left[
    (1+n_\mathrm{ph})\frac{-e^{i\Delta\varepsilon_-(k,q)t}+i\Delta\varepsilon_-(k,q)t+1}{\Delta\varepsilon_-(k,q)^2}+
    n_\mathrm{ph}\frac{-e^{i\Delta\varepsilon_+(k,q)t}+i\Delta\varepsilon_+(k,q)t+1}{\Delta\varepsilon_+(k,q)^2}
    \right],
\end{equation}
\begin{equation}
   \left\langle k\left|\varphi_2(\beta)\mathbbm{1}_\mathrm{e}\right|k\right\rangle=\frac{g^2}{N}\sum_q\left[
   (1+n_\mathrm{ph})\frac{-e^{\beta\Delta\varepsilon_-(k,q)}+\beta\Delta\varepsilon_-(k,q)+1}{\Delta\varepsilon_-(k,q)^2}+
    n_\mathrm{ph}\frac{-e^{\beta\Delta\varepsilon_+(k,q)}+\beta\Delta\varepsilon_+(k,q)+1}{\Delta\varepsilon_+(k,q)^2}
   \right],
\end{equation}
\begin{equation}
\begin{split}
    \langle k|\varphi_3(t,\beta)e^{-\beta(H_\mathrm{e}-\varepsilon_k)}|k\rangle=\frac{g^2}{N}\sum_q\frac{e^{-\beta\varepsilon_{k-q}}}{e^{-\beta\varepsilon_k}}\left\{
    n_\mathrm{ph}\frac{[e^{i\Delta\varepsilon_-(k,q)t}-1][e^{-\beta\Delta\varepsilon_-(k,q)}-1]}{\Delta\varepsilon_-(k,q)^2}+ \right. \\ \left.
    (1+n_\mathrm{ph})\frac{[e^{i\Delta\varepsilon_+(k,q)t}-1][e^{-\beta\Delta\varepsilon_+(k,q)}-1]}{\Delta\varepsilon_+(k,q)^2}
    \right\},
\end{split}
\end{equation}
where
\begin{equation}
\label{Eq:def_delta_epsilon_pm_k_q}
    \Delta\varepsilon_\pm(k,q)=\varepsilon_k-\varepsilon_{k-q}\pm\omega_0.
\end{equation}

\section{Evaluation of $C_{jj}(t)$ and $C_{jj}^\mathrm{bbl}(t)$ in the $t_0\to 0$ limit}
\label{App:polaronic_matsubara}
Here, we rewrite the Holstein Hamiltonian in the site representation and partition it into the zeroth-order term and the perturbation term as appropriate in the limit $t_0\to 0$:
\begin{equation}
    H=H_0+H_1,
\end{equation}
where ($p$ enumerates lattice sites)
\begin{equation}
    H_0=\omega_0\sum_p b_p^\dagger b_p+g\sum_p c_p^\dagger c_p\left(b_p+b_p^\dagger\right),
\end{equation}
while
\begin{equation}
    H_1=-t_0\sum_p\sum_{\gamma=\pm 1}c_{p+\gamma}^\dagger c_p.
\end{equation}
We perform a unitary transformation of the Hamiltonian
\begin{equation}
    \widetilde{H}=e^SHe^{-S}
\end{equation}
with
\begin{equation}
    S=-\frac{g}{\omega_0}\sum_p c_p^\dagger c_p\left(b_p-b_p^\dagger\right).
\end{equation}
The action of the unitary transformation on the electron operators is
\begin{equation}
    e^Sc_pe^{-S}=c_pX_p,
\end{equation}
with
\begin{equation}
    X_p=\exp\left[\frac{g}{\omega_0}\left(b_p-b_p^\dagger\right)\right],
\end{equation}
while its action on the phonon operators is
\begin{equation}
    e^Sb_pe^{-S}=b_p-\frac{g}{\omega_0}c_p^\dagger c_p.
\end{equation}
Limiting the discussion on the Hilbert space of states that contain one electron, the transformed Hamiltonian takes the form $\widetilde{H}=\widetilde{H}_0+\widetilde{H}_1$, with
\begin{equation}
\label{Eq:H_0_LF}
    \widetilde{H}_0=-\frac{g^2}{\omega_0}+\omega_0\sum_p b_p^\dagger b_p,
\end{equation}
\begin{equation}
\label{Eq:H_1_LF}
    \widetilde{H}_1=-t_0\sum_{p\gamma}X_{p+\gamma,p}c_{p+\gamma}^\dagger c_p.
\end{equation}
We introduced the notation
\begin{equation}
\label{Eq:def_X_p_plus_gamma_p}
    X_{p+\gamma,p}=X_{p+\gamma}^\dagger X_p=\exp\left[\frac{g}{\omega_0}\left(b_p-b_p^\dagger-b_{p+\gamma}+b_{p+\gamma}^\dagger\right)\right].
\end{equation}
The transformed current operator reads as
\begin{equation}
\label{Eq:transformed_j_LF}
    \widetilde{j}=e^Sje^{-S}=-it_0\sum_{p\gamma}\gamma c_{p+\gamma}^\dagger c_p X_{p+\gamma,p}.
\end{equation}
We denote by $z$ the (real or imaginary) time (where $z=t$ for real time and $z=-i\tau$ for imaginary time).
We make use of the identity
\begin{equation}
    \langle j(z)j(0)\rangle_K=\langle\widetilde{j}(z)\widetilde{j}(0)\rangle_{\widetilde{K}}
\end{equation}
and we use Eq.~\eqref{Eq:transformed_j_LF} to obtain
\begin{equation}
\label{Eq:C_jj_LF_1}
    C_{jj}(z)=-\frac{1}{\expval{N_\mathrm{e}}_{\widetilde{K}}} t_0^2\sum_{\substack{ps\\\gamma,\delta=\pm 1}}\gamma\delta\left\langle c_p^\dagger(z)c_{p+\gamma}(z)X_{p,p+\gamma}(z)c_{s}^\dagger(0)c_{s+\delta}(0)X_{s,s+\delta}(0)\right\rangle_{\widetilde{K}}.
\end{equation}
All the averages and time evolutions in Eq.~\eqref{Eq:C_jj_LF_1} should in principle be taken with respect to the operator $\widetilde{K}$.
We are, however, interested in the limit of small $t_0$, and we would like to obtain the first nonzero term with respect to $t_0$.
We note that the factor in front of the sum gives us the $\sim t_0^2$ term.
The expansion of the time evolution operator for real $z$ (the situation is similar in other cases) is of the form
\begin{equation}
    e^{-i\widetilde{K}z}=e^{-i\widetilde{K}_0z}T\exp\left[-i\int_0^z ds\:e^{i\widetilde{K}_0s}\widetilde{H}_1e^{-i\widetilde{K}_0s}\right]=e^{-i\widetilde{K}_0z}\left[1+O(t_0)\right],
\end{equation}
that is, its first nonzero term is of order $\sim 1$ and the remaining terms are of order $\sim t_0$ and smaller.
The same is true for the $e^{-\beta\widetilde{K}}$ operator.
Therefore, to obtain the first nonzero term in Eq.~\eqref{Eq:C_jj_LF_1}, it is sufficient to take the terms of order $\sim 1$ in the expansion of all operators $e^{\pm i\widetilde{K}z}$ and $e^{-\beta \widetilde{K}}$.
This is equivalent to replacing the $\widetilde{K}$ operator with the $\widetilde{K}_0$ operator.
All the averages in Eq.~\eqref{Eq:C_jj_LF_1} can then be obtained by applying the Wick’s theorem.
We thus obtain
\begin{equation}
\begin{split}
    &\left\langle c_p^\dagger(z)c_{p+\gamma}(z)X_{p,p+\gamma}(z)c_{s}^\dagger(0)c_{s+\delta}(0)X_{s,s+\delta}(0)\right\rangle_{\widetilde{K}}\approx\\
    &\left\langle c_p^\dagger(z)c_{p+\gamma}(z)X_{p,p+\gamma}(z)c_{s}^\dagger(0)c_{s+\delta}(0)X_{s,s+\delta}(0)\right\rangle_{\widetilde{K}_0}=\\
    &\left\langle c_p^\dagger(z)c_{p+\gamma}(z)c_s^\dagger(0)c_{s+\delta}(0)\right\rangle_{\widetilde{K}_0}\left\langle X_{p,p+\gamma}(z)X_{s,s+\delta}(0)\right\rangle_{\widetilde{K}_0}.
\end{split}
\end{equation}
By applying the Wick’s theorem to the term with electronic operators we obtain
\begin{equation}
\label{Eq:38}
\begin{split}
    \left\langle c_p^\dagger(z)c_{p+\gamma}(z)c_s^\dagger(0)c_{s+\delta}(0)\right\rangle_{\widetilde{K}_0}&=\left\langle c_p^\dagger(z)c_{s+\delta}(0)\right\rangle_{\widetilde{K}_0}\left\langle c_{p+\gamma}(z)c_s^\dagger(0)\right\rangle_{\widetilde{K}_0}\\
    &+\left\langle c_p^\dagger(z)c_{p+\gamma}(z)\right\rangle_{\widetilde{K}_0}\left\langle c_s^\dagger(0)c_{s+\delta}(0)\right\rangle_{\widetilde{K}_0}\\
    &\approx\delta_{p,s+\delta}\delta_{s,p+\gamma}\left\langle c_p^\dagger(z)c_{p}(0)\right\rangle_{\widetilde{K}_0}\left\langle c_{p+\gamma}(z)c_{p+\gamma}^\dagger(0)\right\rangle_{\widetilde{K}_0}.
\end{split}
\end{equation}
The first term on the right hand side of the first equality is proportional to the number of carriers, while the second term is proportional to the square of the number of carriers. Hence, in the limit of small carrier concentration it is the first term that dominates.
Eventually,
\begin{equation}
\label{Eq:41}
    C_{jj}(z)=\frac{1}{\expval{N_\mathrm{e}}_{\widetilde{K}_0}}t_0^2\sum_{p\gamma}\left\langle c_p^\dagger(z)c_{p}(0)\right\rangle_{\widetilde{K}_0}\left\langle c_{p+\gamma}(z)c_{p+\gamma}^\dagger(0)\right\rangle_{\widetilde{K}_0}\left\langle X_{p,p+\gamma}(z)X_{p+\gamma,p}(0)\right\rangle_{\widetilde{K}_0}.
\end{equation}

We now show that the result of Eq.~\eqref{Eq:41} is recovered in the bubble approximation in which, upon neglecting the term proportional to the square of carrier density, one obtains
\begin{equation}
\begin{split}
    C_{jj}^\mathrm{bbl}(z)&=-\frac{1}{\expval{N_\mathrm{e}}_{\widetilde{K}}}t_0^2\sum_{\substack{ps\\\gamma,\delta=\pm 1}}\gamma\delta\left\langle c_p^\dagger(z)c_{s+\delta}(0)\right\rangle_K\left\langle c_{p+\gamma}(z)c_s^\dagger(0)\right\rangle_K\\
    &=-\frac{1}{\expval{N_\mathrm{e}}_{\widetilde{K}}}t_0^2\sum_{\substack{ps\\\gamma,\delta=\pm 1}}\gamma\delta\left\langle c_p^\dagger(z)c_{s+\delta}(0)X_p^\dagger(z)X_{s+\delta}(0)\right\rangle_{\widetilde{K}}\left\langle c_{p+\gamma}(z)c_s^\dagger(0)X_{p+\gamma}(z)X_s^\dagger(0)\right\rangle_{\widetilde{K}}.
\end{split}
\end{equation}
Following the same reasoning as above, the leading term in the expansion of $C_{jj}^\mathrm{bbl}(z)$ in powers of $t_0\to 0$ is obtained by replacing all averages and time evolutions with respect to $\widetilde{K}$ with those with respect to $\widetilde{K}_0$.
Similarly as in Eq.~\eqref{Eq:38}, we obtain
\begin{equation}
\label{Eq:49}
    C_{jj}^\mathrm{bbl}(z)=\frac{1}{\expval{N_\mathrm{e}}_{\widetilde{K_0}}}t_0^2\sum_{p\gamma}\left\langle c_p^\dagger(z)c_p(0)\right\rangle_{\widetilde{K}_0}\left\langle c_{p+\gamma}(z)c_{p+\gamma}^\dagger(0)\right\rangle_{\widetilde{K}_0}\left\langle X_p^\dagger(z)X_p(0)\right\rangle_{\widetilde{K}_0}\left\langle X_{p+\gamma}(z)X_{p+\gamma}^\dagger(0)\right\rangle_{\widetilde{K}_0}.
\end{equation}
We finally note that
\begin{equation}
\label{Eq:50}
\begin{split}
    \left\langle X_{p,p+\gamma}(z)X_{p+\gamma,p}(0)\right\rangle_{\widetilde{K}_0}&=\left\langle X_{p}^\dagger(z)X_{p+\gamma}(z)X_{p+\gamma}^\dagger(0)X_p(0)\right\rangle_{\widetilde{K}_0}\\
    &=\left\langle X_p^\dagger(z)X_p(0)\right\rangle_{\widetilde{K}_0}\left\langle X_{p+\gamma}(z)X_{p+\gamma}^\dagger(0)\right\rangle_{\widetilde{K}_0},
\end{split}
\end{equation}
where the first equality stems from Eq.~\eqref{Eq:def_X_p_plus_gamma_p}, while the second equality follows from the fact that phonon operators acting on different sites $p$ and $p+\gamma$ commute.
From Eqs.~\eqref{Eq:41},~\eqref{Eq:49}, and~\eqref{Eq:50}, we conclude that $C_{jj}(z)=C_{jj}^\mathrm{bbl}(z)$ as $t_0\to 0$.

In the remainder of this Appendix, we provide explicit expressions for $C_{jj}^\mathrm{bbl}(z)$ to show that the corresponding DC mobility is finite. The phonon term from Eq.~\eqref{Eq:50} is given as
\begin{equation}
\label{Eq:phonon_term_1}
    \left\langle X_{p,p+\gamma}(z)X_{p+\gamma,p}(0)\right\rangle_{\widetilde{K}_0}=\exp\left\{-2\frac{g^2}{\omega_0^2}\left[2n_\mathrm{ph}+1-(n_\mathrm{ph}+1)e^{-i\omega_0z}-n_\mathrm{ph}e^{i\omega_0z}\right]\right\}.
\end{equation}
In the limit of low carrier density, the electronic term from Eq.~\eqref{Eq:49} reads
\begin{equation}
\label{Eq:electronic_term_1}
    \left\langle c_p^\dagger(z)c_{p}(0)\right\rangle_{\widetilde{K}_0}\left\langle c_{p+\gamma}(z)c_{p+\gamma}^\dagger(0)\right\rangle_{\widetilde{K}_0}=n_{\mathrm{F}}.
\end{equation}
where $n_\mathrm{F}$ is the occupation of single-particle electronic state given by the Fermi--Dirac function.
It follows from Eqs.~\eqref{Eq:49}, \eqref{Eq:phonon_term_1} and~\eqref{Eq:electronic_term_1} that $C_{jj}^\mathrm{bbl}(z)$ does not decay to zero as real time goes to infinity. Hence, one would obtain infinite dc mobility by integrating $C_{jj}^\mathrm{bbl}(z)$.
To circumvent this issue, we make use again of the fact that the leading term in the limit of small $t_0$ is approximately the same when all averages and time evolutions are taken either with respect to $\widetilde{K}_0$ or $\widetilde{K}$. Therefore, we now make use of
\begin{equation}
\label{Eq:55-56}
    \left\langle c_p^\dagger(z)c_p(0)\right\rangle_{\widetilde{K}_0}\approx\left\langle c_p^\dagger(z)c_p(0)\right\rangle_{\widetilde{K}},\quad
    \left\langle c_p(z)c_p^\dagger(0)\right\rangle_{\widetilde{K}_0}\approx
    \left\langle c_p(z)c_p^\dagger(0)\right\rangle_{\widetilde{K}}.
\end{equation}
To evaluate the terms $\left\langle c_p^\dagger(z)c_p(0)\right\rangle_{\widetilde{K}}$ and $\left\langle c_p(z)c_p^\dagger(0)\right\rangle_{\widetilde{K}}$ for the Hamiltonian given by Eqs.~\eqref{Eq:H_0_LF} and~\eqref{Eq:H_1_LF} (where the irrelevant shift $-g^2/\omega_0$ in $\widetilde{H}_0$ is neglected), we make use of the Matsubara Green’s function formalism to evaluate the self-energy stemming from the perturbation $\widetilde{H}_1$ to the Hamiltonian $\widetilde{H}_0$.
The overall approach is very similar to that developed in Ref.~\onlinecite{PhysRevB.99.104304}, see in particular its Appendix~B. To first order in interaction we obtain the self-energy
\begin{equation}
    \Sigma_{p\pm 1,p}^{(1)}(\omega)=-t_0e^{-\frac{g^2}{\omega_0^2}(2n_\mathrm{ph}+1)},
\end{equation}
which in the momentum representation reads
\begin{equation}
\label{Eq:first_order_sigma}
    \Sigma_k^{(1)}(\omega)=-2t_\mathrm{eff}\cos(k)
\end{equation}
with
\begin{equation}
    t_\mathrm{eff}=t_0e^{-\frac{g^2}{\omega_0^2}(2n_\mathrm{ph}+1)}.
\end{equation}
It can be seen from Eq.~\eqref{Eq:first_order_sigma} that first-order self-energy describes the formation of bands due to renormalized electronic coupling $t_\mathrm{eff}$.
This term, however, does not have an imaginary part and therefore it does not lead to energy level broadening. Hence it is not expected that it will provide energy dissipation of electronic system, which is a requirement for finite dc mobility to occur. For this reason, we proceed to evaluate the self-energy term arising from second order terms in interaction. We obtain that the dominant second order term is the local term
\begin{equation}
\label{Eq:Sigma_2_R_omega}
    \Sigma^{(2)}\qty(\omega)=2t_0^2\frac{i}{2\pi}\int_{-\infty}^{+\infty}d\omega_1 X^>(\omega_1)G^{R}(\omega-\omega_1),
\end{equation}
where $G^{R}$ is the retarded Green's function, and $X^>(\omega)=\int dt\:e^{i\omega t}X^>(t)$ with
\begin{equation}
    X^>(t)=-i\exp\left\{-2\frac{g^2}{\omega_0^2}\left[2n_\mathrm{ph}+1-(n_\mathrm{ph}+1)e^{-i\omega_0t}-n_\mathrm{ph}e^{i\omega_0 t}\right]\right\}.
\end{equation}
We can further transform Eq.~\eqref{Eq:Sigma_2_R_omega} into the form (we now omit superscripts $R, 2$ for brevity)
\begin{equation}
\label{Eq:70}
    \Sigma(\omega)=2t_0^2\frac{i}{2\pi}\int_{-\infty}^{+\infty}d\omega_1 X^>(\omega_1)G(\omega-\omega_1)
\end{equation}
that is amenable to the self-consistent treatment in conjunction with the Dyson equation
\begin{equation}
\label{Eq:Dyson_appendix}
    [\omega-\Sigma(\omega)]G(\omega)=1.
\end{equation}
Making use of the identities
\begin{equation}
    e^{a\cos\theta}=\sum_{l=-\infty}^{+\infty}I_l(a)e^{il\theta}
\end{equation}
where $I_l$ is the modified Bessel function of the first kind of order $l$ and
\begin{equation}
  (n_\mathrm{ph}+1)e^{-i\omega_0t}+n_\mathrm{ph}e^{i\omega_0 t}=2\sqrt{n_\mathrm{ph}(n_\mathrm{ph}+1)}\cos\left[\omega_0\left(t+i\frac{\beta}{2}\right)\right],
\end{equation}
we recast Eq.~\eqref{Eq:70} as
\begin{equation}
\label{Eq:76}
    \Sigma(\omega)=\sum_{l=-\infty}^{+\infty}c_lG(\omega+l\omega_0),
\end{equation}
with $c_l$ given by
\begin{equation}
\label{Eq:def_c_l}
    c_l=2t_0^2e^{-2\frac{g^2}{\omega_0^2}(2n_\mathrm{ph}+1)}e^{-l\frac{\beta\omega_0}{2}}I_l\left[4\frac{g^2}{\omega_0^2}\sqrt{n_\mathrm{ph}(n_\mathrm{ph}+1)}\right].
\end{equation}
Equations~\eqref{Eq:Dyson_appendix} and~\eqref{Eq:76} can be solved using a self-consistent procedure. Alternatively,
these equations can be solved analytically by introducing an additional approximation that the dominant term in the sum in Eq.~\eqref{Eq:76} is the $l=0$ term.
This is a reasonable assumption as one might expect that both $G(\omega)$ and $\Sigma(\omega)$ should have maximal values in the region around $\omega=0$.
This assumption can always be checked by comparing the result with the self-consistent solution of Eqs.~\eqref{Eq:Dyson_appendix} and~\eqref{Eq:76}.
Under this assumption, Eqs.~\eqref{Eq:Dyson_appendix} and~\eqref{Eq:76} reduce to a single quadratic equation for $G(\omega)$ and we obtain the corresponding spectral function as
\begin{equation}
\label{Eq:78}
    A(\omega)=-\frac{1}{\pi}\Im G(\omega)=\frac{\sqrt{4c_0-\omega^2}}{2\pi c_0}\theta(4c_0-\omega^2).
\end{equation}
From the relation between the spectral function and the correlation functions of creation and annihilation operators, we obtain
\begin{equation}
\label{Eq:79}
    \frac{1}{\expval{N_\mathrm{e}}_{\widetilde{K}}}\left\langle c_p^\dagger(z)c_{p}(0)\right\rangle_{\widetilde{K}}\left\langle c_{p+\gamma}(z)c_{p+\gamma}^\dagger(0)\right\rangle_{\widetilde{K}}=\frac{\int d\omega\:e^{-\beta(\omega-iz)}A(\omega)}{\int d\omega\:e^{-\beta\omega}A(\omega)}\int d\omega\:e^{-i\omega z}A(\omega).
\end{equation}
The integrals in the previous equation are all of the form
\begin{equation}
    f(z)=\frac{2}{\pi a^2}\int_{-a}^a d\omega\:e^{-z\omega}\sqrt{a^2-\omega^2}=-\frac{2}{az}I_1(-az),
\end{equation}
with $a=2\sqrt{c_0}$.
The integral is solved by introducing the substitution $\omega=a\cos t$, by making use of $I_n(u)=\frac{1}{\pi}\int_0^\pi d\theta\:e^{u\cos\theta}\cos(n\theta)$, and by using the identity $I_\nu(z)-I_{\nu+2}(z)=2\frac{\nu+1}{z}I_{\nu+1}(z)$.
Combining Eqs.~\eqref{Eq:41},~\eqref{Eq:55-56},~\eqref{Eq:phonon_term_1},~\eqref{Eq:78}, and~\eqref{Eq:79}, we finally obtain the expression given in Eq.~\eqref{Eq:no_vtx_t_0_to_0}, where $J_n$ denotes the Bessel function of the first kind of order $n$.

It is worth investigating the asymptotic behavior of the prefactor in front of the exponential term in Eq.~\eqref{Eq:no_vtx_t_0_to_0} when real time $t$ tends to infinity $z=t\to+\infty$. The time dependence of this prefactor is determined by the term
\begin{equation}
    g(t)=\frac{I_1[-2(\beta-it)\sqrt{c_0}]J_1(2t\sqrt{c_0})}{t(\beta-it)}.
\end{equation}
As $t\to+\infty$, making use of $J_1\qty(x)=-iI_1\qty(-ix)$, we have
\begin{equation}
    g(t)\sim\frac{J_1(-2t\sqrt{c_0})J_1(2t\sqrt{c_0})}{t^2}.
\end{equation}
The asymptotic behavior of the Bessel function when $x\to+\infty$ is
\begin{equation}
    J_1(x)\sim\sqrt{\frac{2}{\pi x}}\cos\left(x-\frac{3\pi}{4}\right).
\end{equation}
Consequently, $g(t)\sim t^{-3}$ as $t\to+\infty$. This is sufficiently fast convergence to make the time integral of the $C_{jj}\qty(t)$  finite, which then leads to finite dc mobility.

\section{Evaluation of $C_{jj}(t)$ and $C_{jj}^\mathrm{bbl}(t)$ in the $\beta\to 0$ limit}\label{App:beta_to_0}
To prove that there are no vertex corrections in the infinite-temperature limit, it is instrumental to first analyze in more detail the single-site limit of the Holstein Hamiltonian in which phonons can be treated as classical harmonic oscillators.
The appropriate Hamiltonian reads (the oscillator mass is set to unity)
\begin{equation}
    h=\underbrace{\frac{p^2}{2}+\frac{\omega_0^2}{2}x^2}_{h_\mathrm{ph}(x,p)}+Cxc^\dagger c,
\end{equation}
where $x$ and $p$ are respectively the classical coordinate and momentum of the oscillator, while $C=g\sqrt{2\omega_0}$.
The greater Green's function can be expressed as [see also Eq.~\eqref{Eq:G_gtr_app}]
\begin{equation}
\label{Eq:G_gtr_single_site_beta_to_0}
    \mathcal{G}^>(t)=-ie^{i\mu_\mathrm{F}t}\frac{\int dx\:dp\:e^{-\beta h_\mathrm{ph}(x,p)}e^{-iCxt}}{\int dx\:dp\:e^{-\beta h_\mathrm{ph}(x,p)}}=-ie^{i\mu_\mathrm{F}t}\frac{\int dx\:e^{-\beta\frac{\omega_0^2x^2}{2}-iCxt}}{\int dx\:e^{-\beta\frac{\omega_0^2x^2}{2}}}=
    -ie^{i\mu_\mathrm{F}t}e^{-\frac{\sigma^2t^2}{2}}.
\end{equation}
In the same vein, the lesser Green's function can be expressed as [see also Eq.~\eqref{Eq:G_less_app}]
\begin{equation}
\label{Eq:G_less_single_site_beta_to_0}
    \mathcal{G}^<(t)=ie^{i\mu_\mathrm{F}t}e^{\beta\mu_\mathrm{F}}\frac{\int dx\:dp\:e^{-\beta h_\mathrm{ph}}e^{-\beta Cx}e^{-iCxt}}{\int dx\:dp\:e^{-\beta h_\mathrm{ph}}}=ie^{i\mu_\mathrm{F}t}e^{\beta\mu_\mathrm{F}}\frac{\int dx\:e^{-\beta\frac{\omega_0^2x^2}{2}}e^{-\beta Cx}e^{-iCxt}}{\int dx\:e^{-\beta\frac{\omega_0^2x^2}{2}}}=
    ie^{i\mu_\mathrm{F}t}e^{\beta\mu_\mathrm{F}}e^{\frac{\sigma^2}{2}(\beta+it)^2}.
\end{equation}
The electron number is
\begin{equation}
\label{Eq:N_e_per_site_beta_to_0}
    \langle c^\dagger c\rangle_K=e^{\beta\mu_\mathrm{F}}e^{\frac{\sigma^2\beta^2}{2}}.
\end{equation}

Let us now consider an $N$-site chain and start from Eq.~\eqref{Eq:C_jj_bubble} determining $C_{jj}^\mathrm{bbl}(t)$.
Because of the locality of single-particle correlation functions at high temperatures, we can replace $\mathcal{G}^>(k,t)$ and $\mathcal{G}^<(k,t)$ by the single-site expressions derived in Eqs.~\eqref{Eq:G_gtr_single_site_beta_to_0} and~\eqref{Eq:G_less_single_site_beta_to_0}, respectively.
The remaining sum in the numerator of Eq.~\eqref{Eq:C_jj_bubble} is readily evaluated using Eq.~\eqref{Eq:j_k}, $\sum_k j_k^2=2t_0^2N$.
The total electron number $\langle N_\mathrm{e}\rangle_K$ is $N$ times the electron number per site, which is derived in Eq.~\eqref{Eq:N_e_per_site_beta_to_0}.
Collecting all pieces together, we obtain the result embodied in Eq.~\eqref{Eq:C_jj_beta_to_0_no_vtx}.

We continue by applying the same approximations to Eq.~\eqref{Eq:C_jj_t_low_density_limit} defining $C_{jj}(t)$.
Because of the assumed locality and classicality of phonons, we can approximate the Hamiltonian [Eq.~\eqref{Eq:def_H}] as
\begin{equation}
    H\approx\underbrace{\sum_r\left(\frac{p_r^2}{2}+\frac{\omega_0^2}{2}x_r^2\right)}_{H_\mathrm{ph}(\mathbf{x},\mathbf{p})}+C\sum_r x_r c_r^\dagger c_r,
\end{equation}
where $\mathbf{x}$ and $\mathbf{p}$ respectively denote (classical) coordinates and momenta of oscillators.
Since the traces in Eq.~\eqref{Eq:C_jj_t_low_density_limit} are to be evaluated over the single-electron subspace, we can replace $c_r^\dagger c_r\to|r\rangle\langle r|$, so that
\begin{equation}
\label{Eq:e_to_alpha_H_high_T}
    e^{-\alpha H}\approx e^{-\alpha H_\mathrm{ph}(\mathbf{x},\mathbf{p})}\sum_r e^{-\alpha Cx_r}|r\rangle\langle r|.
\end{equation}
Inserting Eq.~\eqref{Eq:e_to_alpha_H_high_T} (with $\alpha=\beta,\pm it$) into Eq.~\eqref{Eq:C_jj_t_low_density_limit}, one obtains
\begin{equation}
\label{Eq:C_jj_full_beta_to_0_appendix}
\begin{split}
    C_{jj}(t)=\frac{\int d\mathbf{x}\:d\mathbf{p}\:e^{-\beta H_\mathrm{ph}(\mathbf{x},\mathbf{p})}\sum_{r_1r}|\langle r_1|j|r\rangle|^2\:e^{-itCx_{r_1}}e^{-(\beta-it)Cx_{r}}}{\int d\mathbf{x}\:d\mathbf{p}\:\sum_r e^{-\beta H_\mathrm{ph}(\mathbf{x},\mathbf{p})}e^{-\beta Cx_r}}.
\end{split}
\end{equation}
Because of the current operator in the real-space representation is $j=-it_0\sum_{r\gamma}|r+\gamma\rangle\langle r|$, we have $r_1=r+\gamma$, where $\gamma=\pm 1$, and Eq.~\eqref{Eq:C_jj_full_beta_to_0_appendix} is recast as
\begin{equation}
\label{Eq:C_jj_full_beta_to_0_appendix_penult}
    C_{jj}(t)=\frac{t_0^2\sum_r\sum_{\gamma=\pm 1}\left[\int dx_r\:e^{-\beta\frac{\omega_0^2x_r^2}{2}}e^{-(\beta-it)Cx_r}\right]\left[\int dx_{r+\gamma}\:e^{-\beta\frac{\omega_0^2x_{r+\gamma}^2}{2}}e^{-itCx_{r+\gamma}}\right]\prod_{s\notin\{r,r+\gamma\}}\int dx_s\:e^{-\beta\frac{\omega_0^2x_s^2}{2}}}{\sum_r\left[\int dx_r\:e^{-\beta\frac{\omega_0^2x_r^2}{2}}e^{-\beta Cx_r}\right]\prod_{s\neq r}\int dx_s\:e^{-\beta\frac{\omega_0^2x_s^2}{2}}}.
\end{equation}
The integrals in the numerator have been evaluated in the single-site limit; see Eqs.~\eqref{Eq:G_gtr_single_site_beta_to_0} and~\eqref{Eq:G_less_single_site_beta_to_0}.
Upon inserting the corresponding results in Eq.~\eqref{Eq:C_jj_full_beta_to_0_appendix_penult}, and performing the remaining sums (which produce the factor of $2N$ in the numerator and $N$ in the denominator), one immediately obtains Eq.~\eqref{Eq:C_jj_beta_to_0_no_vtx}.
\end{widetext}

\bibliography{apssamp}

\end{document}


\title{Supplemental Material for\\
Vertex corrections to conductivity in the Holstein model:\\
A numerical--analytical study}
\author{Veljko Jankovi\'c}
 \email{veljko.jankovic@ipb.ac.rs}
 \affiliation{Institute of Physics Belgrade, University of Belgrade, Pregrevica 118, 11080 Belgrade, Serbia}
\author{Petar Mitri\'c}
 \email{petar.mitric@ipb.ac.rs}
 \affiliation{Institute of Physics Belgrade, University of Belgrade, Pregrevica 118, 11080 Belgrade, Serbia}
\author{Darko Tanaskovi\'c}
 \email{darko.tanaskovic@ipb.ac.rs}
 \affiliation{Institute of Physics Belgrade, University of Belgrade, Pregrevica 118, 11080 Belgrade, Serbia}
\author{Nenad Vukmirovi\'c}
 \email{nenad.vukmirovic@ipb.ac.rs}
 \affiliation{Institute of Physics Belgrade, University of Belgrade, Pregrevica 118, 11080 Belgrade, Serbia}

\maketitle


\section{HEOM}
For the sake of completeness, here, we present the equations that we solve to obtain numerically exact and bubble-approximation results for the current--current correlation function and the dynamical-mobility profile.
\subsection{Dynamical equations of the HEOM for $C_{jj}(t)$}
As we have discussed in Ref.~\onlinecite{JChemPhys.159.094113}, the totally symmetric ($q=0$) phonon mode does not affect the dynamics of $C_{jj}=\mathrm{Tr}_\mathrm{e}\{j\iota(t)\}$, which follows from the evolution of the purely electronic operator
\begin{equation}
    \iota(t)=Z^{-1}\mathrm{Tr}_{\mathrm{ph}}\{e^{-iHt}je^{-\beta H}e^{iHt}\}
\end{equation}
The operator $\iota(t)$ is at the root of the hierarchy of dynamical equations, whose higher-order members $\iota_\mathbf{n}^{(n)}(t)$ describe the dynamics of $n-$phonon assisted processes determined by the vector $\mathbf{n}=\{n_{qm}|q\neq 0;m=0,1\}$ of non-negative integers $n_{qm}$ that obey $\sideset{}{'}\sum_{qm}n_{qm}=n$.
The total momentum exchanged between the electron and phonons in the phonon-assisted process described by $\mathbf{n}$ is $k_\mathbf{n}=\sideset{}{'}\sum_{qm}qn_{qm}$.
In the following, primed sums over $q$ exclude the $q=0$ term.
By virtue of the translational symmetry, the only non-zero matrix elements of $\iota_\mathbf{n}^{(n)}(t)$ are $\langle k|\iota_\mathbf{n}^{(n)}(t)|k+k_\mathbf{n}\rangle$, and their time evolution is governed by
\begin{equation}
    \label{Eq:real-time-HEOM-before-closing-paper}
    \begin{split}
        &\partial_{t}\langle k|{\iota}_\mathbf{n}^{(n)}({t})|k+k_\mathbf{n}\rangle=\\
        &-i(\varepsilon_k-\varepsilon_{k+k_\mathbf{n}}+{\mu}_\mathbf{n})\langle k|{\iota}_\mathbf{n}^{(n)}({t})|k+k_\mathbf{n}\rangle\\
        &+i\sideset{}{'}\sum_{qm}\sqrt{(1+n_{qm})c_{m}}\:\langle k-q|{\iota}_{\mathbf{n}_{qm}^+}^{(n+1)}({t})|k+k_\mathbf{n}\rangle\\
        &-i\sideset{}{'}\sum_{qm}\sqrt{(1+n_{qm})c_{m}}\:\langle k|{\iota}_{\mathbf{n}_{qm}^+}^{(n+1)}({t})|k+k_\mathbf{n}+q\rangle\\
        &+i\sideset{}{'}\sum_{qm}\sqrt{n_{qm}c_{m}}\:\langle k+q|{\iota}_{\mathbf{n}_{qm}^-}^{(n-1)}({t})|k+k_\mathbf{n}\rangle\\
        &-i\sideset{}{'}\sum_{qm}\sqrt{n_{qm}}\frac{c_{\overline{m}}}{\sqrt{c_{m}}}\:\langle k|{\iota}_{\mathbf{n}_{qm}^-}^{(n-1)}({t})|k+k_\mathbf{n}-q\rangle\\
        &+\left[\partial_{t}\langle k|{\iota}_\mathbf{n}^{(n)}({t})|k+k_\mathbf{n}\rangle\right]_\mathrm{close}.
    \end{split}
\end{equation}
In Eq.~\eqref{Eq:real-time-HEOM-before-closing-paper}, ${\mu}_\mathbf{n}=\omega_0\sum'_{q}(n_{q0}-n_{q1})$.
The operator $\iota_\mathbf{n}^{(n)}(t)$ couples to analogous operators at depths $n\pm 1$, which are characterized by vectors $\mathbf{n}_{qm}^\pm$ whose components are $\left[\mathbf{n}_{qm}^\pm\right]_{q'm'}=n_{q'm'}\pm\delta_{q'q}\delta_{m'm}$.
The coefficients $c_m$ are defined in Eqs.~\textcolor{red}{(A8) and (A9) of Appendix~A}.
The last term on the right-hand side of Eq.~\eqref{Eq:real-time-HEOM-before-closing-paper} represents the closing term, which renders the HEOM truncated at the maximum depth $D$ numerically stable.
In Ref.~\onlinecite{JChemPhys.159.094113}, we have checked that the following closing term
\begin{equation}
\label{Eq:closing_strategy}
\begin{split}
    &\left[\partial_{{t}}\langle k|{\iota}_\mathbf{n}^{(n)}({t})|k+k_\mathbf{n}\rangle\right]_\mathrm{close}=\\&-\delta_{n,D}\frac{1}{2}\left({\tau}_k^{-1}+{\tau}_{k+k_\mathbf{n}}^{-1}\right)\langle k|{\iota}_\mathbf{n}^{(n)}({t})|k+k_\mathbf{n}\rangle
\end{split}
\end{equation}
stabilizes the HEOM in Eq.~\eqref{Eq:real-time-HEOM-before-closing-paper} without compromising final results for the dynamical-mobility profile.
The closing term comprises the rates at which the electron is scattered out of the free-electron state $|k\rangle$
\begin{equation}
\label{Eq:def_tau_k}
\begin{split}
    \tau_k^{-1}=2\pi\frac{g^2}{N}\sideset{}{'}\sum_q\left[(1+n_\mathrm{ph})\delta(\varepsilon_k-\varepsilon_{k-q}-\omega_0) \right. \\ \left. +n_\mathrm{ph}\delta(\varepsilon_k-\varepsilon_{k-q}+\omega_0)\right],
\end{split}
\end{equation}
which can be computed analytically in the limit $N\to\infty$.
\subsection{Dynamical equations of the HEOM for $G^\gtrless(k,t)$}
The dynamics of the current--current correlation function in the bubble approximation follows from the dynamics of $G^>(k,t)$ and $G^<(k,t)$ as defined in Eqs.~\textcolor{red}{(A13) and (A14) of Appendix A}.
Both quantities can be computed by solving the same set of dynamical equations that have been presented in Ref.~\onlinecite{PhysRevB.105.054311} and read as
\begin{equation}
\label{Eq:HEOM-greater-scaled}
 \begin{split}
 &\partial_{t}{G}^{(\gtrless,n)}_\mathbf{n}\left(k-k_\mathbf{n},{t}\right)=\\&-i(\varepsilon_{k-k_\mathbf{n}}+\mu_\mathbf{n})G^{(\gtrless,n)}_\mathbf{n}\left(k-k_\mathbf{n},{t}\right)\\
 &+i\sideset{}{'}\sum_{qm}\sqrt{1+n_{qm}}\sqrt{c_{m}}\:{G}^{(\gtrless,n+1)}_{\mathbf{n}_{qm}^+}\left(k-k_\mathbf{n}-q,{t}\right)\\
 &+i\sideset{}{'}\sum_{qm}\sqrt{n_{qm}}\sqrt{c_{m}}\:{G}^{(\gtrless,n-1)}_{\mathbf{n}_{qm}^-}\left(k-k_\mathbf{n}+q,{t}\right)\\
 &+[\partial_{t}{G}^{(\gtrless,n)}_\mathbf{n}\left(k-k_\mathbf{n},{t}\right)]_\mathrm{close}.
 \end{split}
\end{equation}

To ensure that $G^\gtrless$ decays to zero at sufficiently long times, we use the closing term $[\partial_{t}{G}^{(\gtrless,n)}_\mathbf{n}\left(k-k_\mathbf{n},{t}\right)]_\mathrm{close}$.
Its form after truncation of Eq.~\eqref{Eq:HEOM-greater-scaled} at the maximum depth $D$ can be derived along the lines presented in Ref.~\onlinecite{JChemPhys.159.094113}.
The final expression for the closing term is analogous to that presented in Eq.~\eqref{Eq:closing_strategy} and reads as
\begin{equation}
\begin{split}
    &[\partial_{t}{G}^{(\gtrless,n)}_\mathbf{n}\left(k-k_\mathbf{n},{t}\right)]_\mathrm{close}=\\&-\delta_{n,D}\frac{1}{2}{\tau}_{k-k_\mathbf{n}}^{-1}{G}^{(\gtrless,n)}_\mathbf{n}\left(k-k_\mathbf{n},{t}\right),
\end{split}
\end{equation}
where $\tau_k$ is defined in Eq.~\eqref{Eq:def_tau_k}.

The propagation of Eq.~\eqref{Eq:HEOM-greater-scaled} is additionally stabilized by transferring to the rotating-wave frame and solving equations for the envelope $\widetilde{G}_\mathbf{n}^{(\gtrless,n)}(t)$ defined as 
\begin{equation}
\label{Eq:def-envelope}
\begin{split}
 {G}^{(\gtrless,n)}_\mathbf{n}(k-k_\mathbf{n},{t})=&\exp\left[-i(\varepsilon_{k-k_\mathbf{n}}+\mu_\mathbf{n}){t}\right]\times\\&\widetilde{{G}}^{(\gtrless,n)}_\mathbf{n}(k-k_\mathbf{n},{t}).
\end{split}
\end{equation}

We note that the HEOM in Eq.~\eqref{Eq:HEOM-greater-scaled} has a smaller number of equations than the HEOM we solved in Ref.~\onlinecite{PhysRevB.105.054311}.
This reduction in HEOM size is possible because the part of the dynamics governed by the zero-momentum phonon mode can be solved analytically.
Additionally, this implies $G^{\gtrless}(k,t)\neq G^{(\gtrless,0)}_\mathbf{0}(k,t)$, and we now establish the relationship between the quantity of our interest $G^{\gtrless}(k,t)$ and the quantity at the root of the hierarchy $G^{(\gtrless,0)}_\mathbf{0}(k,t)$. 
The $q=0$ phonon mode couples to the unit operator in the subspace containing a single electron, i.e., $V_{q=0}=\mathbbm{1}_\mathrm{1e}$, meaning that the action of the corresponding $q=0$ components of the influence phases $\varphi_1(t)$ and $\varphi_3(t,\beta)$ \textcolor{red}{[see Eqs.~(A16) and (A18) of Appendix A]} can be evaluated analytically.
The analytical procedure is, up to prefactors $N^{-1}$, identical to that presented in Sec.~\ref{Sec:vanish-vertex-t0-to-0}, and produces the following final results for $G^\gtrless(k,t)$:
\begin{equation}
\begin{split}
    &G^{>}(k,t)=\widetilde{{G}}^{(>,0)}_\mathbf{0}(k,{t})e^{-i\varepsilon_kt}\\&\times\exp\left[-\sum_m c_m\frac{e^{-\mu_m t}+\mu_m t-1}{\mu_m^2}\right]\\
    &=\widetilde{{G}}^{(>,0)}_\mathbf{0}(k,{t})e^{-i\varepsilon_kt}\exp\left[-\frac{g^2}{\omega_0^2N}(1+2n_\mathrm{ph})\right]\\
    &\times\exp\left[\frac{g^2}{\omega_0^2N}\left((1+n_\mathrm{ph})e^{-i\omega_0t}+n_\mathrm{ph}e^{i\omega_0t}+i\omega_0t\right)\right],
\end{split}
\end{equation}

\begin{equation}
    \begin{split}
      &G^{<}(k,t)=\widetilde{{G}}^{(<,0)}_\mathbf{0}(k,{t})e^{-i\varepsilon_kt}\\&\times\exp\left[-\sum_m \frac{c_m}{\mu_m^2}\left(e^{-\mu_m t}+\mu_m t-1\right)\right]\\
      &\times\exp\left[\sum_m\frac{c_m}{\mu_m^2}(e^{-\mu_m t}-1)(1-e^{i\beta\mu_m})\right]\\
      &=\widetilde{{G}}^{(<,0)}_\mathbf{0}(k,{t})e^{-i\varepsilon_kt}\exp\left[-\frac{g^2}{\omega_0^2N}(1+2n_\mathrm{ph})\right]\\
    &\times\exp\left[\frac{g^2}{\omega_0^2N}\left(n_\mathrm{ph}e^{-i\omega_0t}+(1+n_\mathrm{ph})e^{i\omega_0t}+i\omega_0t\right)\right].
    \end{split}
\end{equation}
\subsection{Initial conditions}
The initial condition under which the HEOM for $G^>$ [Eq.~\eqref{Eq:HEOM-greater-scaled}] is solved reads as~\cite{PhysRevB.105.054311}
\begin{equation}
    G^{(>,n)}_\mathbf{n}(k-k_\mathbf{n},0)=-i\delta_{n,0}.
\end{equation}

The initial conditions under which the HEOM for $\iota(t)$ [Eq.~\eqref{Eq:real-time-HEOM-before-closing-paper}] and $G^<$ [Eq.~\eqref{Eq:HEOM-greater-scaled}] is solved is determined by the equilibrium state of the interacting electron--phonon system~\cite{JChemPhys.159.094113,PhysRevB.105.054311}.
In Ref.~\onlinecite{PhysRevB.105.054311}, we derived that the hierarchical representation of this state is obtained by propagating the following imaginary-time HEOM:
\begin{equation}
    \label{Eq:im-time-HEOM-eq-paper}
    \begin{split}
        &\partial_{{\tau}}\langle k|\sigma_\mathbf{n}^{(n)}({\tau})|k+k_\mathbf{n}\rangle=\\
        &-(\varepsilon_k+{\mu}_\mathbf{n})\langle k|\sigma_\mathbf{n}^{(n)}({\tau})|k+k_\mathbf{n}\rangle\\
        &+\sideset{}{'}\sum_{qm}\sqrt{(1+n_{qm})c_{m}}\langle k-q|\sigma_{\mathbf{n}_{qm}^+}^{(n+1)}({\tau})|k+k_\mathbf{n}\rangle\\
        &+\sideset{}{'}\sum_{qm}\sqrt{n_{qm}c_{m}}\langle k+q|\sigma_{\mathbf{n}_{qm}^-}^{(n-1)}({\tau})|k+k_\mathbf{n}\rangle.
    \end{split}
\end{equation}
Equations~\eqref{Eq:im-time-HEOM-eq-paper} are propagated from $\tau=0$ to $\tau=\beta$ with the initial condition $\langle k|\sigma_\mathbf{n}^{(n)}({\tau})|k+k_\mathbf{n}\rangle=\delta_{n,0}$, which is representative of the electron--phonon system at infinite temperature.
The initial condition for the propagation of Eq.~\eqref{Eq:real-time-HEOM-before-closing-paper} then reads as
\begin{equation}
\label{Eq:init_cond_Cjj}
\begin{split}
    &\langle k|{\iota}_\mathbf{n}^{(n)}(0)|k+k_\mathbf{n}\rangle=
    Z_\mathrm{e}^{-1}\times\\&(-2J)\sin(k)\langle k|\sigma_\mathbf{n}^{(n)}(\beta)|k+k_\mathbf{n}\rangle,
\end{split}
\end{equation}
while the initial condition for the propagation of Eq.~\eqref{Eq:HEOM-greater-scaled} governing the dynamics of $G^<$ is
\begin{equation}
\label{Eq:init_cond_g_less}
 G^{(<,n)}_{\mathbf{n}}(k-k_\mathbf{n},0)=Z_\mathrm{e}^{-1}\times i\left\langle k\left|\sigma_\mathbf{n}^{(n)}(\beta)\right|k+k_\mathbf{n}\right\rangle.
\end{equation}
The so-called electronic partition sum $Z_\mathrm{e}=Z/Z_\mathrm{ph}$ entering Eqs.~\eqref{Eq:init_cond_Cjj} and~\eqref{Eq:init_cond_g_less} reads as
\begin{equation}
    \label{Eq:el_part_sum}
    Z_\mathrm{e}=\sum_p\langle p|\sigma_\mathbf{0}^{(0)}(\beta)|p\rangle.
\end{equation}
\subsection{Propagation algorithm}
The HEOM embodied in Eq.~\eqref{Eq:real-time-HEOM-before-closing-paper} is propagated using the scheme originally proposed in Ref.~\onlinecite{JChemTheoryComput.11.3411}.
We use the fourth-order Wilkins--Dattani scheme with the time step $\omega_0\Delta t=(1-2)\times 10^{-2}$, depending on the values of model parameters.

When the HEOM embodied in Eq.~\eqref{Eq:HEOM-greater-scaled} is recast as the HEOM for the envelope $\widetilde{G}^\gtrless$ [Eq.~\eqref{Eq:def-envelope}], it becomes a system of first-order linear differential equations with time-dependent coefficients.
Because of its non-constant coefficients, the resulting HEOM for the envelope is propagated using the fourth-order Runge--Kutta algorithm~\cite{PresTeukVettFlan92} with the time step $\omega_0\Delta t=(1-2)\times 10^{-2}$, depending on the values of model parameters.

\section{Vanishing vertex corrections in the limit $t_0\to 0$: Insights from formally exact expressions}
\label{Sec:vanish-vertex-t0-to-0}
Starting from the formally exact expressions of Appendix~A, this section presents the proof of the equality $C_{jj}(t)=C_{jj}^\mathrm{bbl}(t)$ in the limit $t_0\to 0$.

As argued in the main text, the dominant term in the expansion of $C_{jj}(t)$ and $C_{jj}^\mathrm{bbl}(t)$ in powers of small $t_0$ is proportional to $t_0^2$.
This term can be obtained from the formally exact expressions by replacing all operators $e^{-\alpha H_\mathrm{e}}$ by the unit operator $\mathbbm{1}_{1\mathrm{e}}$ in the subspace containing a single electron.
In particular, this means that all the real-time/imaginary-time hyperoperators entering the expressions for influence phases become time-independent, which makes the time-ordering sign ineffective and permits us to perform the real-time and/or imaginary-time integrals and sums over $q$ analytically. 

Let us first consider how these simplifications are reflected in the influence phases $\varphi_1(t),\varphi_2(\beta),$ and $\varphi_3(t,\beta)$ governing the single-particle dynamics.
The phase $\varphi_1(t)$ \textcolor{red}{[Eq.~(A16) of Appendix~A]} simplifies to
\begin{equation}
    \varphi_1(t)=\left[\sum_q V_q^CV_{-q}^C\right]\left[\sum_m\frac{c_m}{\mu_m^2}\left(e^{-\mu_m t}+\mu_m t-1\right)\right].
\end{equation}
Because of $V_qV_{-q}=\mathbbm{1}_{1\mathrm{e}}$, we can replace sum over $q$ by $N$, meaning that $\varphi_1(t)$ acts as a scalar that reads as
\begin{equation}
\label{Eq:varphi_1_final_action}
\begin{split}
    &\varphi_1(t)=-\frac{g^2}{\omega_0^2}\times\\&\left[(1+n_\mathrm{ph})e^{-i\omega_0t}+n_\mathrm{ph}e^{i\omega_0t}+i\omega_0t-(1+2n_\mathrm{ph})\right].
\end{split}
\end{equation}
In the same vein,
\begin{equation}
    \varphi_2(t)=\beta\frac{g^2}{\omega_0},
\end{equation}
while
\begin{equation}
    Z_\mathrm{e}=Ne^{-\beta g^2/\omega_0},
\end{equation}
where the factor $N$ comes from the trace of the unit matrix that replaces the operator $e^{-\beta H_\mathrm{e}}$ in Eq.~\textcolor{red}{(A12) of Appendix~A.}
We thus see that the action of $\varphi_2(\beta)$ is cancelled by the normalization $Z_\mathrm{e}$.
The result for $\varphi_3(t,\beta)$ reads as
\begin{equation}
    \varphi_3(t,\beta)=\left[\sum_q V_q^C\:^CV_{-q}\right]\left[\sum_m\frac{c_m}{\mu_m^2}(e^{-\mu_m t}-1)(e^{i\beta\mu_m}-1)\right].
\end{equation}
Since $\varphi_3(t,\beta)$ acts on the unit operator in Eq.~\textcolor{red}{(A14) of Appendix~A}, it effectively acts as a scalar ($V_q\mathbbm{1}_{1e}V_{-q}=V_qV_{-q}=\mathbbm{1}_{1e}$) whose
value reads as
\begin{equation}
\label{Eq:varphi_3_final_action}
    \varphi_3(t,\beta)=\frac{g^2}{\omega_0^2}\left(e^{-i\omega_0t}-e^{i\omega_0t}\right).
\end{equation}
We finally obtain
\begin{equation}
  C^\mathrm{bbl}_{jj}(t)=\frac{1}{N}\sum_k j_k^2\:e^{-\varphi_1(t)-\varphi_1(t)^*-\varphi_3(t,\beta)^*}. 
\end{equation}
Since the terms in the exponent do not depend on $k$, one uses
\begin{equation}
N^{-1}\sum_k j_k^2=N^{-1}\mathrm{Tr}_\mathrm{1e}\{j^2\}=2t_0^2
\end{equation}
to finally arrive at
\begin{equation}
    C^\mathrm{bbl}_{jj}(t)=2t_0^2\:e^{-\varphi_1(t)-\varphi_1(t)^*-\varphi_3(t,\beta)^*}.
\end{equation}

We now turn to $C_{jj}(t)$, starting from
\begin{equation}
\label{Eq:Phi_1_full_action}
\begin{split}
    \Phi_1(t)=&\frac{g^2}{\omega_0^2}\coth\left(\frac{\beta\omega_0}{2}\right)\left[1-\cos(\omega_0t)\right]\left(\frac{1}{N}\sum_q V_q^\times\:V_{-q}^\times\right)+\\
    &\frac{g^2}{\omega_0^2}\left[\sin(\omega_0t)-i\omega_0t\right]\left(\frac{1}{N}\sum_q V_q^\times\:V_{-q}^\circ\right).
\end{split}
\end{equation}
The action of the hyperoperator $\frac{1}{N}\sum_q V_q^\times\:V_{-q}^\times$ on the current operator $j$ reduces to
\begin{equation}
    \frac{1}{N}\sum_q V_q^\times\:V_{-q}^\times j=2j
\end{equation}
because of
\begin{equation}
\label{Eq:varphi_3_action_bubble}
    \frac{1}{N}\sum_q V_q jV_{-q}=\frac{1}{N}\sum_{qp} j_{p-q}|p\rangle\langle p|=0,
\end{equation}
which follows from $N^{-1}\sum_q e^{iq}=0$ (under PBCs).
In other words, one can replace $\frac{1}{N}\sum_q V_q^\times\:V_{-q}^\times$ by 2 in Eq.~\eqref{Eq:Phi_1_full_action}.
Along the same line, we conclude that $\frac{1}{N}\sum_q V_q^\times\:V_{-q}^\circ$ can be replaced by 0 in Eq.~\eqref{Eq:Phi_1_full_action}.
We can thus replace the original $\Phi_1(t)$ by the scalar that reads as
\begin{equation}
\label{Eq:Phi_1_atomic_full}
    \Phi_1(t)=\frac{g^2}{\omega_0^2}
    (2n_\mathrm{ph}+1)\left[2-
    e^{i\omega_0t}-e^{-i\omega_0t}\right].
\end{equation}
In a similar manner,
\begin{equation}
    \Phi_3(t,\beta)=-2\frac{g^2}{\omega_0^2}i\sin(\omega_0t)\left(\frac{1}{N}\sum_q V_q^\times\:^CV_{-q}\right).
\end{equation}
Using the same reasoning as above, we find that
\begin{equation}
    \frac{1}{N}\sum_q V_q^\times\:^CV_{-q} j=-j,
\end{equation}
so that the influence phase $\Phi_3(t,\beta)$ effectively acts as the following scalar:
\begin{equation}
\label{Eq:Phi_3_atomic_full}
    \Phi_3(t,\beta)=\frac{g^2}{\omega_0^2}\left(e^{i\omega_0t}-e^{-i\omega_0t}\right).
\end{equation}
Collecting all pieces together, we remain with
\begin{equation}
\begin{split}
    &C_{jj}(t)=\frac{1}{N}\mathrm{Tr}_\mathrm{e}\{j^2\}
    e^{-\Phi_1(t)-\Phi_3(t,\beta)}.
\end{split}
\end{equation}
The equality $C_{jj}(t)=C_{jj}^\mathrm{bbl}(t)$ then follows from Eqs.~\eqref{Eq:varphi_1_final_action},~\eqref{Eq:varphi_3_final_action},~\eqref{Eq:Phi_1_atomic_full}, and~\eqref{Eq:Phi_3_atomic_full}.

\newpage

\section{Summary of the parameter regimes examined}
\begin{table*}[htbp!]
\begin{center}
\begin{tabular}{|l||*{5}{c|}}\hline
\backslashbox{$(\omega_0/t_0,\lambda)$}{Method} 
& HEOM & QMC & \shortstack{HEOM\\ bubble} & \shortstack{QMC\\ bubble} & DMFT \\\hline\hline
$\left(1,\frac{1}{100}\right)$ & $[1,10]$ & $[1,10]$ & $[1,10]$ & $[1,10]$ & $[1,10]$\\\hline
$\left(1,\frac{1}{8}\right)$ & $[1,10]$ & $[1,10]$ & $[1,10]$& $[1,10]$ & $[1,10]$\\\hline
$\left(1,\frac{1}{2}\right)$ & $[0.4,10]$ & $[0.2,10]$ & $[0.4,10]$ & $[0.2,10]$ & $[0.2,10]$\\\hline
$\left(1,1\right)$ & $[2,10]$ & $[1,10]$ & $[2,10]$ & $[1,10]$ & $[1,10]$\\\hline
$\left(1,2\right)$ & $\times$ & $[1,10]$ & $\times$ & $[1,10]$ & $[1,10]$\\\hline
$\left(\frac{1}{3},\frac{1}{100}\right)$ & $[1,10]$ & $[1,10]$ & $[1,10]$ & $[1,10]$ & $[1,10]$\\\hline
$\left(\frac{1}{3},\frac{1}{8}\right)$ & $[1,10]$ & $[1,10]$ & $[1,10]$ & $[1,10]$ & $[1,10]$\\\hline
$\left(\frac{1}{3},\frac{1}{2}\right)$ & $[1,10]$ & $[0.1,10]$ & $[1,10]$ & $[0.1,10]$ & $[0.1,10]$\\\hline
$\left(\frac{1}{3},1\right)$ & $[1,5]$ & $[1,10]$ & $[1,5]$ & $[1,10]$ & $[1,10]$\\\hline
$\left(\frac{1}{3},2\right)$ & $\times$ & $[1,10]$ & $\times$ & $[1,10]$ & $[1,10]$\\\hline
$\left(3,\frac{1}{8}\right)$ & $[5,10]$ & $[2,10]$ & $[5,10]$ & $[2,10]$ & $[2,10]$\\\hline
$\left(3,\frac{1}{2}\right)$ & $[2,10]$ & $[1,10]$ & $[2,10]$ & $[1,10]$ & $[1,10]$\\\hline
$\left(3,1\right)$ & $[2,10]$ & $[1,10]$ & $[2,10]$ & $[1,10]$ & $[1,10]$\\\hline
$\left(3,2\right)$ & $\times$ & $[1,10]$ & $\times$ & $[1,10]$ & $[1,10]$\\\hline
\end{tabular}
\caption{Summary of parameter regimes [determined by pairs $(\omega_0/t_0,\lambda)$] and numerical methods that are used to assess the importance of vertex corrections.
For each parameter regime and each method, we provide the minimum ($T_\mathrm{min}/t_0$) and maximum ($T_\mathrm{max}/t_0$) temperature at which we performed computations.
The choice of parameter values is largely dictated by the feasibility of HEOM and HEOM bubble computations.
We emphasize that the DMFT results for the dynamical mobility are available for all values of $\omega_0/t_0,\lambda,$ and $T$, and the same applies to QMC and QMC bubble results for short-time dynamics of $C_{jj}$.
For $\lambda=2$, HEOM and HEOM bubble computations could be performed only on short time scales [comparable to those accessible by QMC (bubble), see also Sec.~III.F of Ref.~\onlinecite{JChemPhys.159.094113}], which is indicated by "$\times$".}
\label{Tab:parameter_regimes}
\end{center}
\end{table*}

\newpage

\section{Extensive comparisons of HEOM and DMFT results}
\subsection{Intermediate-frequency phonons ($\omega_0/t_0=1$)}
\begin{figure}[htbp!]
    \centering
    \includegraphics[width=\textwidth]{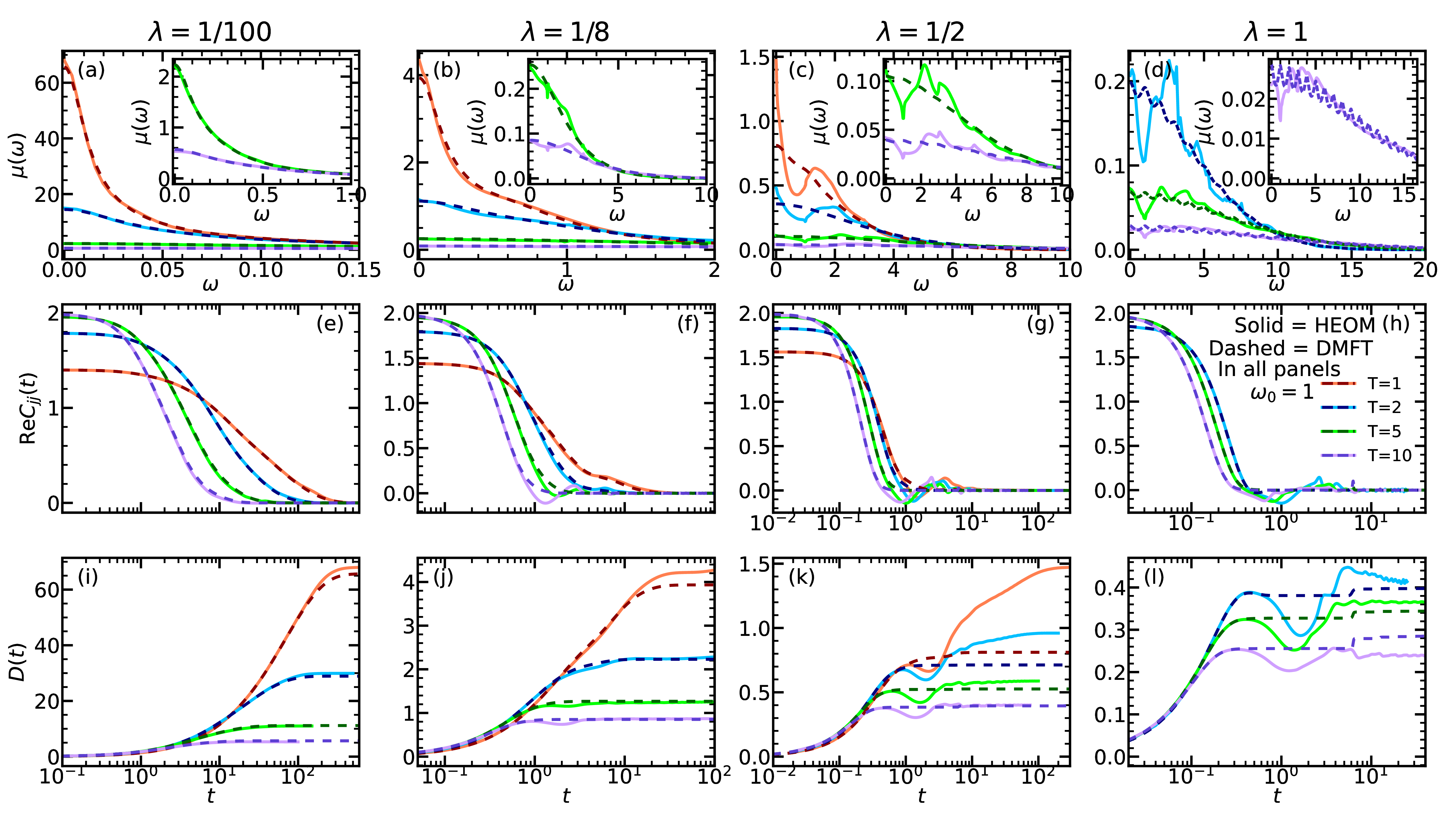}
    \caption{Comparison of HEOM (solid lines) and DMFT (dashed lines) results for (a)--(d) the dynamical-mobility profile, (e)--(h) the real part of the current--current correlation function $C_{jj}(t)$, and (i)--(l) the diffusion constant $\mathcal{D}(t)$. In all panels, $t_0=\omega_0=1$. The strength of the electron--phonon interaction is determined by the cited values of $\lambda$, while the temperatures are $T=1,2,5,$ and 10. The insets in panels (a)--(d) zoom in the dynamical-mobility profiles for the highest temperatures considered ($T=5$ and 10).}
    \label{Fig:w=1}
\end{figure}
\newpage

\subsection{Slow phonons ($\omega_0/t_0=1/3$)}
\begin{figure}[htbp!]
    \centering
    \includegraphics[width=\textwidth]{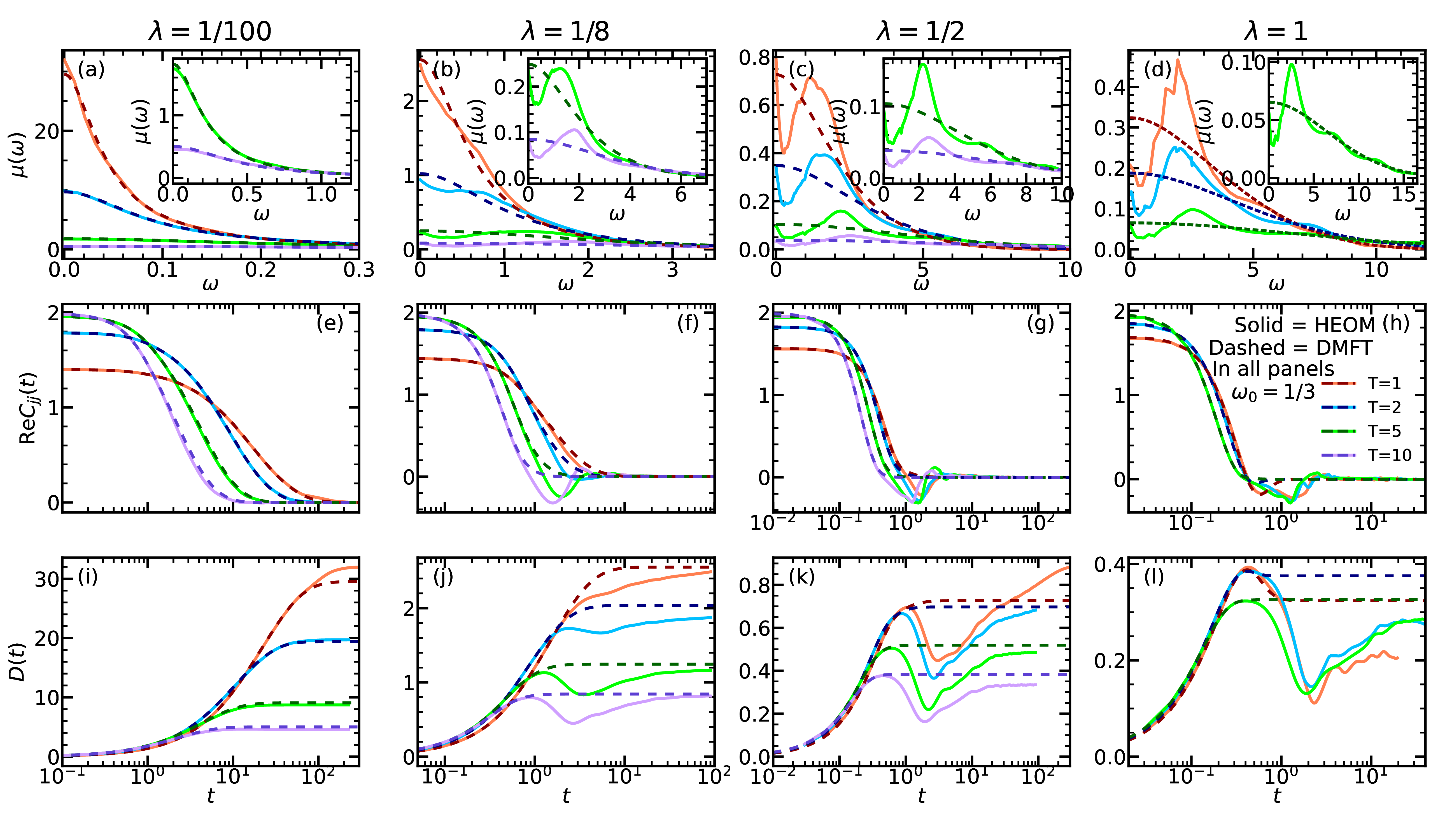}
    \caption{Comparison of HEOM (solid lines) and DMFT (dashed lines) results for (a)--(d) the dynamical-mobility profile, (e)--(h) the real part of the current--current correlation function $C_{jj}(t)$, and (i)--(l) the diffusion constant $\mathcal{D}(t)$. In all panels, $t_0=1,\omega_0=1/3$. The strength of the electron--phonon interaction is determined by the cited values of $\lambda$, while the temperatures are $T=1,2,5,$ and 10. The insets in panels (a)--(d) zoom in the dynamical-mobility profiles for the highest temperatures considered ($T=5$ and 10).}
    \label{Fig:w=0_333}
\end{figure}
\newpage

\subsection{Fast phonons ($\omega_0/t_0=3$)}
\begin{figure}[htbp!]
    \centering
    \includegraphics[width=0.9\textwidth]{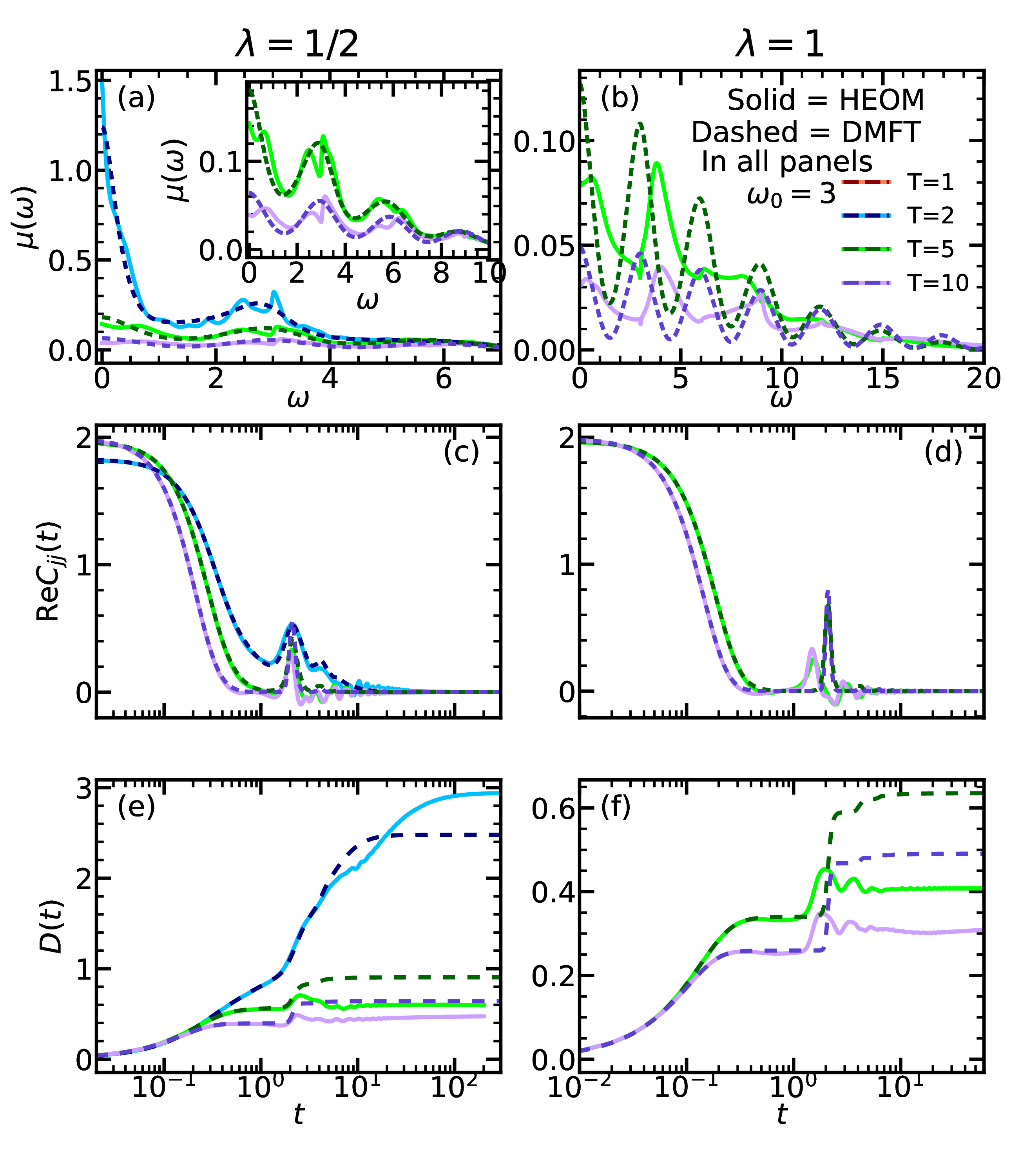}
    \caption{Comparison of HEOM (solid lines) and DMFT (dashed lines) results for (a) and (b) the dynamical-mobility profile, (c) and (d) the real part of the current--current correlation function $C_{jj}(t)$, and (e) and (f) the diffusion constant $\mathcal{D}(t)$. In all panels, $t_0=1,\omega_0=3$. The strength of the electron--phonon interaction is determined by the cited values of $\lambda$, while the temperatures are $T=2,5,$ and 10. The inset in panel (a) zooms in the dynamical-mobility profile for the highest temperatures considered ($T=5$ and 10).}
    \label{Fig:w=3}
\end{figure}
\newpage

\section{Real-time QMC in the limit $\omega_0\to 0$}

Phonon momentum can be neglected in the adiabatic limit. For this reason QMC methodology becomes much simpler in this case and can be efficiently performed to obtain the relevant quantities in real time. In this section, we first derive the relevant equations and then present selected results for $C_{jj}\qty(t)$ and $\mathcal{D}\qty(t)$ in the adiabatic limit.

The Holstein Hamiltonian in this limit takes the form
\begin{equation}
H=\sum_{ij} h_{ij}\qty(\qty{q})c_i^\dagger c_j+\sum_i\frac{1}{2}m\omega_0^2q_i^2,
\end{equation}
where
\begin{equation}
h_{ij}\qty(\qty{q})=-t_0\qty(\delta_{i,j+1}+\delta_{i,j-1})+g\sqrt{2m\omega_0}q_i\delta_{ij}.
\end{equation}
The operator $c_i$ is the electron annihilation operator at site $i$, $q_i$ is the coordinate of 
the phonon at site $i$, $m$ is the oscillator mass, the symbol $\qty{q}$ denotes all phonon 
coordinates, while phonon momenta $p_i$ and the corresponding kinetic energy $\frac{p_i^2}{2m}$ were neglected. 
With the substitution of variables $x_i=q_i\omega_0\sqrt{m}$, the Hamiltonian reduces to
\begin{equation}
H=\sum_{ij} h_{ij}\qty(\qty{x})c_i^\dagger c_j+\sum_i\frac{1}{2}q_i^2,
\end{equation}
with
\begin{equation}
h_{ij}\qty(\qty{x})=-t_0\qty(\delta_{i,j+1}+\delta_{i,j-1})+2\sqrt{\lambda t_0}q_i\delta_{ij}.
\end{equation}
We evaluate the correlation function \textcolor{red}{[see Eq. (10) of the main paper]}
\begin{equation}
C_{jj}\qty(t)=\frac{\Tr\qty(e^{-z_1 H} j e^{-z_2 H}j) }{\Tr e^{-\beta H}}
\label{eq:at177cjj}
\end{equation}
(with $z_1=\beta-\mathrm{i}t$, $z_2=\mathrm{i}t$) by expressing the trace in the basis $\ket{\qty{x}n_{\qty{x}}}$, where $\ket{n_{\qty{x}}}$ are the eigenstates of $h\qty(\qty{x})$ with phonon coordinates $\qty{x}$ treated as classical variables.
The matrix element of the Hamiltonian in this basis reads
\begin{equation}
 \mel{\qty{x}n_{\qty{x}}}{H}{\qty{y}m_{\qty{y}}}
 =
 \delta\qty(\qty{x}-\qty{y})
 \delta_{mn}
 \qty[\varepsilon_m\qty(\qty{x})+\sum_i\frac{1}{2}x_i^2],
\end{equation}
where $\varepsilon_m\qty(\qty{x})$ are the eigenvalues of $h\qty(\qty{x})$. Consequently, we find
\begin{equation}
 \mel{\qty{x}n_{\qty{x}}}{e^{-zH}}{\qty{y}m_{\qty{y}}}
 =
 \delta\qty(\qty{x}-\qty{y})
 \delta_{mn}
 e^{-z\qty[\varepsilon_m\qty(\qty{x})+\sum_i\frac{1}{2}x_i^2]}.
\end{equation}
The trace in the numerator in Eq.~\eqref{eq:at177cjj} reads
\begin{equation}
 \Tr\qty(e^{-z_1 H} j e^{-z_2 H}j) =
 \int \dd\qty{x} \sum_{m_{\qty{x}}}
 \mel{\qty{x} m_{\qty{x}}}
 {e^{-z_1 H} j e^{-z_2 H}j}
 {\qty{x} m_{\qty{x}}}
\end{equation}
which leads to
\begin{equation}
 \Tr\qty(e^{-z_1 H} j e^{-z_2 H}j) =
 \int \dd\qty{x} \sum_{m_{\qty{x}}n_{\qty{x}}}
 \mel{\qty{x} m_{\qty{x}}}
 {e^{-z_1 H} j }
 {\qty{x} n_{\qty{x}}}
 \mel{\qty{x} n_{\qty{x}}}
 {e^{-z_2 H} j }
 {\qty{x} m_{\qty{x}}}
\end{equation}
and eventually
\begin{equation}
\begin{split}
 \Tr\qty(e^{-z_1 H} j e^{-z_2 H}j) =
 \int \dd\qty{x}
 e^{-\beta\sum_i\frac{1}{2}x_i^2}
 & \sum_{m_{\qty{x}}n_{\qty{x}}}
 e^{-\beta\varepsilon_{m_{\qty{x}}}}
 e^{\mathrm{i}t\qty(\varepsilon_{m_{\qty{x}}}-\varepsilon_{n_{\qty{x}}})} \times\\
 &\mel{m_{\qty{x}}}{j }{n_{\qty{x}}}
 \mel{n_{\qty{x}}}{j }{m_{\qty{x}}},
 \end{split}
 \label{eq:at182}
\end{equation}
as well as
\begin{equation}
 \Tr\qty(e^{-\beta H}) =
 \int \dd\qty{x}
 e^{-\beta\sum_i\frac{1}{2}x_i^2}
 \sum_{m_{\qty{x}}}
 e^{-\beta\varepsilon_{m_{\qty{x}}}}.
 \label{eq:at183}
\end{equation}
The correlation function given in Eq.~\eqref{eq:at177cjj} can then be evaluated by sampling the phonon coordinates as Gaussians with standard deviation $\sigma^2=\frac{1}{\beta}$ and performing the summation of the terms in the numerator and denominator. These summations are much less demanding than the summations in full Monte Carlo simulations because the number of phonon coordinates in this case is equal to the number of sites, while in full Monte Carlo simulations it is equal to the number of sites times the number of timesteps.
The summations in Eqs.~\eqref{eq:at182} and \eqref{eq:at183} can be interpreted as averages over classical phonon coordinates $\qty{x}$. One should, however, note that the assumption of classical phonons was not introduced in the derivation. It is the neglect of phonon momentum that led to the expression which can be interpreted this way.

Next, we also give expression for the quantities $\expval{N_{\mathrm{e}}}_{\mathrm{K}}$, $\mathcal{G}^<(k,t)$ and $\mathcal{G}^>(k,t)$ which are needed to evaluate $C_{jj}^\mathrm{bbl}(t)$ in accordance with \textcolor{red}{Eq.~(17) of the main paper}. We obtain
\begin{equation}
 \mathcal{G}^>(k,t)
 =-ie^{\mathrm{i}\mu_{\mathrm{F}}t}
 \frac{
  \int \dd\qty{x}
 e^{-\beta\sum_i\frac{1}{2}x_i^2}
 \sum_{m_{\qty{x}}}
 e^{-\mathrm{i}\varepsilon_{m_{\qty{x}}}t}
 \qty|c_{mk}\qty(\qty{x})|^2
 }
 { \int \dd\qty{x}
 e^{-\beta\sum_i\frac{1}{2}x_i^2}},
 \label{eq:tr185_sm}
\end{equation}
\begin{equation}
 \mathcal{G}^<(k,t)
 =ie^{\qty(\beta+\mathrm{i}t)\mu_{\mathrm{F}}}
 \frac{
  \int \dd\qty{x}
 e^{-\beta\sum_i\frac{1}{2}x_i^2}
 \sum_{m_{\qty{x}}}
 e^{-\qty(\beta+\mathrm{i}t)\varepsilon_{m_{\qty{x}}}}
 \qty|c_{mk}\qty(\qty{x})|^2
 }
 { \int \dd\qty{x}
 e^{-\beta\sum_i\frac{1}{2}m\omega_0^2x_i^2}},
 \label{eq:tr186}
\end{equation}
\begin{equation}
\expval{N_{\mathrm{e}}}_{\mathrm{K}}=
e^{\beta\mu_{\mathrm{F}}}
 \frac{
  \int \dd\qty{x}
 e^{-\beta\sum_i\frac{1}{2}x_i^2}
 \sum_{m_{\qty{x}}}
 e^{-\beta\varepsilon_{m_{\qty{x}}}} 
 }
 { \int \dd\qty{x}
 e^{-\beta\sum_i\frac{1}{2}x_i^2}},
\end{equation}
where $c_{mk}\qty(\qty{x})$ is the overlap of the electronic state of momentum $\ket{k}$ and the electronic state $\ket{m_{\qty{x}}}$ given as
$c_{mk}\qty(\qty{x})=\braket{k}{m_{\qty{x}}}$.

\begin{figure}
    \centering
    \includegraphics[width=0.5\textwidth]{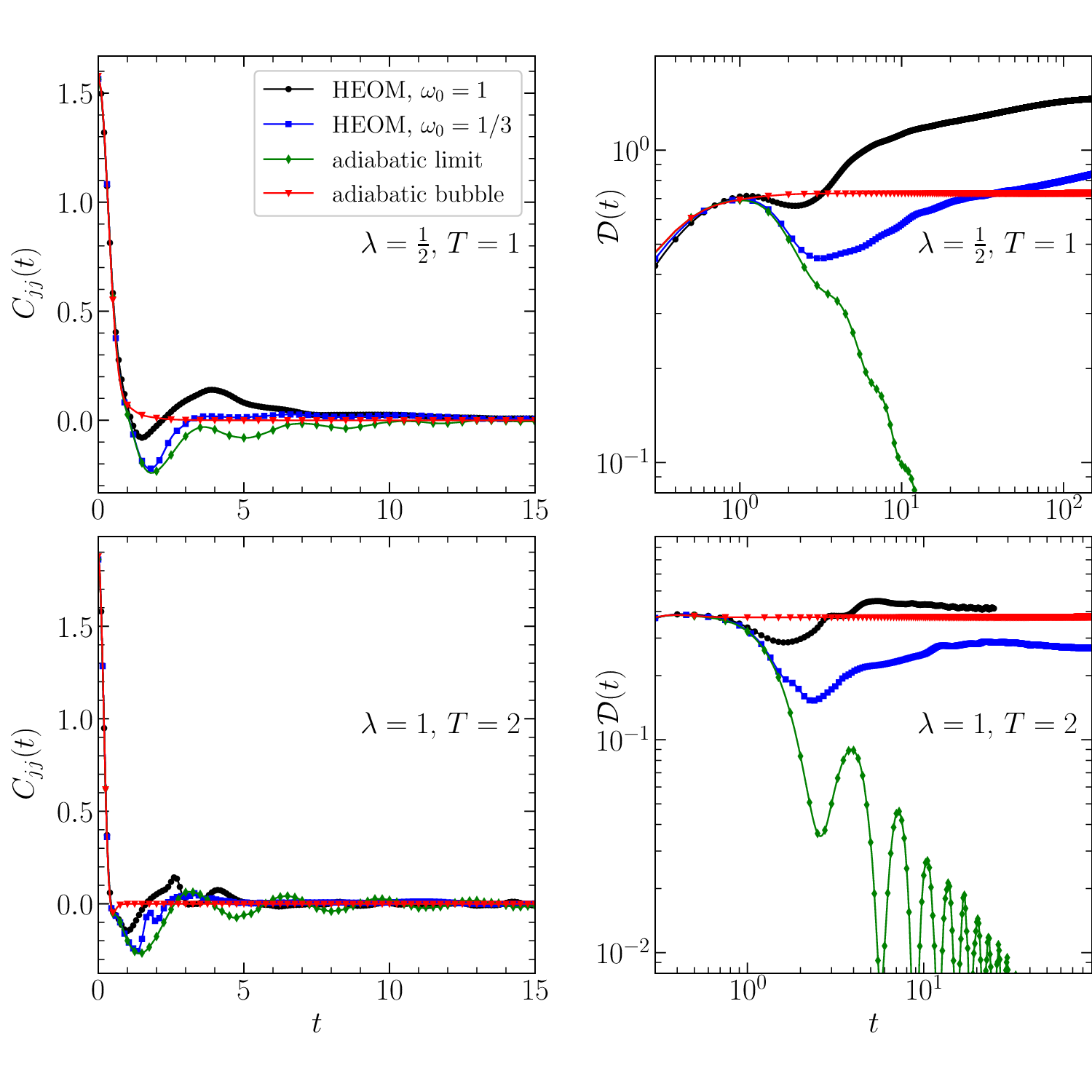}
    \caption{The current-current correlation function $C_{jj}\qty(t)$ and the diffusion constant $\mathcal{D}\qty(t)$ for two parameters sets $\qty(\lambda=\frac{1}{2},\: T=1)$ and $\qty(\lambda=1,\: T=2)$ ($t_0=1$ in both cases). The results labeled as 'HEOM' denote the numerically exact results obtained using the hierarchical equations of motion method for two values of phonon frequncies $\omega_0=1$ and $\omega_0=1/3$, the results labeled as 'adiabatic limit' denoted the results obtained in the adiabatic limit using the methodology described in this section, while the results labeled as 'adiabatic bubble' are the results obtained with the bubble approximation in the adiabatic limit.}
    \label{Fig:fig-al}
\end{figure}

In Fig.~\ref{Fig:fig-al} we present $C_{jj}\qty(t)$ and $\mathcal{D}\qty(t)$ for two parameters sets $\qty(\lambda=\frac{1}{2},\: T=1,\: t_0=1)$ and $\qty(\lambda=1,\: T=2,\: t_0=1)$. The numerically exact results obtained using HEOM are presented, as well as the results in the adiabatic limit obtained as described in this section. As expected, the exact results for $\omega_0=1/3$ are closer to the adiabatic limit results than the exact results for $\omega_0=1$. Nevertheless, both of these sets of results are still quite far from the adiabatic limit results. As discussed in the main paper, this is consistent with the fact that vertex corrections for these parameter sets are not very strong. We also present the bubble approximation results in the adiabatic limit. These results which yield a non-zero mobility strongly differ from the exact results which give a zero mobility, in accordance with expectations.

\newpage

\section{Comparison of the results for dc mobility obtained from real-time and imaginary-time computations}
\begin{figure}[htbp!]
    \centering
    \includegraphics[width=0.5\columnwidth]{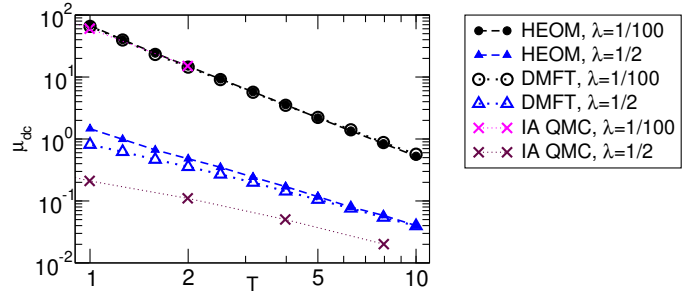}
    \caption{Comparison of HEOM (full symbols), DMFT (empty symbols), and imaginary-axis QMC (crosses) results of Ref.~\onlinecite{PhysRevLett.114.146401} for the temperature dependent dc mobility. The model parameters are $t_0=\omega_0=1$ and $\lambda=1/100$ and $1/2$.}
    \label{Fig:mishchenko}
\end{figure}

Figure~\ref{Fig:mishchenko} compares some of our HEOM and DMFT results for $\mu_\mathrm{dc}$, both of which follow from real-axis computations, with the corresponding results of Ref.~\onlinecite{PhysRevLett.114.146401}, which were extracted from imaginary-axis data using numerical analytical continuation.
While all the results are virtually the same in the weak-interaction regime $\lambda=1/100$, the results of Ref.~\onlinecite{PhysRevLett.114.146401} seem to severely underestimate $\mu_\mathrm{dc}$ (by approximately an order of magnitude) in the intermediate-interaction regime $\lambda=1/2$.

\newpage

\bibliography{apssamp}